\documentclass[ oneside,openright,titlepage,numbers=noenddot,headinclude,
                footinclude=true,cleardoublepage=empty,abstractoff, 
                BCOR=5mm,paper=a4,fontsize=11pt,
                ngerman,american,%
                ]{scrreprt}
\usepackage{scalerel}
\usepackage{setspace}

\PassOptionsToPackage{eulerchapternumbers,listings,%
				 pdfspacing,
				 subfig,beramono,parts}{classicthesis}										

\usepackage{ifthen}
\newboolean{enable-backrefs} 
\setboolean{enable-backrefs}{false} 

\newcommand{\myTitle}{Probing the central engine and early emission of Gamma-Ray Bursts}

\newcommand{\myName}{Vidushi Sharma\xspace}

\newcommand{\myFaculty}{Put data here\xspace}

\newcommand{\myUni}{Put data here\xspace}

\newcommand{\myVersion}{version 4.1\xspace}

\newcommand{\ber}{\begin{eqnarray}}
\newcommand{\eer}{\end{eqnarray}}

\def\beq{\begin{equation}}
\def\eeq{\end{equation}}
\def\ber{\begin{eqnarray}}
\def\eer{\end{eqnarray}}
\def\benu{\begin{enumerate}}
\def\eenu{\end{enumerate}}

\def\sq{\lower.25ex\hbox{\large$\Box$}}
\def \lleq {\lower0.9ex\hbox{ $\buildrel < \over \sim$} ~}
\def \ggeq {\lower0.9ex\hbox{ $\buildrel > \over \sim$} ~}

\def\om0{\Omega_{0m}}

\PassOptionsToPackage{latin9}{inputenc}	
 \usepackage{inputenc}				

 \usepackage{babel}					

\PassOptionsToPackage{square,numbers}{natbib}
 \usepackage{natbib}				

\PassOptionsToPackage{fleqn}{amsmath}		
 \usepackage{amsmath}

\PassOptionsToPackage{T1}{fontenc} 
	\usepackage{fontenc}     
\usepackage{textcomp} 
\usepackage{scrhack} 
\usepackage{xspace} 
\usepackage{mparhack} 
\usepackage{fixltx2e} 
\usepackage{setspace} 
\PassOptionsToPackage{printonlyused,smaller}{acronym}
	\usepackage{acronym} 
\usepackage[parfill]{parskip}    
\usepackage{graphicx}
\usepackage{epstopdf}
\usepackage{hyperref}
\usepackage{mathrsfs}
\usepackage{paralist}
\usepackage{varioref}  
\usepackage{multirow}

\usepackage{lscape}
\usepackage{booktabs}
\usepackage{longtable}
\usepackage{supertabular,booktabs}
\usepackage{adjustbox}

\usepackage[bottom]{footmisc}
\usepackage[caption=false]{subfig}

\usepackage[normalem]{ulem}
\usepackage{float}
\usepackage{epsfig}
\usepackage{epstopdf}
\usepackage{hyperref}
\hypersetup{colorlinks=True,citecolor=blue}
\usepackage{amsfonts}
\usepackage{amssymb}
\usepackage{caption}
\usepackage{bm}
\usepackage{graphicx}
\usepackage{color}
\usepackage{verbatim}
\usepackage{blindtext, multicol}
\usepackage{setspace}
\usepackage{multirow}

\usepackage{textcomp}
\usepackage{enumerate}

\usepackage{titlesec}

\usepackage{color}
\usepackage[normalem]{ulem}

\expandafter\ifx\csname package@font\endcsname\relax\else
\expandafter\expandafter
\expandafter\usepackage
\expandafter\expandafter
\expandafter{\csname package@font\endcsname}%
\fi
\usepackage{graphics}
\usepackage{comment}
\includecomment{comment}
\titleclass{\subsubsubsection}{straight}[\subsection]

\newcounter{subsubsubsection}[subsubsection]
\renewcommand\thesubsubsubsection{\thesubsubsection.\arabic{subsubsubsection}}
\renewcommand\theparagraph{\thesubsubsubsection.\arabic{paragraph}} 

\titleformat{\subsubsubsection}
  {\normalfont\normalsize\bfseries}{\thesubsubsubsection}{1em}{}
\titlespacing*{\subsubsubsection}
{0pt}{3.25ex plus 1ex minus .2ex}{1.5ex plus .2ex}

\makeatletter
\renewcommand\paragraph{\@startsection{paragraph}{5}{\z@}%
  {3.25ex \@plus1ex \@minus.2ex}%
  {-1em}%
  {\normalfont\normalsize\bfseries}}
\renewcommand\subparagraph{\@startsection{subparagraph}{6}{\parindent}%
  {3.25ex \@plus1ex \@minus .2ex}%
  {-1em}%
  {\normalfont\normalsize\bfseries}}
\def\toclevel@subsubsubsection{4}
\def\toclevel@paragraph{5}
\def\toclevel@paragraph{6}
\def\l@subsubsubsection{\@dottedtocline{4}{7em}{4em}}
\def\l@paragraph{\@dottedtocline{5}{10em}{5em}}
\def\l@subparagraph{\@dottedtocline{6}{14em}{6em}}
\makeatother

\usepackage{tabularx} 
\usepackage{caption}
\usepackage{subfig}  

\usepackage{listings} 
\PassOptionsToPackage{pdftex,hyperfootnotes=false,pdfpagelabels}{hyperref}
	\usepackage{hyperref}  
\PassOptionsToPackage{pdftex}{graphicx}
	\usepackage{graphicx} 

\newcommand{\backrefnotcitedstring}{\relax}
\newcommand{\backrefcitedsinglestring}[1]{(Cited on page~#1.)}
\newcommand{\backrefcitedmultistring}[1]{(Cited on pages~#1.)}
\ifthenelse{\boolean{enable-backrefs}}%
{%
		\PassOptionsToPackage{hyperpageref}{backref}
		\usepackage{backref} 
		   \renewcommand*{\backref}[1]{}  
		   \renewcommand*{\backrefalt}[4]{
		      \ifcase #1 %
		         \backrefnotcitedstring%
		      \or%
		         \backrefcitedsinglestring{#2}%
		      \else%
		         \backrefcitedmultistring{#2}%
		      \fi}%
}{\relax}    

\hypersetup{%
    colorlinks=true, linktocpage=true, pdfstartpage=3, pdfstartview=FitV,%
    breaklinks=true, pdfpagemode=UseNone, pageanchor=true, pdfpagemode=UseOutlines,%
    plainpages=false, bookmarksnumbered, bookmarksopen=true, bookmarksopenlevel=1,%
    hypertexnames=true, pdfhighlight=/O,
    urlcolor=Red, linkcolor=RoyalBlue, citecolor=webgreen, 
    pdftitle={\myTitle},%
    pdfauthor={\textcopyright\ \myName, \myUni, \myFaculty},%
    pdfsubject={},%
    pdfkeywords={},%
    pdfcreator={pdfLaTeX},%
    pdfproducer={LaTeX with hyperref and classicthesis}%
}   

\makeatletter
\@ifpackageloaded{babel}%
    {%
       \addto\extrasamerican{%
				}%
       \addto\extrasngerman{%
				}%
    }{\relax}
\makeatother

\usepackage{classicthesis} 
\usepackage{lmodern}
\usepackage{tikz}
\usepackage{threeparttable}
\usetikzlibrary{calc}
\usetikzlibrary{mindmap}
\usepackage{pifont}

\ifthenelse{\boolean{@linedheaders}}%
    {
    \titleformat{\chapter}[display]%
        {\relax}
        {\raggedleft{\color{halfgray}\chapterNumber\thechapter} \\ }
        {0pt}%
        {\color{maroon!70!black}\huge\titlerule\vspace*{.9\baselineskip}\raggedright\spacedallcaps}
        [\normalcolor\normalsize\vspace*{.8\baselineskip}\titlerule]%
    }
    {
    \titleformat{\chapter}[display]%
        {\relax}
        {\mbox{}\oldmarginpar{\vspace*{-3\baselineskip}\color{halfgray}\chapterNumber\thechapter}}
        {0pt}%
        {\color{blue!70!black}\huge\raggedright\spacedallcaps}
        [\normalcolor\normalsize\vspace*{.8\baselineskip}\titlerule]%
    }

\begin{document}

\frenchspacing 

\raggedbottom 

\selectlanguage{american} 


\pagenumbering{roman} 

\pagestyle{plain} 

\begin{titlepage}
	\begin{addmargin}[-2cm]{-1.5cm}
	
	\begin{tikzpicture}[remember picture, overlay]
  \draw[line width = 2.4pt] ($(current page.north west) + (1in,-1in)$) rectangle ($(current page.south east) + (-1in,1in)$); 
        \end{tikzpicture} 
        \begin{tikzpicture}[remember picture, overlay]
  \draw[line width = 1.2pt] ($(current page.north west) + (1.05in,-1.05in)$) rectangle ($(current page.south east) + (-1.05in,1.05in)$);
    \end{tikzpicture}
    
    \begin{center}
        \large  
        \hfill
        \vfill

        \begingroup
        \begin{center}
            \color{midnightblue}\spacedallcaps{\Large {\bf Study of Compact Stars}\xspace} 
        \end{center} 
        \endgroup
        \begingroup
        \begin{center}
            \color{midnightblue}\spacedallcaps{\Large \bf{ and their properties based on}\xspace} 
         \end{center}    
        \endgroup  
        \begingroup
        \begin{center}
            \color{midnightblue}\spacedallcaps{\Large{\bf  General Relativistic Core-Envelope Models}\xspace} 
         \end{center}    
        \endgroup  

        \medskip
        \bigskip
        \bigskip
        \bigskip
        \textls[90]{\scshape{{\LARGE \textbf{Ankitkumar Chaturbhai Khunt}}}}
        \medskip
        \bigskip
        \bigskip
        \bigskip
        \begingroup
        \begin{center}
            {\textbf{Thesis}\xspace} 
        \end{center} 
        \endgroup
        \begingroup
        \begin{center}
            {submitted for the degree of\xspace} 
        \end{center}
        \endgroup  
        \begingroup
        \begin{center}
            {\textit{\textbf{Doctor of Philosophy} }(in Physics)\xspace} 
         \end{center}    
        \endgroup  
        \begin{center}
         to
        \end{center}
        \begin{center}
         {\Large Sardar Patel University}
         
        \end{center}

        \bigskip

    \begin{center}
        Under the supervision of 
    \end{center} 
    \begin{center}
        Prof. P. C. Vinodkumar
    \end{center}
        \bigskip
        \bigskip
        \bigskip

        \vfill
    \bigskip


        \begin{center}
        Department of Physics, Sardar Patel University 
        \end{center}
        \begin{center}
        Vallabh Vidyanagar -388120, Gujarat (India)
        \end{center}
        \begin{center}
            July 2023
        \end{center}
        \medskip
        \bigskip
        \bigskip
        \bigskip
        \bigskip
        \bigskip
        \bigskip
        \bigskip

    \end{center}  
  \end{addmargin}       
\end{titlepage}   
\setstretch{1.5}
\cleardoublepage

\begin{titlepage}
  
\
\vfil
\hfil \Large \textit{To all who struggle to understand this strange world}\hfil
\vfil



\end{titlepage}   
\cleardoublepage
\pdfbookmark[1]{Certificate}{Certificate}


\bigskip

\begingroup
\let\clearpage\relax
\let\cleardoublepage\relax
\let\cleardoublepage\relax

\,\;$\;$
\bigskip
\bigskip
\bigskip
\bigskip
\bigskip
\bigskip
\bigskip

\chapter*{ Certificate}

This is to certify that the thesis entitled `\textbf{Study of Compact Stars and their Properties based on General Relativistic Core-Envelope Models}' submitted by Mr. Ankitkumar Chaturbhai Khunt for the award of the degree of Doctor of Philosophy of Sardar Patel University, Vallabh Vidyanagar is his
original work. This has not been published in or submitted to any other University for any
other Degree or Diploma.


\bigskip
\bigskip
\bigskip 
\bigskip
\bigskip 
  

%
\bigskip

\def\s#1#2{\vbox{\hsize=4.5cm
    \kern2cm
    \hrule\kern2ex
    \hbox to \hsize{\strut\hfil #1 \hfil}
    \hbox to \hsize{\strut\hfil #2 \hfil}}}

\hbox to \hsize{\s{Prof. V. M. Pathak}{(Head, Department of Physics)}\hfil
    \s{Prof. P. C. Vinodkumar}{(Thesis Supervisor)}}



\endgroup
\cleardoublepage
\pdfbookmark[1]{Declaration}{Declaration}


\bigskip

\begingroup
\let\clearpage\relax
\let\cleardoublepage\relax
\let\cleardoublepage\relax

\,\;$\;$
\bigskip
\bigskip

\chapter*{ Declaration }

This thesis is a presentation of my original research work. Wherever contributions of
others are involved, every effort is made to indicate this clearly, with due reference to
the literature, and acknowledgement of collaborative research and discussions.

The work was done under the guidance of \textbf{ Prof. P. C. Vinodkumar} at the Department of Physics, Sardar Patel University,
Vallabh-Vidyanagar.

 \def\s#1#2{\vbox{\hsize=4.5cm
    \kern2cm
    \hrule\kern1ex
    \hbox to \hsize{\strut\hfil #1 \hfil}
   \hbox to \hsize{\strut\hfil #2 \hfil}}}

\smallskip

\hbox to \hsize{\s{Ankit C. Khunt}{(Ph.D. Candidate)}\hfil
    }


\bigskip
\bigskip
\bigskip 

In my capacity as supervisor of the candidate's thesis, I certify that the above statements are true to the best of my knowledge.
  

%
\bigskip

\def\s#1#2#3{\vbox{\hsize=4.5cm
    \kern2cm
    \hrule\kern1ex
    \hbox to \hsize{\strut\hfil #1 \hfil}
    \hbox to \hsize{\strut\hfil #2 \hfil}
    \hbox to \hsize{\strut\hfil #3 \hfil}
    }}

\hbox to \hsize{\s{Prof. P. C. Vinodkumar}{(Thesis Supervisor)}  {\textbf{Date : 23/06/2023 }}\hfil
    }



\endgroup

\tableofcontents
\listoftables
\listoffigures
\cleardoublepage
\pdfbookmark[1]{Publications}{Publications}


\bigskip

\begingroup
\let\clearpage\relax
\let\cleardoublepage\relax
\let\cleardoublepage\relax

\bigskip
\chapter*{{\bf Publications}}
\section*{ Included in this thesis}
\subsection*{{\bf Published}}

\begin{enumerate}

\item \textbf{A. C. Khunt},  V. O. Thomas and P. C. Vinodkumar,  ``\textit{Distinct classes of compact stars based on geometrically deduced equations of state},''   Int. J. Mod. Phys. D \textbf{30}, 2150029  (2021) 
[arXiv:2101.00415 [gr-qc]].\\
\href{https://arxiv.org/abs/2101.00415v1}{{\ttfamily arXiv:2101.00415}}

\item \textbf{A. C. Khunt},  V. O. Thomas and P. C. Vinodkumar, 
``\textit{Relativistic stellar modeling with perfect fluid core and anisotropic envelope fluid},''
Indian J. Phys  (2023)
[arXiv:2303.07238 [gr-qc]].\\
\href{https://arxiv.org/abs/2303.07238}{{\ttfamily arXiv:2303.07238
}}

\end{enumerate}

\section*{Presentations at conferences}
\begin{enumerate}
\item \textbf{A. C. Khunt}, V. O. Thomas and P. C. Vinodkumar ,
 `` \textit{The properties of massive compact stars using a geometrically deduced equation of state}'', 64th DAE-BRNS Symposium on Nuclear Physics, Lucknow (UP), India, (2019).

 \item \textbf{A. C. Khunt},
 ``\textit{Compact stars as inner engines for Gamma-ray Bursts?}'', Prof. P. C. Vaidya National Conference on Mathematical Sciences. Vallabh Vidyanagar (Gujarat), India,  (2022).
 
\end{enumerate}

\endgroup
\cleardoublepage\pdfbookmark[1]{Abstract}{Abstract}

\begingroup
\let\clearpage\relax
\let\cleardoublepage\relax
\let\cleardoublepage\relax

\chapter*{Abstract}

This thesis presents a comprehensive exploration of compact objects, specifically focusing on neutron stars, by investigating their properties, classification, and stability in the context of general relativity. Two distinct research studies are conducted, each contributing valuable insights into the properties, classification, and stability of these superdense astrophysical entities.

In the first study, we compute the properties of compact objects, including neutron stars, by utilizing an equation of state  derived from a core-envelope model of superdense stars. By solving Einstein's equation with pseudo-spheroidal and spherically symmetric space-time geometries, we analyze these superdense stars and compare their properties with those obtained using nuclear matter equations of state. Through the derived mass-radius  relationship, we classify compact stars into three categories: highly compact self-bound stars with radii below 9 km, normal neutron stars with radii ranging from 9 to 12 km, and soft matter neutron stars with radii between 12 and 20 km. Additionally, we compute other important parameters such as Keplerian frequency, surface gravity, and surface gravitational redshift for each category. This research provides valuable insights into the study of highly compact neutron stars with exotic matter compositions.

The second study focuses on investigating the effect of density perturbations and local anisotropy on the stability of stellar matter structures within the framework of general relativity. Using the concept of cracking, we adopt a core-envelope model of a super-dense star and examine its properties and stability conditions by introducing anisotropic pressure to the envelope region. Moreover, we propose self-bound compact stars with an anisotropic envelope as potential progenitors for starquakes. By considering the difference between sound propagation in radial and tangential directions, we identify potentially stable regions within a configuration. As the anisotropic parameter increases, strain energy accumulates in the envelope region, which becomes a prime candidate for the buildup of starquake-like situations. The stored stress-energy in the envelope region, computed as a function of the magnitude of anisotropy at the core-envelope boundary, reveals that it can reach values as high as $10^{50}$ erg when the tangential pressure slightly outweighs the radial pressure. Remarkably, this energy level is comparable to that associated with giant $\gamma$-ray bursts. Therefore, this investigation contributes to the correlation studies between starquakes and $\gamma$-ray bursts, opening up new avenues for research in this field.

Collectively, these studies significantly advance our knowledge of compact objects, including neutron stars, by comprehensively analyzing their properties, classifying them based on their radii, and investigating the stability of their matter structures. The findings from these investigations provide crucial insights into the behavior and nature of these astrophysical entities, paving the way for further advancements in the study of highly compact neutron stars with exotic matter compositions, as well as establishing connections between starquakes and $\gamma$-ray bursts.

\endgroup
\cleardoublepage

\vfill

\cleardoublepage\pdfbookmark[1]{Notation}{Notation}

\begingroup
\let\clearpage\relax
\let\cleardoublepage\relax
\let\cleardoublepage\relax

\chapter*{Notation}
\begin{singlespace}
\begin{tabbing}
xxxxxxxxxxx \= xxxxxxxxxxxxxxxxxxxxxxxxxxxxxxxxxxxxxxxxxxxxxxxx \kill
$g_{\mu\nu}$  \> Metric Tensor \\
$g \equiv \text{det} (g_{\mu\nu})$  \> Determinant of metric tensor \\
$T_{\mu\nu}$  \> Energy-momentum tensor \\
$\Gamma^{\sigma}_{\mu\nu}$  \> Christoffel symbol of the second kind \\
$\mathcal{R}^{\tau}_{\mu\nu\sigma}$  \> Riemann-Christoffel curvature tensor\\
$\mathcal{R}_{\mu\nu}$  \> Ricci tensor \\
$\mathcal{R}$  \> Ricci scalar (scalar curvature) \\
$\mathcal{J}$ \> The full contraction of the Ricci tensor\\
$\mathcal{K}$ \> The full contraction of the Riemann tensor (Kretschmann scalar)\\
$\mathcal{W}$ \> The full contraction of the Weyl tensor\\
$G_{\mu\nu}$  \> Einstein tensor \\
$\tau$     \> Observer's proper time \\
$u^{\mu}$   \> Four-velocity \\
$(\mathrm{ds})^{2}$  \> Line element \\
$\nu, \mu$   \> Metric functions of spherically symmetric star\\
$K, \mathfrak{R}$ \>  Geometric parameters \\
$M_{\odot}$  \> Solar mass \\
$R_{\odot}$ \> Solar radius \\
$r_{g}$   \> Schwarzschild radius \\
$\eta$   \>  Compactness parameter \\
$\Gamma$   \> Adiabatic index \\
$M$  \> Gravitational mass \\
$R$  \> Stellar radius (equatorial) \\
$\rho$   \> Mass density \\
$\rho_{c}$  \> Central density \\
$\rho_{*}$ \> Uniform density \\
$\rho_{n}$ \> Nuclear saturation density \\
$\rho_{s}$ \> Surface density \\
$P$   \> Pressure \\
$P_{c}$ \> Central pressure \\
$v_{s}$  \> Sound speed \\
$I$   \> Moment of inertia \\
$E_{g}$   \> Gravitational energy \\
$z$  \> Gravitational redshift \\ 
$M_{H}$ \> Mass of black hole \\
$\Omega$   \>  Star's angular velocity relative to infinity\\
$\Omega_{\text{K}}$  \> Kepler (mass-shedding) frequency \\
$\bar \omega$  \>  Star's angular velocity relative to a local inertial frame \\
\end{tabbing}
\end{singlespace}

\endgroup
\cleardoublepage\pdfbookmark[1]{Acronyms}{Acronyms}

\begingroup
\let\clearpage\relax
\let\cleardoublepage\relax
\let\cleardoublepage\relax

\chapter*{Acronyms}
\begin{singlespace}
\begin{tabbing}
xxxxxxxxxxx \= xxxxxxxxxxxxxxxxxxxxxxxxxxxxxxxxxxxxxxxxxxxxxxxx \kill
GR  \> General Relativity \\
WD \> White Dwarf \\
NS  \> Neutron Star \\
BH  \> Black Hole \\
PBH \> Primodial Black Hole \\
SBH \> Stellar Black Hole \\
SMBH \> Supper Massive Black hole \\
IMBH \> Intermediate Mass Black hole \\
LIGO \>  Laser Interferometer Gravitational-Wave Observatory \\
LVC \> LIGO/ Virgo Collaboration \\
GW  \> Gravitational Wave \\
NICER \> Neutron Star Interior Composition Explorer \\
RXTE \> Rossi X-ray Timing Explorer \\
INTEGRAL \> INTErnational Gamma-Ray Astrophysics Laboratory \\
LMXM \> Low-Mass X-Ray Binary System \\
GRB \> Gamma-Ray Burst \\
EoS  \> Equation of State \\
QGP \>  Quark-Gluon Plasma \\
QCD  \>  Quantum Chromodynamics \\
SQM  \>  Strange Quark Matter \\
TOV \> Tolman-Oppenheimer-Volkoff \\
LMXB \> Low Mass X-ray Binary \\
SGR \> Soft Gamma Repeater \\
MSP \> Millisecond Radio Pulsar \\
sGRB \> Short Gamma-Ray Burst \\

\end{tabbing}
\end{singlespace}
\endgroup
\cleardoublepage

\manualmark
\markboth{\spacedlowsmallcaps{\bibname}}{\spacedlowsmallcaps{\bibname}} 

\addtocontents{toc}{\protect\vspace{\beforebibskip}} 
\addcontentsline{toc}{chapter}{\tocEntry{\bibname}}
\label{app:bibliography} 
\bibliographystyle{unsrt}

\bibliography{references/references}

\pagenumbering{arabic} 
\newpage

\chapter{Introduction: compact astrophysical objects and their properties}
\label{chap:Introduction}

``\textit{Science cannot solve the ultimate mystery of nature. And that is because, in the last analysis, we ourselves are a part of the mystery that we are trying to solve.}''                                
\begin{flushright}
-- Max Planck
\end{flushright}

\bigskip


It is known that stars do have a life cycle, it takes birth, live and die; they are not everlasting. Astronomers have discovered a lot about the life of the stars through observations and the methodical construction of detailed models based on knowledge of the laws of physics \cite{soderblom_2010}. The sort of life an individual star has, as well as the circumstances of its death, are determined by its mass, and to a lesser extent by its chemical composition \cite{burbidge_1957}. Less massive stars, such as the Sun, consume their fuel slowly and live a long time; when they run out of fuel, they fade away as slowly cooling white dwarfs\footnote{A compact stellar remnant supported by electron degeneracy pressure and shining only by the diffusion of light from its interior. }.
More massive stars live fast and die young, ending their lives in a dramatic explosion that leaves behind compact and incredibly dense ashes known as neutron stars\footnote{A dead ``star'' supported by neutron degeneracy pressure}. Massive stars might explode into existence in a supernova or, if a core is left behind, they can collapse until they split themselves off from the rest of the cosmos.

In this Chapter we  describe some of the  important properties of compact objects. After describing some of the most significant events in the life cycle of stars in general, compact objects like the neutron star, its  structure  as well as their internal compositions and the physical parameters that enable us to classify them will  be discussed. We also discuss extremely diverse phenomenology of other astrophysical compact objects. Towards the end, we  discuss the Gamma-ray bursts, causes of starquakes and other observational constraints.

\section{Compact stars as Astrophysical Laboratories}

We can describe the wondrous universe that natural world created and in which we live by pointing to four fundamental forces. The gravitational force, which is the most amazing of these forces$-$along with the electromagnetic, strong, and weak$-$becomes the inevitable ruler of everything around us when large masses or vast distances are taken into consideration. It is exceptionally weak over short distances. Each of these four forces governs our universe, and understanding the relationships between them is the only way to understand the origin and evolution of our universe. The neutron stars are a remarkable fusion of these four forces. With a radius of around $10$ km, a mass of about $1-2 M_{\odot}$, and a matter density of about $10^{15}$ g cm$^{-3}$, neutron stars are the densest stars known. Their density is higher than that of nuclear matter ($\rho_{n}\approx 2.8 \times 10 ^{14}$ g cm$^{-3}$). As more than just a consequence, any of the forces mentioned above could well be tested using neutron stars as astrophysical laboratories without being constrained by the same technological constraints that hold to experiments performed on Earth.

Understanding the behavior of matter under extreme conditions of temperature and/or density and determining the equation of state (EoS) \footnote{An equation of state, or EoS for short, is a relationship that describes the connection between pressure $P$ and matter density $\rho$ (or energy density via the relation $\mathcal{E}=\rho/c^2$), temperature $T$ (even though the temperature is mainly negligible for a system largely composed of degenerate fermions as neutron stars are), and other parameters depending on the star composition.} associated with it is one of the most difficult but also most complex problems of modern physics, both experimental and theoretical fronts. Knowing the characteristics of all kinds of matter are crucial for laboratory physics as well as our understanding of the physics of the early Universe and its evolution. 
In subsection \ref{chap_1_sub_sec_1} we discuss various types of compact objects in brief based on their physical characteristics.

\subsection{Different classes of compact objects}\label{chap_1_sub_sec_1}

\begin{itemize}
    \item[\ding{202}] \bf White Dwarfs :
\end{itemize}

The discovery of white dwarfs and the successful characterization of their characteristics by Fermi-Dirac statistics, assuming that they are allowed to hold against gravitational collapse by the degeneracy pressure of the electrons, an idea first put forward by Fowler in 1926 \cite{fowler_1926}, mark the beginning of the study of compact stars. Due to relativistic considerations, Chandrasekhar's comprehensive study in 1930 discovered that white dwarfs have a maximum mass \cite{chandrasekhar_1931}.

\begin{itemize}
    \item[\ding{217}] White dwarfs cannot be more massive than $1.4 M_{\odot}$ \footnote{The upper limit to the mass of a white dwarf, equals $(5.87/ \mu^{2})$, where $\mu$ is the mean number of nucleons per electron. For Fe we have $\mu=56/26$.}, and their radii are on the order of $R_{WD} \sim 10,000$  kms.
    \item[\ding{217}] This means that white dwarfs are extremely dense, with densities ranging from $2\times 10^{5} \; \text{to} \; 1\times10^{8}$ g cm$^{-3}$. Note that the density of lead is 11.3  g cm$^{-3}$. At the Earth's surface, a sugar cube made of white dwarf material would weigh anywhere from 200 kg to few tonnes!
\end{itemize}

\begin{itemize}
    \item[\ding{203}] \bf Neutron Stars :
\end{itemize}

Immediately after the discovery of neutron by
Chadwick in 1932 \cite{chadwick_1932}, existence of neutron star was predicted in 1934 by  W. Baade and F. Zwicky \cite{baade_1934}. Neutron stars originally anticipated to be a new kind of compact objects with its core consist of degenerate neutrons \footnote{One generally connects neutron stars with neutrons, their primary constituents. On February 27, 1932, J. Chadwick published an article in Nature reporting his discovery of a neutron \cite{chadwick_1932}. However, Lev Landau \cite{landau_1932} had published a study on dense stars a year before .} . Oppenheimer and Volkoff \cite{oppenheimer_1939} and Tolman \cite{tolman1939static} presented the first NS model  in 1939, which described the materials in such a star as an ideal degenerate neutron gas. Importantly, their calculations revealed that stars have a limiting mass at which they become unstable and fall into black holes.

The mass limit for neutron stars was estimated by Oppenheimer and Volkoff to be $0.7 M_{\odot}$ \cite{oppenheimer_1939,rhoades_1974}. But, how is it conceivable that white dwarfs have a higher maximum mass limit of $1.4 M_{\odot}$ neutron star? As a result of its failure, several physicists attempted  more computations while making numerous new assumptions. Rhoades and Remo Ruffini's calculation of $3.2 M_{\odot}$ is one of the well-known limits that are currently recognised \cite{rhoades_1974}. Nauenberg and Chapline  \cite{nauenberg_1973} , however, approximated the limit of a neutron star at its greatest value, which they estimated to be $3.6 M_{\odot}$. The real mass limit for neutron stars is still unknown since various assumptions produced conflicting results. However, neither the neutron star's physics nor the observations are consistent with mathematical predictions based on the assumptions physicists make. There is a definite need for a more accurate computation that is compatible with the observations.

The first neutron star was discovered in 1967, almost 30 years after it was initially proposed. It was really a peculiar object that was pulsing in the radio spectrum (a radio pulsar), but it was immediately determined to be a rapidly spinning neutron star \cite{bell_2017}. Hulse and Taylor  made the initial observation of the pulsar PSR 1913+16 in a binary system in 1974 \cite{hulse_1975}. This resulted in the ability to calculate its mass precisely, which turned out to be $1.44 M_{\odot}$. Therefore, this mass measurement disproved the straightforward hypothesis that the core of this star composed of an ideal gas of neutrons. It indicates the need of accounting for the interactions between the nucleons. Theorists speculated about the possibility of quark matter within neutron stars shortly after the quark model for nucleons was introduced. Gerlach showed in his PhD thesis with Wheeler in 1968 \cite{gerlach_1968}, in addition to white dwarfs and neutron stars, a third family of compact stars could exist in Nature. He developed general equation of state conditions for such  new kind of star to exist, namely that a significant softening in the equation of state, similar to phase transitions, must occur in neutron stars. Some astrophysicists assume that the very ground state of matter is strange quark matter (made up of $u$, $d$, and $s$ quarks). Since the mid-1980s, such objects have been studied and are known as strange stars \cite{haensel_1986,alcock_1986}.

Neutron stars as a compact star have supranuclear matter density ($\rho \sim 2.8 \times 10^{14}$  g cm$^{-3}$)  in their interior (presumably with a large fraction of neutrons). They have $M \sim 1.4 M_{\odot}$ average masses and $R \sim 10$ km radii. Consequently, their masses are near to the solar mass $M_{\odot} = 1.989 \times 10^{33}$ g, but their radius $R_{\odot} = 6.96 \times 10^5$ km is $10^5$ times less. As a result, neutron stars have a massive gravitational energy $E_{g}$ and surface gravity $g_{s}$, 

\begin{eqnarray}
E_{g} \;\sim \; G M^2/ R \sim 5 \times 10^{53} \;\text{erg} \sim 0.2 \; M c^2,\\
g_{s}\;\sim \;   G M / R^2 \sim 2 \times 10^{11} \; \text{cm s}^{-2},
\end{eqnarray}

where $c$ is the speed of light and $G$ is the gravitational constant. Neutron stars are evidently very dense. Their average mass density is

\begin{eqnarray}
    \bar \rho \simeq 3 M /(4\pi R^3) \simeq 7 \times 10^{14} \;\text{g cm}^{-3} \sim (2-3) \rho_{n},
\end{eqnarray}

Neutron star central density is even higher, reaching $(10-20)\rho_{n}$. All these characterises, neutron stars as one of the most compact astrophysical objects known in the Universe.

The following are some historical events that took place along the route that lead to the conception of neutron stars:

\begin{itemize}
    \item [\ding{102}] 1932 :  Chadwick confirmed the neutron discovery.
    \item [\ding{102}]  1934 : Neutron stars originally theorized by Baade and Zwicky.
    \item [\ding{102}]  1939 :  First calculations of neutron star models were done by Tolman, Oppenheimer, and Volkoff.
    \item [\ding{102}]  1962 : Our galaxy has distinct X-ray sources, which Giacconi discovered by launching a rocket with three Geiger counters (e.g. Sco X-1, Cyg X-1). There were ten recognised sources of celestial X-rays by the year 1965. The number now reaches $10^5$ at this time. 
    \item [\ding{102}]  1964 : The crab nebula has been found as an X-ray source.
\end{itemize}
This period was followed by events that finally validated the existence of neutron stars in nature:
\begin{itemize}
    \item [\ding{102}] 1967 : Hewish et al \cite{hewish_1968}. reported the first pulsar discovery. Pacini discussed the energy emission from a neutron star in respect to supernova remnants like Crab about the same time \cite{pacini_1967,pacini_1969}.
    \item [\ding{102}]  1968 : Pulsars, according to Gold, are rotating, strongly magnetised neutron stars \cite{gold_1968}. Soon after, Gunn \cite{gunn_1969} and Ostriker and Gunn \cite{ostriker_1969,gunn_1970} provided strong support for this view.

    \item [\ding{102}]  1968 : A pulsar exists in the Crab nebula.
    \item [\ding{102}]  1970s : NASA launched the first astronomy satellite, Uhuru. It has identified over 300 distinct X-ray sources until its destruction in March 1973.
    \item [\ding{102}]  1975 : Hulse and Taylor discover the first binary radio pulsar, PSR 1913+16 \cite{hulse_1975}.
    \item [\ding{102}]  1982 : Backer, Kulkarni, Heiles, Davis, and Goss discovered the first millisecond pulsar, PSR 1937+215 \cite{backer_1982}.
    \item [\ding{102}]  1990s : A flood of new observed pulsar data has been produced by the launch of a new generation of satellites, including COMPTON, ROSAT, ASCA, and the Hubble Space Telescope (HST).
    \item [\ding{102}]  2007 :  The International Space Station (ISS) mission known as the Neutron star Interior Composition Explorer (NICER) is used to investigate neutron stars using soft X-ray timing.
     \item [\ding{102}] 2017 : The Advanced LIGO and Virgo observatories detected the gravitational-wave signal GW170817 on August 17, 2017. The first signal linked to the merger of two neutron stars is this one \cite{abbott2017gw170817}.
    
    \item [\ding{102}]  2018 : An X-ray pulsar in the fastest stellar orbit ever was found in May 2018 by NICER.  It was discovered that the pulsar and its companion star orbit each other every 38 minutes \cite{strohmayer_2018}.
    \item [\ding{102}] 2019 : The brightest  known X-ray burst was discovered on August 21 by NICER. It originated from the Sagittarius neutron star SAX J1808.43658, which is located around 11,000 light-years from Earth \cite{bult_2019}.

    \item [\ding{102}] 2022 : The PSR J0952-0607 neutron star was found in 2017 in the constellation Sextans, roughly 3,000 light-years from Earth. The star is the known heaviest neutron star, weighing 2.35 times as much as the sun, according to recent measurements \cite{romani_2022}.
   \end{itemize}

 The main goal of this thesis is to investigate the models and properties of various classes of neutron stars. Thus, we  provide more details about the neutron star as a compact  astrophysical object in this section.  
\subsection*{The Anatomy of a Neutron star}
\begin{figure}[H]
    \centering
    \includegraphics[scale=0.5]{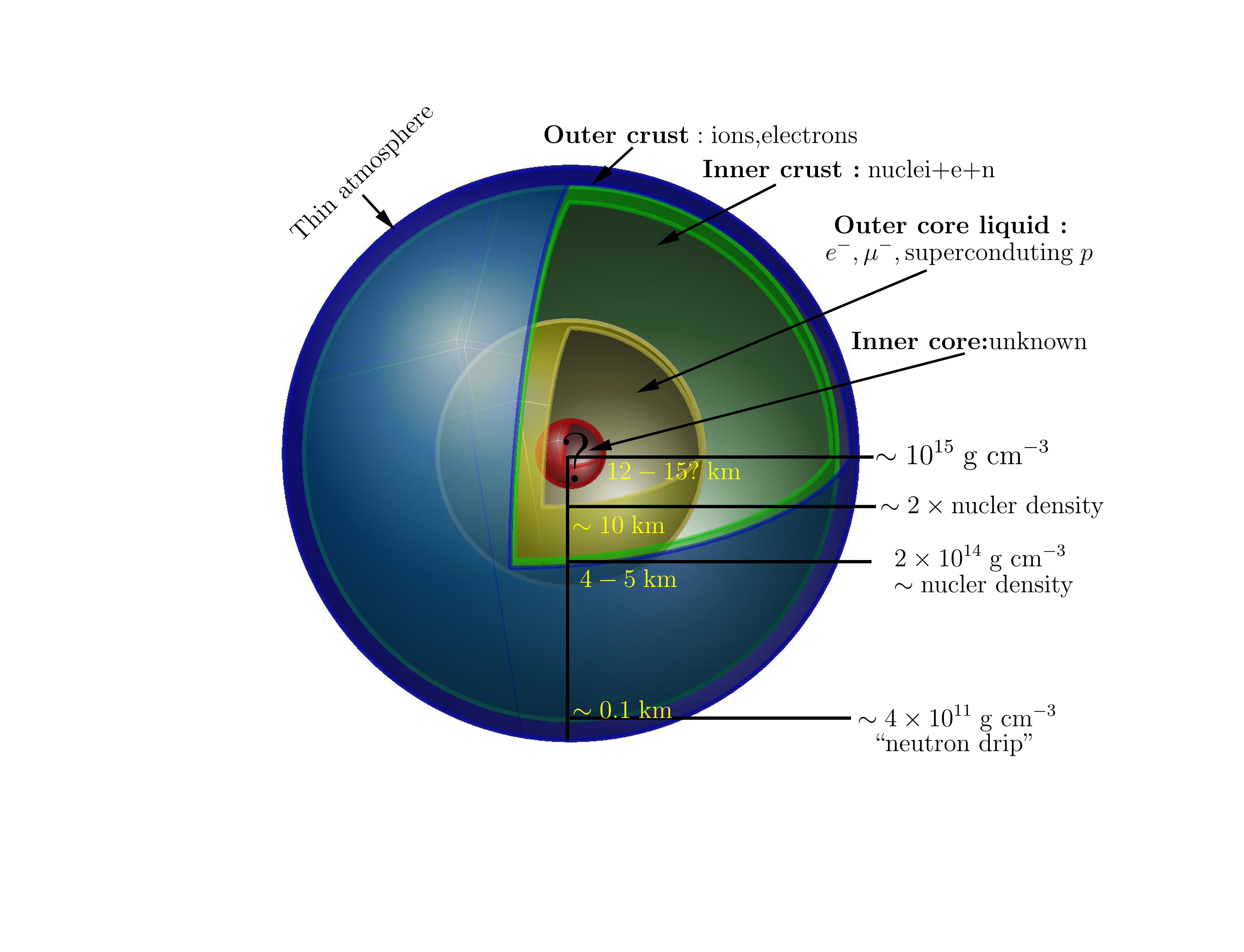}
    \caption{cross-section of a neutron star's interior. A thin atmosphere and an outer crust of heavy nuclei and electrons surround the neutron star. The inner crust is made up of nuclei, neutrons, and electrons, which undergoes transition to a neutron matter fluid at nuclear density. The central core's composition is still unexplained, however it is certain that the outer layer is made up entirely of neutrons, protons, electrons, and muons. }
    \label{fig:1_chap_1}
\end{figure}

A neutron star's cross-section can be generally split into four distinct regions (see Fig. \ref{fig:1_chap_1}):

\begin{itemize}
    \item [\ding{226}]\textit{Surface:} The thin atmosphere, only a few cm thick. Normal nuclei and non-relativistic electrons comprise matter at mass densities between $10^{4}$ g cm$^{-3}$ and $10^{6}$ g cm$^{-3}$.
    
    \item[\ding{226}]\textit{Outer crust :} It is composed of an atomic nucleus lattice and a Fermi liquid of relativistic degenerate electrons. It's mostly white dwarf matter. The electrons become relativistic and form a relativistic electron gas at densities of $7 \times 10^{6} \;\text{g cm}^{-3} <\rho < 4.3 \times 10^{11} \;\text{g cm}^{-3}$. When becoming further neutron-rich, the atomic nuclei (lighter metals) form a solid Coulomb lattice.
    
    \item[\ding{226}] \textit{Inner crust:} It has densities ranging from $4.3 \times 10^{11} \;\text{g cm}^{-3}\; \text{to} \;2 \times 10^{14}\; \text{g cm}^{-3}$. Around $4.3 \times 10^{11}\; \text{g cm}^{-3}$, neutrons start leaking out of neutron-saturated nuclei and populating free states outside the inner crust. As a result, this density value is known as neutron drip density. Matter clusters into extremely neutron-rich nuclei (heavy metals) that are formed on a lattice and immersed in a gas of neutrons and relativistic electrons to increase density.
\end{itemize}

\begin{itemize}
    \item[\ding{226}] \textit{Core region :} Beyond the transition density ($\rho_{tr}\simeq 1.7\times \times 10^{14}\; \text{g cm}^{-3}$), one reaches the core, which has dissolved all atomic nuclei into their components, neutrons and protons. The core may also include hyperons, more massive baryon resonances, and perhaps a gas of free up, down, and strange quarks due to the high Fermi pressure. Eventually, $\pi$- and K-meson condensates may also  be formed.
\end{itemize}

Reportedly, neutron stars are classified according to their core composition. In this perspective, we now refer to classical neutron stars (or hadronic stars), whose core is mostly composed of neutrons. But, at high densities, heavier baryons are excited, and the neutron star transforms into a hyperon star. Since these baryons are so tightly packed, the quark bags might diffuse, and a phase of quarks and gluon emerge with a possible   color-superconducting phase to occur. Eventually, pion and K- meson  condensates may develop \cite{yagi_2017,kumari_2021}. All of these distinct internal structures result in various mass-radii relationships. The typical neutron star has the largest radius for a given mass, however neutron stars with quark cores are found to be more compact. The radii of strange stars are the smallest\cite{haensel_1986}. Fig. \ref{fig:2_chap_1} shows a summary of various models of compact star inner structure. Whereas strange stars (strange quark stars and strangeon stars) are self-bound on the surface by strong force, typical neutron stars (hadron stars and hybrid/mixed stars) are bounded by gravity.

 In addition, distinct class of neutron stars could be recognized according to their physical nature and properties : 

 \begin{figure}[H]
    \centering
    \includegraphics[width=15cm, height=9cm]{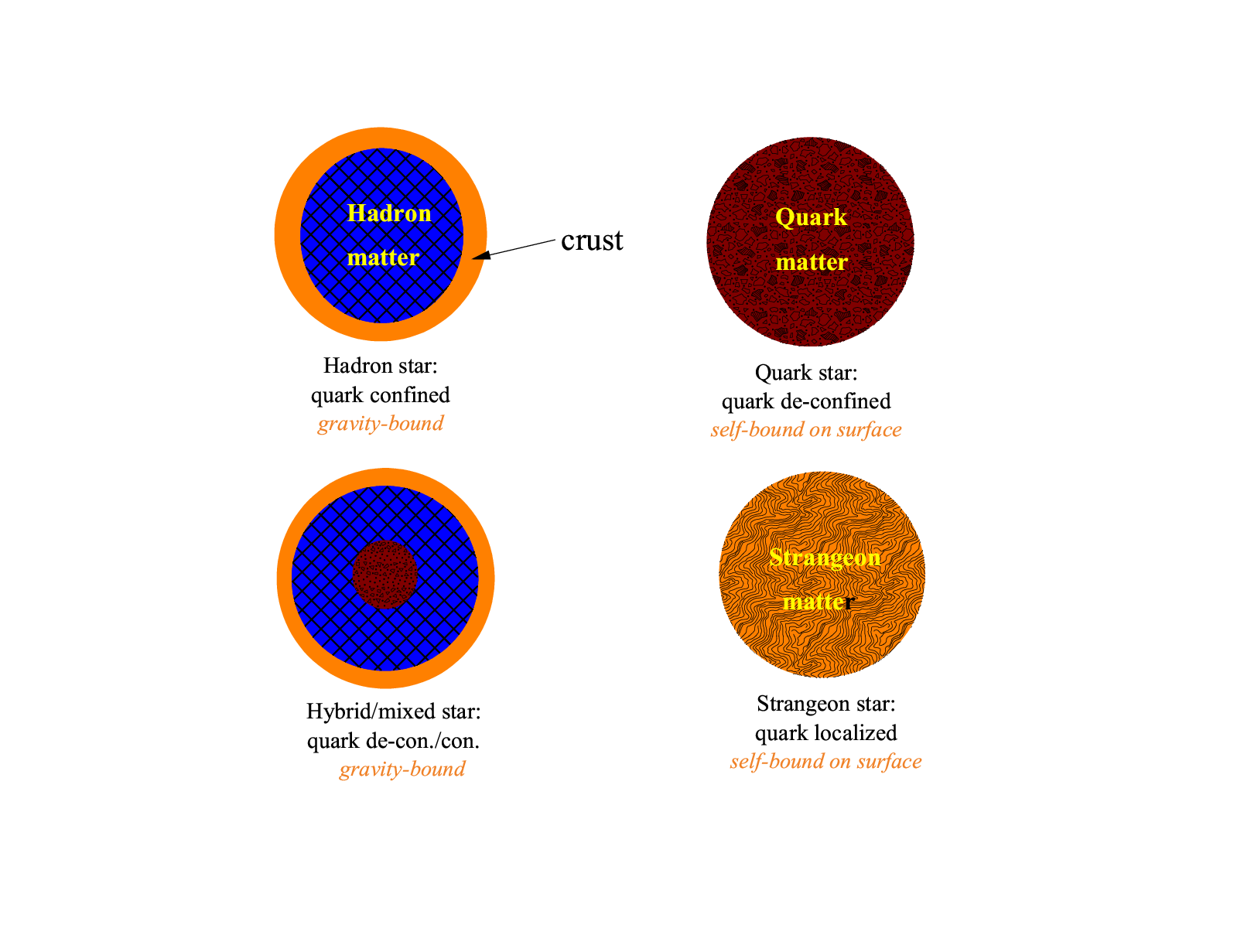}
    \caption{Different models of neutron stars. Conventional neutron stars comprise hadron stars and hybrid/mixed stars. Strangeness is necessary for strange quark and strangeon stars because of three-light-flavor ($u$, $d$, and $s$) symmetry. }
    \label{fig:2_chap_1}
\end{figure}
 \begin{itemize}
     \item [\ding{250}] \textbf{\textit{ Rotating Neutron stars (pulsars)}} : A radiation-emitting object with a characteristic pulse period and duration that emits radiation in the form of rapid pulses. Used to characterize the pulsed radiation that comes from rotating neutron stars.  Pulsars, in some cases, spin extremely fast, taking just milliseconds to complete a single rotation. These ``millisecond pulsars'' remain elusive to scientists. The fastest known pulsar, PSR J17482446ad, spins at 716 Hz, or 716 times per second. On November 10, 2004, Jason W. T. Hessels of McGill University observed this pulsar, which was confirmed on January 8, 2005 \cite{hessels_2006}. Even though radio pulsars were first identified almost 50 years ago, astronomers and scientists are still unsure of the actual nature of this compact object.

     \item[\ding{250}]\textbf{\textit{ Magnetars (magnetized neutron stars)}}: A magnetar is a neutron star that has a magnetic field that is 1,000 times greater than that of a typical neutron star. Numerous high-energy electromagnetic radiation, including X-rays and gamma-rays, are released as a result of various instabilities in their surfaces. A neutron star that releases huge bursts of X-rays and gamma-rays at irregular intervals is known as a soft-gamma ray repeater (SGR). They are also thought to be a form of magnetar. An unusual gamma-ray burst was recorded on March 5, 1979, and it was previously thought to be the result of a supernova remnant \cite{qiu_1990}. Gamma-ray photons were less energetic, and it was not a regular one. Eventually, it was discovered that the bursts originated in a single place, indicating the existence of a soft-gamma ray repeater. Anomalous X-ray pulsars (AXP) are neutron stars that are young and strongly magnetised. Magnetars are now largely assumed to be what they are. They have modest rotation periods ranging from 2 to 12 seconds. The observation of anomalous X-ray pulsars was driven by the occurrence of soft-gamma ray repeaters. There are 12 confirmed and two prospective AXPs as of 2017 \cite{bisnovatyi_2014}.
     
 \end{itemize}
\begin{itemize}
    \item[\ding{204}] \bf Exotic stars :
\end{itemize}
A hypothetical exotic star is a compact star made of elements other than electrons, protons, and neutrons that is prevented from collapsing by degeneracy pressure or other quantum attributes.

A quark star is a hypothetical compact object that is thought to form when neutrons break down into their up and down quark constituents as a consequence of gravitational pressure. In the event that no additional mass is transferred, it may remain in this new state constantly and be smaller and denser than a neutron star. In short, it is a single, massive hadron. Strange stars are quark stars that contain strange matter.

There is no significant proof that quark stars exist at the moment; however, several observations suggest that they might be present. Two objects, RX J1856.53754 and 3C 58, were suggested as quark star candidates based on measurements provided by the Chandra X-Ray Observatory on April 10, 2002 \cite{drake_2002,slane_2002}. The former appear to be significantly smaller and far colder than anticipated for a neutron star, indicating that they were made of denser material than neutronium. Nevertheless, researchers were doubtful about these results, arguing that the observations were unclear. RX J1856.53754 was eliminated from the list of quark star candidates after further detailed analysis \cite{trumper_2004}.

\begin{itemize}
    \item[\ding{205}] \bf  Black Holes:
\end{itemize}
As nothing can move faster than light, nothing can escape from within a black hole since they are very dense objects from which not even light can escape their gravitational pull. On the other hand, a black hole acts on objects distant from it with the same force that any other object of the same mass would experience. For instance, if our Sun were to be mysteriously shrunk down to a size of roughly 3 km, it would turn into a black hole, but the Earth would continue to circle the Sun as usual \cite{bambi_2019}.

Beginning in the 1970s, scientists began looking for possible black holes in the universe. The possibility of stellar-mass black holes ($M\approx  3- 100\; M_{\odot}$) and supermassive black holes (SMBH) ($M\approx  10^{5}- 10^{10}\; M_{\odot}$) is now supported by a large percentage of observational data.
While their origin is still debatable and they lack accurate mass measurements, intermediate mass black holes (IMBH) are likely a third class of compact astrophysical objects with a mass that lies between stellar-mass and supermassive ones. Black holes were suggested as the ultimate quasar energy source as early as 1964. At the meantime, the presence of black holes has been confirmed in the cores of massive elliptical galaxies with masses ranging from around $3 M_{\odot}$ to $10^7 M_{\odot}$.

Here, a remark is required. The existence of what we refer to as black holes is not proven beyond a reasonable doubt. The reason for this, in principle, is that physics is an experimental science. All we can do is criticize a model or hypothesis. We only get some limitations; we are unable to confirm a model or theory. For the present being, all observable evidence supports the notion that these objects are black holes, thus we refer to them as black holes. However, at the moment we are in the opposite situation, and all recent data, including gravitational waves and black hole imaging, are confirming what we expected from the theory. 

According to their mass, black holes in the universe are naturally divided into the following types :

\begin{itemize}
    \item[\ding{224}] \textbf{ Primordial black holes (PBHs)} having masses between that of the Earth and the Sun. While it is absolutely unknown how they formed, they could be products of the early Universe. An exceptional way to study the early Universe, gravitational collapse, high-energy physics, and quantum gravity could be such as through PBHs \cite{carr_2005}.
     \item[\ding{224}] \textbf{Stellar black holes (SBHs)} that range in mass from $2.5 \;M_{\odot}$ to $50 \;M_{\odot}$ \cite{narayan_2013}. These objects are the result of neutron star accretion and the development of massive stars. Over 60 stellar black holes have been discovered in X-ray binary systems as of 2015 \cite{corral_2015}.The event GW150914 corresponds to a binary black hole merger in the first gravitational wave detection by LIGO and Virgo \cite{abbott_2016}. These results demonstrate that binary SBH systems are exist. The initial black hole masses in the source frame are $36^{+5}_{-4} M_{\odot}$ and $29^{+4}_{-4}M_{\odot}$, while the final black hole mass is $62^{+4}_{-4}M_{\odot}$, with $3.0^{+5}_{-5}M_{\odot}c^2$ of gravitational waves radiated \cite{abbott_2016}. This marks the inaugural observation of a binary black hole merger and the first direct detection of gravitational waves by humankind.
     
     \item[\ding{224}] \textbf{Intermediate mass black hole}  with masses ranging from $50\; M_{\odot}$  to $10^4\; M_{\odot}$ \cite{coleman_2004}. Whether objects originate in this intermediate mass range is still unexplained. Ultraluminous X-ray sources (ULXs) are a distinct class of objects that emit 10 to $10^3$ times the X-ray power of neutron stars and stellar mass black holes \cite{mezcua_2017,greene_2020}.
     \item[\ding{224}] \textbf{Supermassive black holes} have masses of $10^{6} M_{\odot} \leq M_{H} \leq 10^{10} M_{\odot}$ \cite{narayan_2013}. These objects evolve in the centres of spheroidal galaxies and have been spotted in almost all galaxies. More than $10^{6}$ SMBHs in quasars have been discovered in major galaxy surveys \cite{begelman_2006,volonteri_2010}.
\end{itemize}

\begin{itemize}
    \item[\ding{206}] \bf Gravastars : 
\end{itemize}

This is a strange idea that is both fascinating and odd. Mazur and Mottola presented this hypothesis for the first time in 2004 \cite{mazur2004gravitational}.  Interpreted as ``Gravitational Vacuum Condensate Star '', Gravastar is proposed as a characteristic of gravitational systems and is (theoretically) an extension of the Bose-Einstein Condensate. In the end, it's meant to act as a black hole alternative. In general, a gravastar is a highly compact matter object with a radius that is very close to the Schwarzschild radius in the absence of an event horizon or a central singularity \cite{chirenti_2007}. To make a gravastar, it is theorised that a phase transition occurs at or near the region where the event horizon would have normally occurred during the gravitational collapse that signifies the end of the life of a massive star \cite{rezzolla2013relativistic}.

\begin{figure}[h]
    \centering
    \includegraphics[scale=0.7]{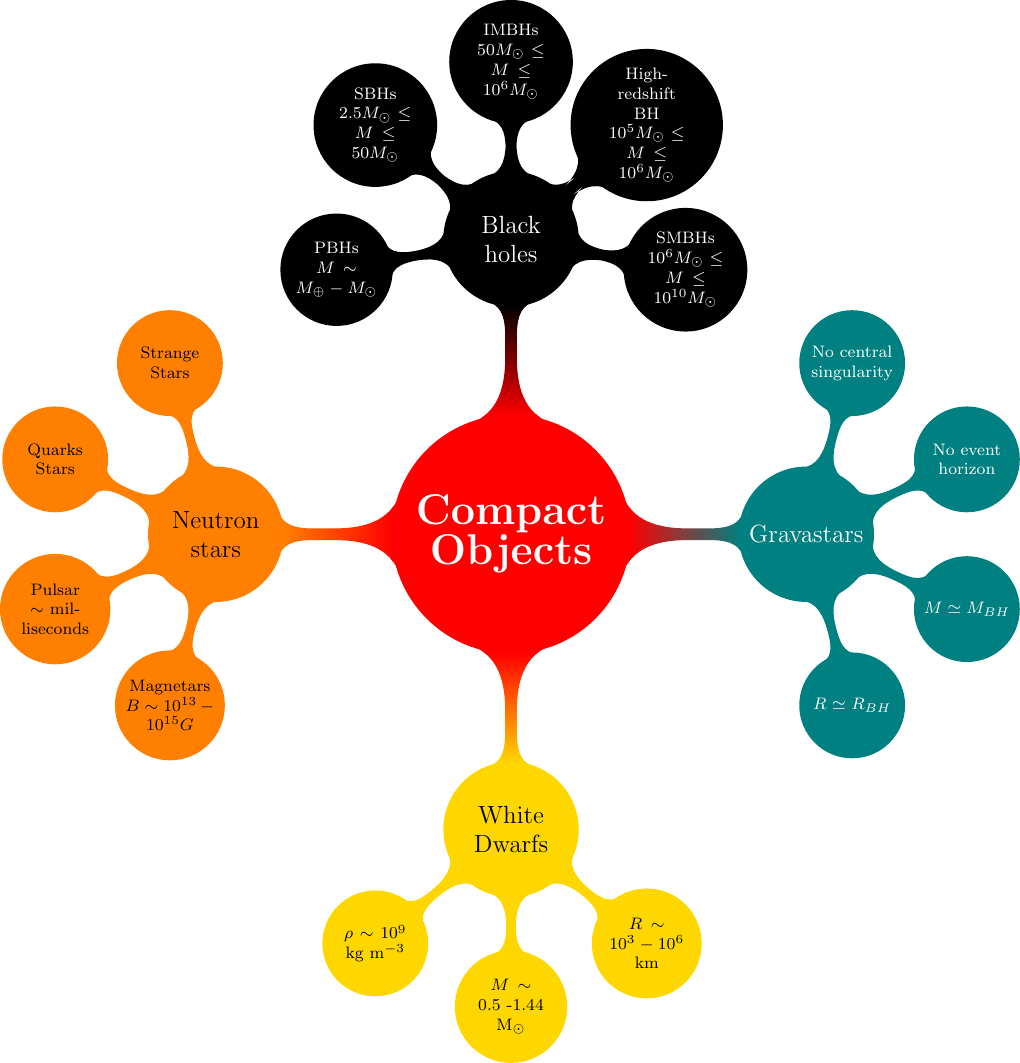}
    \caption{Listing different classes of compact objects in the universe based on observations and theoretical predictions}
    \label{fig:3_chap_1}
\end{figure}

Without looking deeper into the physical processes that would causes the formation of a gravastar or the astronomical supporting evidence their origin, the gravastar model, even as ingenious, also challenges one of modern astrophysics most adored foundations: the existence of astrophysical black holes. Gravastars could, in fact, be constructed to be extremely compact, with an outermost surface that is only infinitesimally bigger than the horizon of a black hole of the same mass \cite{visser_2004,sakai_2014,ray_2020}. Fig. \ref{fig:3_chap_1} provides a comprehensive listing of these objects, considering both observational evidence and theoretical predictions.

\section{ Pulsar Glitches and Starquakes overview}
Pulsars are among the finest clocks in the universe, providing incredibly accurate periodic pulses as a consequence of the revolution of a neutron star with a strong, inclined dipolar magnetic field. Its spin frequency can be estimated using pulsar timing, which is the periodic monitoring of a neutron star's rotation by recording the time of arrival (TOA) of a pulse obtained by folding a few minute-time measurements \cite{miller2019psr} . Since a pulsar's rotating period is nearly constant, the spin-down model can be constructed by fitting these pulses-after accounting for systematic influences such as Earth motion or pulsar proper motion - with a Taylor expansion of the pulsation frequency around a reference time $t_{0}$:

\begin{eqnarray}\label{chap1_eq_4}
    v_{\text{spin-down}}(t) = v_{0}+ \dot{v_{0}} (t-t_{0}) + \frac{1}{2}\ddot{v_{0}}(t-t_{0})^2.
\end{eqnarray}

This equation usually offers a satisfactory fit to the pulsar data. Any gradual and stochastic divergence from this formula is referred to as timing noise (see, e.g., \cite{hobbs_2010}), its origin of which is still unknown. It can be attributed to a variety of phenomena such as neutron star precession, turbulence in the interior superfluid \cite{melatos_2014},  or changes in magnetic fields \cite{kramer_2006,lyne_2010} . The details of such phenomena are  beyond the scope of the present study.

\begin{figure}[h]
    \centering
    \includegraphics[scale=0.3]{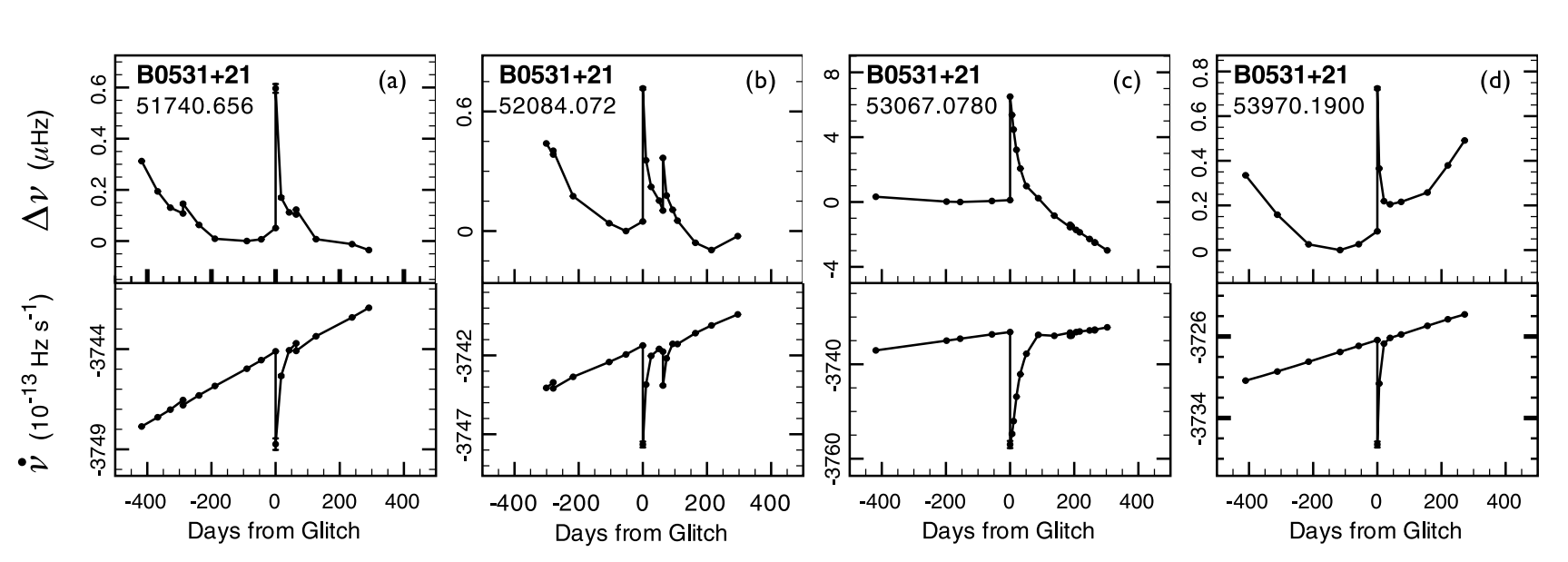}
    \caption{Espinoza et al.\cite{espinoza_2011} estimated four glitches in the Crab pulsar (PSR B0531+21 or PSR J0534+2200). Each plot shows the frequency time derivative and residuals against time. The glitch epoch, which is denoted underneath the pulsar name in Modified Julian Day (MJD), is represented by the zero on the primary axis.}
    \label{fig:4_chap_1}
\end{figure}

\begin{figure}[h]
    \centering
    \includegraphics[scale=0.50]{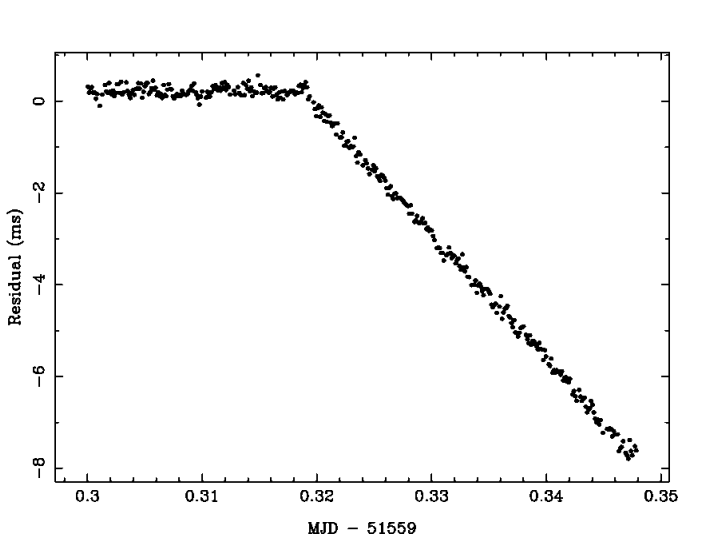}
    \caption{Timing residuals of the 2000 Vela glitch, figure taken from Dodson et al. \cite{dodson_2001}.}
    \label{fig:5_chap_1}
\end{figure}

Changes in the pulsar frequency and frequency time derivative, on the other hand, are glitches (see Fig. \ref{fig:4_chap_1}). Pulsar glitches occur when a spinning neutron star suddenly spins up, followed by a period of slow recovery during which the values return back to pre-glitch ranges. We used the word ``sudden'' because the exact timeframe of a glitch cannot be determined at this time (see  \cite{dodson_2001,ashton_2019,montoli_2020}). Timing irregularities can be investigated as residuals from Eq. (\ref{chap1_eq_4}):

\begin{eqnarray}
    \phi(t) = 2\pi  \int_{t_{0}}^{t} (v{t}- v_{\text{spin-down}}(t)) \mathrm{dt}, 
\end{eqnarray}

where $v(t)$ represents the actual pulse frequency (thus, including all the deviations from Eq. (\ref{chap1_eq_4})).
$\phi(t)$ are the phase residuals that reflect a pulsar's advance (if positive) or delay (if negative) with relative to the spin down model in units of radians.
Therefore, $v(t) = v_{\text{spin-down}}(t))$ and $\phi(t) = 0$ for all $t$ if there are no deviations from the spin down model in Eq. (\ref{chap1_eq_4}). By scaling the residuals by the star's angular velocity ($\Omega$), we can study them as  
 \begin{eqnarray}
     r(t) = - \frac{\phi(t)}{\Omega_{0}},
 \end{eqnarray}

where $\Omega _{0} = 2\pi v(t = t_{0})$ is the assumed constant angular velocity of the star at the reference time. This function indicates in seconds how early (if negative) or late (if positive) a new neutron star pulse is detected in relation to the spin down (\ref{chap1_eq_4}). If a pulsar glitch is identified, this function decreases steadily over time (see Fig. \ref{fig:5_chap_1}).

\begin{figure}[h]
    \centering
    \includegraphics[width=0.5\textwidth]{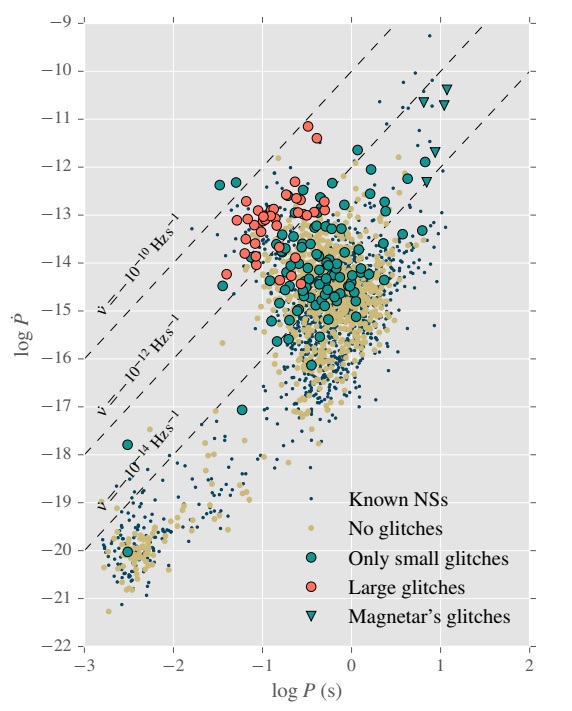}
    \caption{$P-\dot P$ plot of all known neutron star, with green and saffron markers for glitches. Figure taken from Fuentes et al \cite{fuentes_2017}}
    \label{fig:7_chap_1}
\end{figure}

Only a few years after pulsars were discovered, in 1969 , the Vela pulsar showed the first glitch \cite{radhakrishnan_1969,reichley_1969}. Presently, 190 pulsars have seen more than 550 glitches\footnote{see \href{https://www.jb.man.ac.uk/pulsar/glitches.html}{https://www.jb.man.ac.uk/pulsar/glitches.html} for an up-do-date catalogue of pulsar
glitches \cite{espinoza_2011}} (see Fig. \ref{fig:7_chap_1}). With the significant exception of a small glitch in an accreting pulsar, all glitches have been observed in isolated stars \cite{serim_2017}. Glitch sizes usually vary from $\Delta v / v \approx 10^{-12}$ to  $\Delta v / v \approx 10^{-5}$ orders of magnitude and are measured with respect to the pulsar frequency \cite{espinoza_2011}.

Since the observation of the first glitches \cite{hewish_1968} , scientists have attempted to comprehend their physical mechanism. When a pulsar glitch was discovered in the Vela pulsar, the first model that was suggested was that of crust quakes \cite{ruderman_1969}. The outer crust in this model, which is made up of a crystalline lattice that can sustain stress, is where a glitch in the mechanism first develops. Of course, the rotational frequency of the star impacts the shape of the crust; we anticipate that a rotating star cannot be entirely spherical but instead become rather oblate due to centrifugal forces. In the Newtonian framework, a basic example is the Maclaurin ellipsoid, an axisymmetric figure of equilibrium for a self-gravitating fluid with constant density. When the star spins down, we anticipate it to change its shape in order to establish a new equilibrium shape, but because the crust is solid, stress builds up and is eventually released uncontrollably as crust quakes. This starquake (Originally, Ruderman \cite{ruderman_1969} and Baym et al \cite{baym_1969} interpreted them as ``starquakes'', sudden fractures of the elastic neutron star crust) distorts the structure of the crust, changes the moment of inertia, and speeds up the spin of the star. $L = \Omega I $, where $L$ is the total angular momentum of the star, $I$ is the total moment of inertia and    $\Omega$ the angular velocity, and angular momentum ($L$) conservation can be used to express this in a simple manner:

\begin{eqnarray*}
    \Delta L = \Delta I \Omega + I \Delta \Omega
\end{eqnarray*}

The variances have been estimated between the after- and before-glitch amounts.
Angular momentum conservation provides  $\Delta L=0$, therefore giving:

\begin{eqnarray}
    \frac{\Delta \Omega}{\Omega}= - \frac{\Delta I}{I}
\end{eqnarray}

this shows that the magnitude of the glitch can vary in directly proportional to the moment of inertia during the quake. It would take roughly ten million years for an ordinary neutron star with mass $M \approx 1.4M_{\odot}$ and radius $R \approx 10$ km to develop enough stress to produce in the significant glitches in Vela pulsar \cite{smoluchowski_1970,baym_1971b}.

\begin{figure}[h]
    \centering
    \includegraphics[scale=0.5]{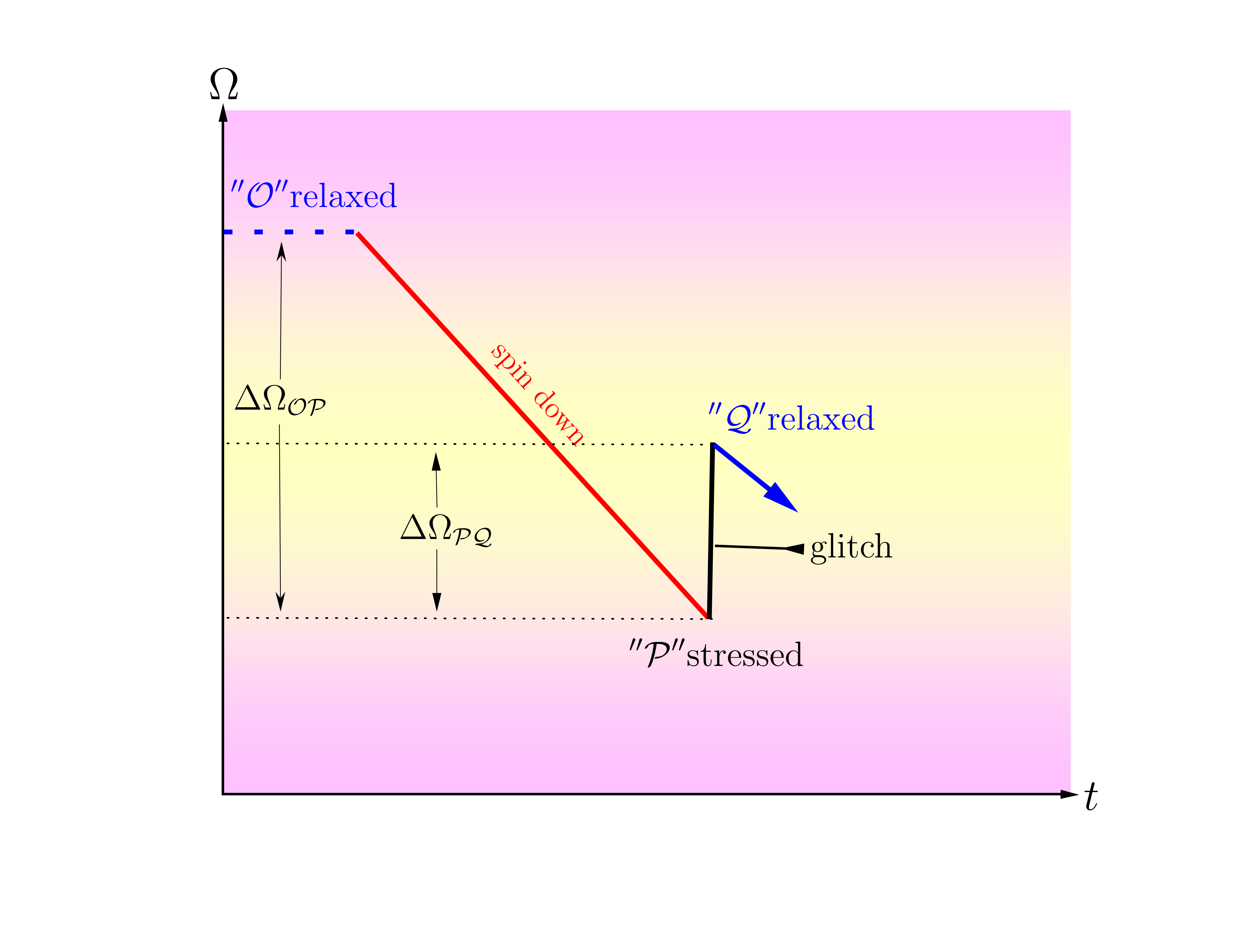}
    \caption{A starquake driven by spin-down is shown schematically, with time $t$ on the horizontal axis and rotation rate $\Omega$ on the vertical axis.
The star is relaxed in state "$\mathcal{O}$". It spins down from "$\mathcal{O}$" to "$\mathcal{P}$" with its crust deforming and stress stacking up. The crust reaches a critical stress and breaks at "$\mathcal{P}$", the moment of inertia decreases, and the star spins up to a new relaxed state at "$\mathcal{Q}$".}
    \label{fig:6_chap_1}
\end{figure}

On the other side, Link et al. \cite{link1998starquake} and Franco et al. \cite{franco2000quaking}, assumed that glitches are induced by starquakes that fracture the crust abnormally (distorted by the structure of the stellar magnetic field), causing a decelerated rotation motion of the star, resulting in an increase in the angle between the rotational and magnetic axes and thus explaining the observation that each glitch increases the spin-down rate.

The strain accumulated between successive glitches of the Vela pulsar is always much too small to break the crust, so if starquakes are used to induce glitches, the crust must always remain close to the  breaking strain. Recently, Giliberti et al. \cite{giliberti_2020} numerically calculated the centrifugal deformation of a (Newtonian) polytropic star with different adiabatic indices and confirmed that although the value of the adiabatic index has the centrifugal deformation of a star with uniform but potentially variable densities in the fluid core and the elastic crust was analytically determined by the same authors \cite{giliberti_2020, giliberti_2019}.

In the starquake situation, the neutron star is distorted towards a more spherical shape by the decreasing centrifugal force. Such stresses the elastic crust, which finally gives way and permits an abrupt, further deformation that decreases the moment of inertia and causes the star to spin-up. All elastic stresses are dissipated after an extreme quake. Thus, if, as shown in Fig. \ref{fig:6_chap_1}, the star starts from a relaxed state "$\mathcal{O}$" (at birth or following a previous, large quake), we can regard the state "$\mathcal{P}$" just before and the state "$\mathcal{Q}$" just after the quake as perturbations of "$\mathcal{O}$" under the same applied force (the change in the centrifugal force), in which the process $\mathcal{O} \rightarrow \mathcal{P}$ is subject to the elastic stresses in the crust but $\mathcal{O}\rightarrow \mathcal{Q}$ is not, that is, in the latter process the star behaves as if completely fluid. As a result, we must investigate the displacement fields that explain how an elastic solid and a fluid deform as the force field changes \cite{rencoret_2021}.

\section{Compact stars as inner engines for Gamma ray bursts?}

The unknown nature of the engine that powers Gamma-ray bursts (GRBs) is a central problem contributing to the riddle of GRBs \cite{kouveliotou_1996,kulkarni_1999,piran_1999}. Several mechanisms  have been proposed, but it is probable that we are still a long way from a comprehensive understanding of their origin. Regardless of the form of the engine, it is widely believed that the observed GRBs originate from the conversion of the kinetic energy of ultra-relativistic particles to radiation in an optically thin environment. Particles are being transported by a fireball mechanism (or radiation explosion) near the central engine \cite{goodman_1986,shemi_1990}.

The astrophysical system that comes just before  GRB and undergoes catastrophic self-destruction to trigger the burst is the progenitor of the GRB. By definition, the progenitor system has already disintegrated when a GRB is observed. Which is also why, it is quite challenging to directly observe the progenitor system of a GRB. As an outcome, the progenitor system of a GRB may only be deduced indirectly by logical reasoning, theoretical modelling, and indirect observational evidence. Several progenitor systems for various types of GRBs are addressed in this section.

\subsection*{General Observational Constraints}
Any GRB progenitor system must adhere to the following observational constraints:

\begin{itemize}
    \item[\ding{226}] \textbf{Energetics :}  The collimation-corrected gamma-ray emission energy of GRBs is determined by observations to be $E_{\gamma}\sim 10^{51}$ erg (generally in the range of $10^{49} - 10^{52}$ erg) \cite{ghirlanda_2004}. Thus, the GRB progenitor must release an event of this amount of energy .
    \item[\ding{226}] \textbf{Variability time scale:} The temporal scale of the observed variability, $\delta t$, could only be a few milliseconds long.
Thus, the core engine's size must be less than $c\delta t \sim 3\times 10^7$ cm \cite{piran_1997,nakar_2002,bernardini_2012}. This suggests that the primary engine is a compact stellar-size object, such as a black hole or neutron/quark star. One demands that after the catastrophic event, the progenitor leaves behind some sort of object. Therefore, the progenitor system must be of stellar size.
\item[\ding{226}] \textbf{Collimation:} There are several reasons to believe that GRBs are collimated. For this to be possible, the progenitor system must be able to release a collimated jet \cite{panaitescu_2001,kobayashi_2002}.
\end{itemize}

Diverse GRB data obtained over the decades, specifically during the afterglow phase starting in 1997 \cite{costa_1997} , have supported the recognition of at least two significant groups of progenitors structures:
those causing massive star deaths (i.e. \textit{ massive star} GRBs or Type II GRBs) and those causing compact object but not massive stars (i.e. \textit{compact object} GRBs or Type I GRBs). The earlier have durations more than 2 seconds, whilst the latter have durations less than 2 seconds. Many such categories of progenitor systems may also be constructed within each class. Here are  some examples:

\begin{itemize}
    \item[\ding{108}] Massive star GRBs (Type II) :
    \begin{itemize}
        \item[\ding{223}] The collapse of a single rapidly rotating Wolf-Rayet star\footnote{Wolf-Rayet stars are massive stars that are at an early stage of stellar evolution and are shedding mass at a rapid rate.};
        \item[\ding{223}] Mergers involving binary systems with a single massive star, such as those involving helium stars and neutron stars or black holes;
        \item[\ding{223}] Blue supergiant candidates for ultra-long GRBs?
        \item[\ding{223}] Population III stars?
    \end{itemize}
\end{itemize}
\begin{itemize}
    \item[\ding{108}]Compact object GRBs (Type I):
    \begin{itemize}
        \item[\ding{223}] NS-NS mergers;
        \item[\ding{223}] BH-NS mergers;
        \item[\ding{223}] NS-NS and NS-BH collisions;
        \item[\ding{223}]BH-WD mergers;
        \item[\ding{223}]Accretion-induced NS collapse;
       \item[\ding{223}] BH-BH mergers?
       \item[\ding{223}] Starquakes?
    \end{itemize}
\end{itemize}

The Advanced LIGO and Virgo observatories recorded the gravitational-wave signal GW170817 on August 17, 2017 \cite{abbott2017gw170817}. The first signal attributed to the merger of two neutron stars is this one. The Anticoincidence Shield for the Spectrometer for the Fermi Gamma-ray Burst Monitor (INTEGRAL SPI-ACS) and the Fermi Gamma-ray Burst Monitor (GBM) only detected the gravitational wave signal 1.7 seconds later. A short GRB, designated GRB 170817A, was identified by this event. For so many years, astronomers believed that the merging of two neutron stars or a neutron star and a black hole was what caused short GRBs. The two events, GW170817 and GRB 170817A, provide the first definite proof that neutron stars in collision might  generate short GRBs \cite{abbott_2016,murguia_2017}.

The eruption recorded from the object SGR 1806-20 on December 27, 2004, represents a good piece of evidence for starquakes. The specifics of this observation are reported in \cite{hurley_2005}.

While the equation of state of neutron star matter is not completely known, it is believed that the core is made up of a fluid that is mostly composed of neutrons and is held in place by the pressure of neutron degeneracy \cite{potekhin_2010}. A solid crust of neutron matter, often referred to as ``neutronium'' by science fiction writers, surrounds its core. Similar to how the Sun's magnetic field permeating its photosphere, the neutron star's magnetic field is linked through its solid surface.

When the crust, which is typically stiff, experiences abrupt movement, a starquake occurs. Usually, the neutron star's approach to a more spherical form as it slows down from fast revolution causes this motion (the energy of rotation is lost to gravitational waves). The changes are dramatic, like an earthquake, since the crust material is so stiff. 

SGR 1806-20 is a magnetar, a kind of neutron star with a huge (teragauss)  magnetic field. It was reported that it produced a surprising, absolutely huge gamma ray emission on December 27, 2004 \cite{boggs_2007}. INTEGRAL and four other missions, including Swift, all collected on this flare. Energy of order $10^{42} $ erg was released by the 142 s flare.

The physical explanation for this flare that best fits the data is a starquake on the detected neutron star. Similar to how solar flares are produced, the massive burst of energy is caused by magnetic reconnection when the crust changes. The light curve for this event has a number of temporal characteristics that are comparable with these magnetic and dynamical scales. Starquakes might, in fact, account for at least part of the shorter gamma ray bursts. We used the core-envelope superdense star model \cite{thomas_2005} to study the impact of anisotropy in the evelope region based on these viewpoints. specifically how differences in anisotropy magnitude can affect the star's physical properties. We anticipate observing whether defects, quakes, or fracture-like phenomena occurred in the envelope region due to the different nature of pressure in the two regions: the core and the envelope. Whether any activity occurred as a result of an anisotropy pressure that is different from the radial one, then something extremely catastrophic must have happened. We may be able to determine how much energy can be released during this dramatic event.  Based on this motivation, we have investigated  simple starquakes phenomena in the general relativistic framework in  this thesis study.

As General Theory of Relativity (GTR)  framework is important for the study of compact astrophysical objects, we employ the GTR framework for the present study.

In general relativity, the study of compact stars, notably neutron stars, has several objectives, including:
 \begin{itemize}
     \item[\ding{248}] Understand the composition, attributes, and equation of state of neutron stars, as well as their structure.
   \item[\ding{248}]  Investigate the gravitational effects on space-time around neutron stars.
  \item[\ding{248}]  Investigate how matter responds to extremely high pressures and densities, such as those seen in the cores of neutron stars.
\item[\ding{248}] Examine GR's predictions in the extreme gravity realm to search for any deviations.
\item[\ding{248}] Investigate how the neutron star's crust's anisotropy affects the dynamics of starquakes, which can have an impact on the amount of energy released and the neutron star's subsequent behavior.

\item[\ding{248}] Study the effects of the neutron star's crustal anisotropy on the dynamics of starquakes, which comprises potential cracking and elastic energy release.

\item[\ding{248}] Understand the role of compact stars, such as neutron stars and quark stars, in the production of GRBs and SGRs high-energy astrophysical phenomena. 
\end{itemize}

The following are some ongoing challenges in the study of compact stars:

 \begin{itemize}
     \item[\ding{248}] The inability of an accurate understanding of the physical equation of state for dense matter in the core of a NS is  required for accurately predicting the star's mass and radius.
   \item[\ding{248}] The challenge of figuring out the real nature of compact objects that may not be neutron stars, such as quark stars or strange stars.
  \item[\ding{248}] The lack of observational data on the physical characteristics of compact stars can lead to testing theoretical models and validating predictions challenging.
\end{itemize}

Some of the most recent observations involving compact stars include:
 \begin{itemize}
     \item[\ding{248}] The observation of GWs caused by the merging of two NSs, reveals novel insights about the behavior of matter at extremely high densities.

    \item[\ding{248}] The discovery identified massive neutron stars with masses up to $2.35$ $M_{\odot}$ (PSR J0952-0607), challenging earlier assumptions about the maximum possible mass of NSs.

 \end{itemize}

\section{organisation of the thesis}

In Chapter \ref{chap:2}, the general relativistic hydrodynamic gravity models of non-rotating compact stars are discussed. Using two nuclear EoSs, we measure the curvature strength inside and outside of a neutron star.

In Chapter \ref{chap:3}, based on the EoS deduced from a core-envelope model of superdense stars, we analyzed the properties of compact objects like neutron stars. The calculated star properties are compared to those obtained from nuclear matter EoSs. The mass-radius (M-R) relationship found here allows us to classify compact stars into three categories: (i) highly compact self-bound stars with radius less than 9 km that represent exotic matter compositions; (ii) conventional neutron stars with radius between 9 and 12 km; and (iii) soft matter neutron stars with radius between 12 and 20 km. For all three kinds, additional properties such as Keplerian frequency $\Omega_{k}$, surface gravity $g_{s}$, and surface gravitational redshift $z_{\text{surf}}$ are computed and  compared with observed astrophysical evidence  at the end of the chapter.

In Chapter \ref{chap:4}, using the concept of cracking, we examine whether local anisotropy and density perturbations impact the stability of stellar matter structures in general relativity. We examine the characteristics and stability conditions using a core-envelope model of a super-dense star by providing anisotropic pressure to the envelope region. Moreover, we suggest that self-bound compact stars with an anisotropic envelope could be at the genesis of starquakes.

In Chapter \ref{chap:5}, we  compile  our main results  and end the thesis with a concluding remark while  highlighting  the future  scope of the present study.

The spacetime metric signature $(+,-,-,-)$ and units $G=c=1$ are used throughout the thesis  unless otherwise specified.

\chapter{Gravity and Theory of Compact Stars }
\label{chap:2}

``\textit{Spacetime tells matter how to move; matter tells spacetime how to curve.}''                                
\begin{flushright}
--John Archibald Wheeler
\end{flushright}

\bigskip

The physics of compact stars has evolved as  a multidisciplinary and an active area of research in recent times.
The gravity of compact stellar objects requires a description of gravitational fields that extends well beyond the Newtonian model. Thus, in this chapter, we briefly review  the most significant general relativity theories and approaches adopted for the study of compact stellar objects. Many classical textbooks treat gravity in the general relativistic framework; for example, Misner, Thorne, and Wheeler \cite{misner1973gravitation}, Schutz \cite{schutz2022first}, Carroll \cite{carroll2019spacetime},  Straumann \cite{straumann2012general} and Hobson \cite{hobson2006general}.

 We  emphasize here,  the fundamental concepts required to compute and  interpret various properties of compact stars.


\section{General Relativistic Hydrodynamics} 
\label{sec: I_GRH}
Attempts to understand the internal structures of white dwarfs, neutron stars, and Gravastars\footnote{Pawel O. Mazur and Emil Mottola \cite{mazur2004gravitational} proposed the gravastar as an alternative to the black hole hypothesis in astrophysics.}  based on the hydrostatic approximation, and accretion onto compact objects in general, demand a time-dependent treatment of gas dynamics. Few cases of the many attempts to study  compact objects  are based on a hydrodynamical description of matter. A perfect fluid can be defined as one that is isotropic in local comoving coordinates. The fluid's energy-momentum tensor, presuming that spacetime is Minkowskian \footnote{Three-dimensional Euclidean space and time are combined to create four-dimensional Minkowski space, where the space-time interval between any two occurrences is independent of the inertial frame of reference.}, is given by
\begin{eqnarray}
    T_{tt}= \rho ,\;\;\;\;\;\; T_{xx} = T_{yy} =T_{zz} =P ,
\end{eqnarray}
where $P$ is the pressure and $\rho$ is the  energy density. The equation of the energy-momentum tensor is determined through a Lorentz boost when each fluid element has a spatial velocity $v^{i}$ with respect to some fixed lab frame.
\begin{eqnarray}
    T_{\mu \nu}=  \left(\rho+P \right)u_{\mu} u_{\nu} + P \eta_{\mu \nu}.
\end{eqnarray}
In this case, the fluid four-velocity, $u_{\mu}$, satisfies the equation $u^{\mu} u_{\mu}=-1$. In Minkowski spacetime, the equations for energy and momentum conservation can be expressed as $T^{\mu \nu},_{\nu}=0$. We simply need to substitute the Minkowskian metric $\eta$ for the generic Lorentz metric of the spacetime and the partial derivatives with covariant ones in order to extend this equation to curved spacetime. As a result, for a perfect fluid (plasma) in a curved spacetime, the stress-energy tensor  can be written as
\begin{eqnarray} \label{en_stress_g}
    T_{\mu \nu}=  \left(\rho+P \right)u_{\mu} u_{\nu} + P g_{\mu \nu}.
\end{eqnarray}
Pressures and stresses are generally so high in the strong gravity domain that we cannot assume the fluid is incompressible. Consequently, the pressure contributions to the stress tensor could be of the same order as the energy density contributions. This causes GR plasmas to behave quite differently from other types of plasmas, where the stress-energy tensors are dominated by their rest-mass density.

\section{Relativistic Stellar Structure} \label{rel_stell_stru}
The appearance of Einstein's field equations is entirely generic and straightforward. However, because of their nonlinear nature and the interaction between spacetime and matter, they are very complex.  In  few cases, closed-form solutions may be obtained. The Schwarzschild metric outside of a stationary spherical star is one of the most significant closed-form solutions. Einstein's equations may also be computationally solved as coupled differential equations for the internal structure of a spherical static star, generally known as the Oppenheimer-Volkoff equations for stellar structure \cite{oppenheimer_1939}.

The important task of determining the equations that control spacetime and the distribution of matter in the situation of relativistic, spherical static stars is addressed in this section. The construction of neutron star models in the next chapters of the thesis is based on these basic equations.

\subsection{Spacetime of Relativistic Stars}
The most generic form of the line element in spherical coordinates in static isotropic regions of spacetime, such as the interior and exterior regions of an isolated static star, is
\begin{eqnarray}
     \mathrm{d}s^{2}= U\left(r\right)\mathrm{d}t^{2}-V(r)\mathrm{d}r^{2}-W(r)r^{2}(\mathrm{d}\theta^{2}+\mathrm{sin}^{2}\theta \;\mathrm{d}\phi^{2})
\end{eqnarray}
We may substitute any function of $r$ for  $U(r)$ and $V(r)$ without breaking the spherical symmetry. This is done in such a manner that $W(r)=1$. Therefore, the line element can be expressed as
\begin{eqnarray}\label{schw_met_chap2}
     \mathrm{d}s^{2}= e^{\nu(r)}\mathrm{d}t^{2}-e^{\lambda(r)}\mathrm{d}r^{2}-r^{2}(\mathrm{d}\theta^{2}+\mathrm{sin}^{2}\theta \;\mathrm{d}\phi^{2})
\end{eqnarray}
where $\nu(r)$ and $\lambda(r)$ are only $r$-specific functions. $\mathrm{d}s^{2}=g_{\mu \nu} dx^{\mu}dx^{\nu}$ is the connection between the line element and the metric tensor. This can be compared to (\ref{schw_met_chap2}) to express the metric tensor,

 \begin{gather} \label{metricform1}
 g_{\mu\nu}
 =
  \begin{bmatrix}
   e^{\nu(r)} & 0 & 0 & 0 \\
   0& -e^{\lambda(r)} &0 &0 \\
   0& 0 & -r^{2} &0  \\
   0& 0 & 0 & -r^{2} \sin ^{2}\theta \\
   \end{bmatrix}
  \end{gather}
The definition of Einstein tensor leads to 
\begin{eqnarray}\label{eintensorchap2}
    G_{\mu \nu}= \mathcal{R}_{\mu \nu}-\frac{1}{2}g_{\mu \nu }\mathcal{R} = -\frac{8 \pi G}{c^{4}}T_{\mu \nu}.
\end{eqnarray}
The Einstein curvature tensor, $G_{\mu \nu}$,  and its dependence with the metric, is discussed below. The  ``matter'' composition of space is represented by the energy-momentum tensor, $T_{\mu \nu}$. We follow the units in which $G=c=1$ and the coordinate $x_{\mu}$ as

 \begin{align}
x^{0}&=t   &  x^{1}&=r,                            &  x^{2}&=\theta,              &  x^{3}&=\phi,           
\end{align}

The fundamental tensor components for spacetime metric (\ref{schw_met_chap2}) are
\begin{align}\label{metric_elements}
 g_{00}&=e^{\nu},  &g_{11}&=-e^{\lambda},              &g _{22}&=-r^{2} ,             &  g_{33}&=  -r^{2} \sin^{2}\theta,                     
\end{align}
and
\begin{eqnarray}
    g=g_{00}g_{11}g_{22}g_{33}= -e^{\lambda+\nu} r^{2}\sin ^{2}\theta
\end{eqnarray}
The Christoffel's symbol of second kind is given by
\begin{eqnarray}\label{ch2_chris}
\Gamma^{\rho}_{\mu\nu}=\frac{1}{2} g^{\rho \lambda} (\partial_{\mu}g_{\nu \lambda} +\partial_{\nu} g_{\mu \lambda}  - \partial_{\lambda} g_{\mu \nu}  )           
\end{eqnarray} 
The non-vanishing components of Christoffel's symbol of second kind are shown in Table \ref{table:appex_chris} below.

\begin{table}[htb]
\begin{center}
 \begin{tabular}{c  c  c  c  c  c   c  c  c  c } 
 \hline \hline
 $\Gamma^{0}_{01}$  & $\Gamma^{1}_{00}$  &$\Gamma^{r}_{rr}$& $\Gamma^{1}_{2 2 }$  & $\Gamma^{1}_{33}$      & $\Gamma^{2}_{1 2}$  & $\Gamma^{2}_{33}$ & $\Gamma^{3}_{13}$  & $\Gamma^{3}_{23}$ \\ [1ex] 
 \hline 
 $\frac{\nu'}{2}$      & $\frac{1}{2}e^{\nu-\lambda} \nu'$          & $\frac{1}{2}\lambda'$   &  $r e^{-\lambda} $  & $e^{-\lambda} r \sin ^2 \theta$  & $\frac{1}{r}$  & $-\cos \theta \sin \theta$  & $\frac{1}{r}$   & $\cot \theta $
  \\ [1.2ex] 
 \hline \hline
  
\end{tabular}
\captionsetup{
	justification=raggedright,
	singlelinecheck=false
}
\caption{Christoffel symbols of the diagonal metric (\ref{schw_met_chap2}).}
\label{table:appex_chris}
\end{center}
\end{table}
 The Riemann-Christoffel curvature tensor is the only tensor that can be constructed from the metric and its first- and second-order derivatives (hereafter Riemann tensor for short),

 \begin{eqnarray}
      \mathcal{R}_{\mu \nu}= \mathcal{R}^{\sigma}_{\mu\sigma\nu} = ( \partial_{\sigma} \Gamma^{\sigma}_{\mu \nu} +  \Gamma^{\sigma}_{\kappa \sigma } \Gamma^{\kappa}_{\mu \nu}) -
    (\partial_{\nu} \Gamma^{\sigma}_{\mu \sigma} +  \Gamma^{\sigma}_{\kappa \nu } \Gamma^{\kappa}_{\mu \sigma}) .
 \end{eqnarray}
 The non-vanishing components of Ricci tensor are ( see the details in Appendix \ref{App:AppendixB})
  \begin{eqnarray}\label{ricci_ten1}
  \mathcal{R}_{00} =e^{\nu-\lambda} \left( -\frac{\nu''}{2}+ \frac{\lambda ' \nu '}{4} -\frac{\nu'^{2}}{4}-\frac{\nu '}{r} \right), \\
 \mathcal{R}_{11} =\frac{\nu''}{2}- \frac{\lambda ' \nu '}{4} +\frac{\nu'^{2}}{4}-\frac{\lambda '}{r} ,  \\
    \mathcal{R}_{22} = e^{-\lambda} \left(1-\frac{r \lambda'}{2}+ \frac{r \nu '}{2} \right)-1 ,\\
    \mathcal{R}_{33}= \mathcal{R}_{22}\sin ^{2} \theta .
\end{eqnarray}

The scalar curvature ($\mathcal{R}=g^{\mu \nu} \mathcal{R}_{\mu \nu}$) is obtained as a contraction of the Ricci tensor with the
metric.

\begin{eqnarray} \label{ricci_scal}
    \mathcal{R}= - 2 e^{-\lambda} \left[   -\frac{\nu ''}{2}-\frac{\lambda ' \nu '}{4} + \frac{\nu '^{2}}{4}  -\frac{(\lambda '-\nu ')}{r}  + \frac{1}{r^{2}}  \right]-\frac{1}{r^{2}}.
\end{eqnarray}

Einstein's curvature tensor (\ref{eintensorchap2}) can now be constructed from the Ricci tensor and the scalar curvature (\ref{ricci_scal}) as
\begin{subequations}\label{EFEs_all}
\begin{eqnarray}\label{ein_1}
    G_{00}= e^{\nu-\lambda} \left(    -\frac{\nu''}{2} + \frac{\lambda' \nu'}{4} -\frac{\nu '^{2}}{4}-\frac{\nu'}{r}\right), 
  \end{eqnarray}  
\begin{eqnarray}\label{ein_2}
     G_{11} = \frac{\nu ''}{2}- \frac{\lambda' \nu '}{4}+ \frac{\nu'^{2}}{4}-\frac{\lambda '}{r}, 
     \end{eqnarray}
  \begin{eqnarray}\label{ein_3}
          G_{22}= e^{-\lambda} \left( 1-\frac{r \lambda'}{2}+\frac{r \nu'}{2} \right)-1, 
     \end{eqnarray}
     \begin{eqnarray}\label{ein_4}
          G_{33}=\left\{ e^{-\lambda}\left(  1-\frac{r\lambda'}{2}+ \frac{r \nu '}{2}                 \right)-1\right\}\sin ^{2}\theta .
\end{eqnarray}
\end{subequations}
\section{The Schwarzschild solution} \label{schw_solution}

The energy-momentum tensor disappears in the vacuum around a static compact star, $T_{\mu \nu}=0$. As a result, Einstein's field equation (\ref{eintensorchap2}) is reduced to $G_{\mu \nu}=0$, which implies that
\begin{eqnarray}
    \mathcal{R}_{\mu \nu}=\frac{1}{2}g_{\mu \nu }\mathcal{R}
\end{eqnarray}
 When multiplied by the metric $g^{\sigma \mu}$, these equations can be simplified significantly easier,
\begin{eqnarray}
    \mathcal{R}_{\sigma}^{\nu}=\frac{1}{2}\delta^{\sigma}_{\nu}\mathcal{R}
\end{eqnarray}
and contracted for $\sigma=\nu$,
\begin{eqnarray}
    \mathcal{R}=\frac{1}{2}\mathcal{R} \Longrightarrow  \text{it is possible only if } \;\mathcal{R}=0.
\end{eqnarray}
Therefore, the vanishing of the Einstein tensor requires the elimination of both the scalar curvature and the Ricci tensor. However, the Riemann tensor only vanishes in the limit $r \longrightarrow \infty $, where spacetime is flat.

Setting $\mathcal{R}_{00}=\mathcal{R}_{11}=0$ in (\ref{ricci_ten1}), it follows that
\begin{eqnarray}
     \lambda ' + \nu ' =0 , \\
     \lambda+\nu =0 \implies \lambda =-\nu \label{set}.
     \end{eqnarray}

whose solution is
\begin{eqnarray}
    \lambda+\nu = C_{1},
\end{eqnarray}
where $C_{1}$ is constant of integration.  Inserting this result in $\mathcal{R}_{22}=0$, We get an equation for the function $\lambda(r)$,
\begin{eqnarray} \label{lambda}
    e^{-\lambda} \left( 1-\frac{r\lambda'}{2}+\frac{r \nu'}{2} \right)=1,
\end{eqnarray}
Using equation (\ref{set}) in equation  (\ref{lambda}),
\begin{eqnarray}
    e^{-\lambda}(1+ r\nu')=1 \implies \frac{\mathrm{d}}{\mathrm{d}r}(r e^{\nu})
    =1 \implies r e^{\nu}= r-2m
\end{eqnarray}
where is $m$ is constant, therefore
\begin{eqnarray}
    e^{\nu} = 1-\frac{2m}{r} \;\;\; \text{ and } \;\;\; e^{\lambda} =\left(1-\frac{2m}{r}   \right)^{-1},
    \end{eqnarray}
    therefore complete solution is of the form
    \begin{eqnarray}\label{sch_int_st}
        \mathrm{d}s^{2} = \left (1-\frac{2m}{r} \right)\mathrm{d}t^{2}-\left (1-\frac{2m}{r} \right)^{-1}\mathrm{d}r^{2}-r^{2} (\mathrm{d}\theta^{2}+\sin ^{2}\theta \mathrm{d}\phi^2).
    \end{eqnarray}

 This completes the metric's derivation outside of a static spherically symmetric object. The solution is called in tribute of Schwarzschild, who found it already in 1916, a few months after Einstein published his general theory of relativity,  Birkhoff has proven that this is the most general static, spherically symmetric vacuum solution to Einstein's field equation. Thence, Eq.(\ref{sch_int_st}) describes the  Schwarzschild solution's line element.   \\
\\
It follows then from Eq.(\ref{metric_elements}),
\begin{eqnarray}
    g_{00}(r) = e^{\nu(r)}= 1-\frac{2M}{r}, \\
     g_{11}(r) = -e^{\lambda(r)}= \left(1-\frac{2M}{r}\right)^{-1}, \label{singularity} \\
     g_{22}(r) =-r^{2},\\
      g_{33}(r) =-r^{2} \sin ^{2} \theta .
\end{eqnarray}

Eq.(\ref{singularity}) then provides a  singularity at the radius $r=2M$ and it is called Schwarzschild radius $r_{g}$.  The singularity is not a spacetime singularity and can be removed simply by changing the coordinates. However, an object  becomes a black hole if its $r_{g}$ is greater than the physical radius, $R$, of the object. No future lightcone extends beyond $r_{g}$, hence neither a particle nor a light can escape the region $R< r <r_{g}$. Consider the fact that the Schwarzschild solution only holds true in the empty space above a star's surface. The interior solutions will be derived in the next section.

\section{The Tolman-Oppenheimer-Volkoff equations}
Tolman, Oppenheimer, and Volkoff developed and applied general-relativistic hydrostatic equations to neutron star models in 1939 \cite{oppenheimer_1939}. These equations are derived from Einstein's field equation, assuming that the metric is static and isotropic, and matter is a perfect fluid. In Section \ref{rel_stell_stru}, the metric and associated Einstein curvature tensor were derived.
The energy-momentum tensor should have a vanishing covariant divergence and be a symmetric second-rank tensor. In a perfect fluid, there are no shear stresses and the pressure is isotropic in the fluid element's rest frame. The energy-momentum tensor in this local frame is

 \begin{gather}
 T_{\mu\nu}
 =
  \begin{bmatrix}
   \rho & 0 & 0 & 0 \\
   0& P &0 &0 \\
   0& 0 & P &0  \\
   0& 0 & 0 & P \\
   \end{bmatrix}
  \end{gather}

where $\rho$ is the matter density and $P$ the pressure. We allow  a variable density $\rho (r)$ and a pressure $P(r)$. The energy-momemntum tensor  which enters the field equations (\ref{eintensorchap2}) is determined by Eq.(\ref{en_stress_g}). Since the matter is at rest at each point, the components of the velocity four-velocity $u^{\mu}$ are $(u^{0},0,0,0)$. On trajectory of each particle of matter in the fluid the relation between proper time and coordinate-time is given by

\begin{eqnarray}
    \mathrm{d}s^2= g_{00}(\mathrm{d}x^{0})^2 = g_{00} u^{0} c^2 =\sqrt{g_{00}} \;\;\;\;\;\;\;\;\;\;\;\;\;\;\;\;\; 1= g_{00}(u^{0})^2
\end{eqnarray}
 we have, furthermore, 
 \begin{eqnarray}
    u_{0}= g_{0\mu} u^{\mu} = g_{00}u^{0}=\sqrt{g_{00}}\;\;\;\;\;\;\;\;\;\;\;\;\;\;\;\;\; u_{i}=0
\end{eqnarray}
 
and $T_{\mu\nu}$ as

\[
  T_{\mu\nu} =\rho 
  \begin{bmatrix}
    g_{00} & 0 & 0 & 0\\
    0 & 0 & 0 & 0\\
    0 & 0 & 0 & 0\\
    0 & 0 & 0 & 0
  \end{bmatrix}
  -P
  \begin{bmatrix}
     0 & g_{01} & g_{02} & g_{03} \\
    g_{10} & g_{11} & g_{12}  & g_{13} \\
    g_{20} & g_{21} & g_{22}  & g_{23} \\
    g_{30} & g_{31} & g_{32}  & g_{33}
  \end{bmatrix}
\]

This allows us to write $T_{\mu\nu}$ (hydrostatic density and pressure) in the form, which simplifies, by virtue of 
(\ref{metricform1}), to the final form

 \begin{gather}\label{full_em_matrix}
 T_{\mu\nu} (\text{hydrostatic})
 =
  \begin{bmatrix}
   \rho e^{\nu}  & 0 & 0 & 0 \\
   0& P e^{\lambda} &0 &0 \\
   0& 0 & P r^{2} &0  \\
   0& 0 & 0 & P r^{2} \sin^{2}\theta \\
   \end{bmatrix}
  \end{gather}
  
for the perfect fluid at rest. When we insert this into  the field equations (\ref{eintensorchap2}), we shall naturally obtain a set of relations between the geometric function $\nu(r)$ and $\lambda(r)$ and the fluid parameters $\rho (r)$ and $P(r)$. 

We require the  Riemann tensor $R_{\mu\lambda}$ to derive the left side of the field equations in terms of $\nu(r)$ and $\lambda(r)$. Fortunately, in this situation, we already obtain all  the components of $R_{\mu\lambda}$ associated with the metric tensor (\ref{metricform1}) from our discussion of the Schwarzschild solution in section \ref{schw_solution}.

We now possess explicit forms for both sides of the field equations. We may express the equations, using (\ref{EFEs_all}) and (\ref{full_em_matrix}), as

\begin{subequations}\label{full_EFS_comp}
\begin{eqnarray} \label{full_EFS_comp1}
 \;\;\;   -8\pi \rho \;  & = e^{-\lambda} \left( \frac{1}{r^{2}} -\frac{\lambda^{'}} {r^{2}} \right) -\frac{1}{r^{2}}\\\label{full_EFS_comp2} 
 \;\;\;    -8 \pi P  & = \frac{1}{r^{2}} -e^{-\lambda} \left(   \frac{1}{r^{2}}+ \frac{\nu^{'}}{r} \right)\\ 
 -8\pi P  & = e^{-\lambda} \left[ \frac{1}{4} \nu^{'} \lambda^{'}   -\frac{1}{4} \nu^{'2}  -\frac{1}{2}\nu^{''} - \frac{1}{2} \left( \frac{\nu^{'}-\lambda^{'}}{r}\right)
\right] \label{full_EFS_comp3}
\end{eqnarray}
\end{subequations}

As is evident from (\ref{full_em_matrix}) and (\ref{EFEs_all}), the equation $G_{33}=-8\pi P T_{33}$ is same as $G_{22}=-8\pi P T_{22}$. So,  we have at this stage only three independent equations.

Up to this point, the fluid's density and pressure have been represented as arbitrary independent scalar functions. The set of Eqs. (\ref{full_EFS_comp1})-(\ref{full_EFS_comp3}) provides three ordinary differential equations for the four unknown functions of $r$ that characterize the system's geometry and physics, namely, $\nu(r)$, $\lambda(r)$,  and $\rho(r)$, $P(r)$. In addition, we may persist in making arbitrary assumptions about the physical properties of the fluid that forms up the system. In fluid mechanics, this is commonly accomplished by prescribing the pressure-density relation, often known as the fluid's equation of state $P=P(\rho)$. To identify different model solutions, a straightforward technique is to eliminate $P$ from Eqs. (\ref{full_EFS_comp1}) and (\ref{full_EFS_comp3}).
\begin{eqnarray}\label{42}
    \frac{e^{\lambda}}{r^{2}} = \frac{1}{4} \nu ^{'} \lambda^{'}-\frac{1}{4}\nu^{'2}-\frac{1}{2}\nu^{''}+ \frac{\nu^{'}+\lambda^{'}}{2r}+\frac{1}{r^{2}}
\end{eqnarray}
This equation relates only the geometric functions $\lambda(r)$ and $\nu(r)$.

By using quadratures to solve the simple first-order differential equation, we may arbitrarily prescribe $\nu(r)$ and obtain $\lambda(r)$ from (\ref{42}). The equation of state $P(\rho)$ may then be calculated using $P(r)$ from (\ref{full_EFS_comp2}) and $\rho(r)$ from (\ref{full_EFS_comp1}). This indirect method would provide a specific solution to the field equations for the set of functions, $\lambda$, $\nu$, $P$ and $\rho$. If $P(r)\geq 0$ and $\rho(r) \geq 0$, it would then be necessary to determine if the solution is physically plausible or not.

We add (\ref{full_EFS_comp1}) and (\ref{full_EFS_comp2}) and find
\begin{eqnarray}\label{43}
    -8\pi \left( \rho+ P\right) = \frac{e^{-\lambda}}{r} (\nu^{'}+\lambda^{'})
\end{eqnarray}
which indicates that $\nu^{'}+\lambda^{'}>0$ for any physically acceptable solution. Now we differentiate (\ref{full_EFS_comp1}) with respect to $r$ and obtain
\begin{eqnarray} \label{44}
    -8\pi P^{'}(r) = -\frac{2}{r^{3}} + e^{-\lambda} \left[ \frac{\lambda^{'}}{r^2} + \frac{\lambda ^{'}\nu^{'}}{r}+\frac{2}{r^{3}} +\frac{\nu^{'}}{r^{2}}-\frac{\nu^{''}}{r}
    \right]
\end{eqnarray}
The second derivative term $\nu^{''}$ can be eliminated from (\ref{44}) by means of (\ref{42}). A simple calculation and rearrangement of terms leads to
\begin{eqnarray}
    -8\pi P^{'}= e^{-\lambda} \frac{\nu^{'}}{2r} (\lambda^{'}+\nu^{'})
\end{eqnarray}
We can simply eliminate $\lambda$ by combining this result with the differential relation (\ref{43}) : 
\begin{eqnarray}
    P^{'}= -\frac{1}{2}\nu^{'}(r) \left( \rho(r) +P(r)\right)
\end{eqnarray}
This formula leads directly to the geometry term $\nu(r)$ if the functions $\rho(r)$ and $P(r)$ are known.
We begin with a useful redefinition:
\begin{eqnarray} \label{metric_pot_e}
    e^{-\lambda(r)} \equiv 1-\frac{2m(r)}{r}.
\end{eqnarray}
This only changes out one function, $\lambda(r)$, with another, $m(r)$. Outside of the star, $M$, the entire mass of the star, is a constant value for the new function $m(r)$. For $m(r)$, $\rho(r)$, and $\nu(r)$, the three equations are
 \begin{subequations}
     \begin{eqnarray}
     \label{tov_m}
\frac{\mathrm{d}m(r)}{\mathrm{d}r} = 4\pi r^{2}\rho(r), \\ \label{tov_p}
-\frac{\mathrm{d}P(r)}{\mathrm{d}r} =\frac{m(r) \rho(r)}{r^2} \left( 1+\frac{P(r)}{\rho(r)}\right) \left( 
 1+\frac{4\pi r^3 P(r)}{m(r)}\right) \left( 1-\frac{2m(r)}{r} \right)^{-1},\\
\frac{1}{2} \frac{\mathrm{d}\nu(r)} {\mathrm{d}r}=  -\frac{1}{\rho(r)+P(r)} \frac{\mathrm{d}P(r)}{\mathrm{d}r} = \frac{ m(r)+4\pi r^{3}P(r) }     {r^{2}(1-2m(r)/r)   }.
    \end{eqnarray}
 \end{subequations}

This is the Tolman-Oppenheimer-Volkoff (TOV) equation for a spherically symmetric object's hydrostatic equilibrium. This equation provides the equilibrium solution for the pressure in a compact star when used in combination with the expression for the mass (\ref{tov_m}), a microscopic explanation for the relationship between the pressure and the energy density, and other factors. Where $P=P(r)$ is the pressure, $m = m(r)$ is the mass within radial coordinate $r$, and $\rho =\rho (r)$ is the density. Relativistic corrections are shown in parenthesis in Eq. (\ref{tov_p}). In general relativity, all sorts of energy, not only mass, are used as the source of gravity. Even though $P$ is included in Eq. (\ref{en_stress_g}), which indicates that pressure is a source of gravity, pressure appears on the right side of Equation (\ref{tov_p}). The boundary conditions associated with these parameters are,  $m(0)=0$, $\rho(0)=\rho_{c}$, $P(R)=0$ and  $m(R)=M$. An equation of state ($P\equiv P(\rho$)) must be used to support these equations.

It could be tempting to believe that since $m(0) = 0$, gravity would generally be weak and that general relativistic effects would thus be insignificant close to the center. This is improper since it has a Newtonian perspective on gravity. Near the center, the Newtonian gravitational acceleration, $G m(r)/r^2$, is negligible, but the relativistic correction to it is considerable because pressure contributes to gravity \cite{ekcsi_2016}.

\begin{figure}[H]
\centerline{\includegraphics[scale=0.45]{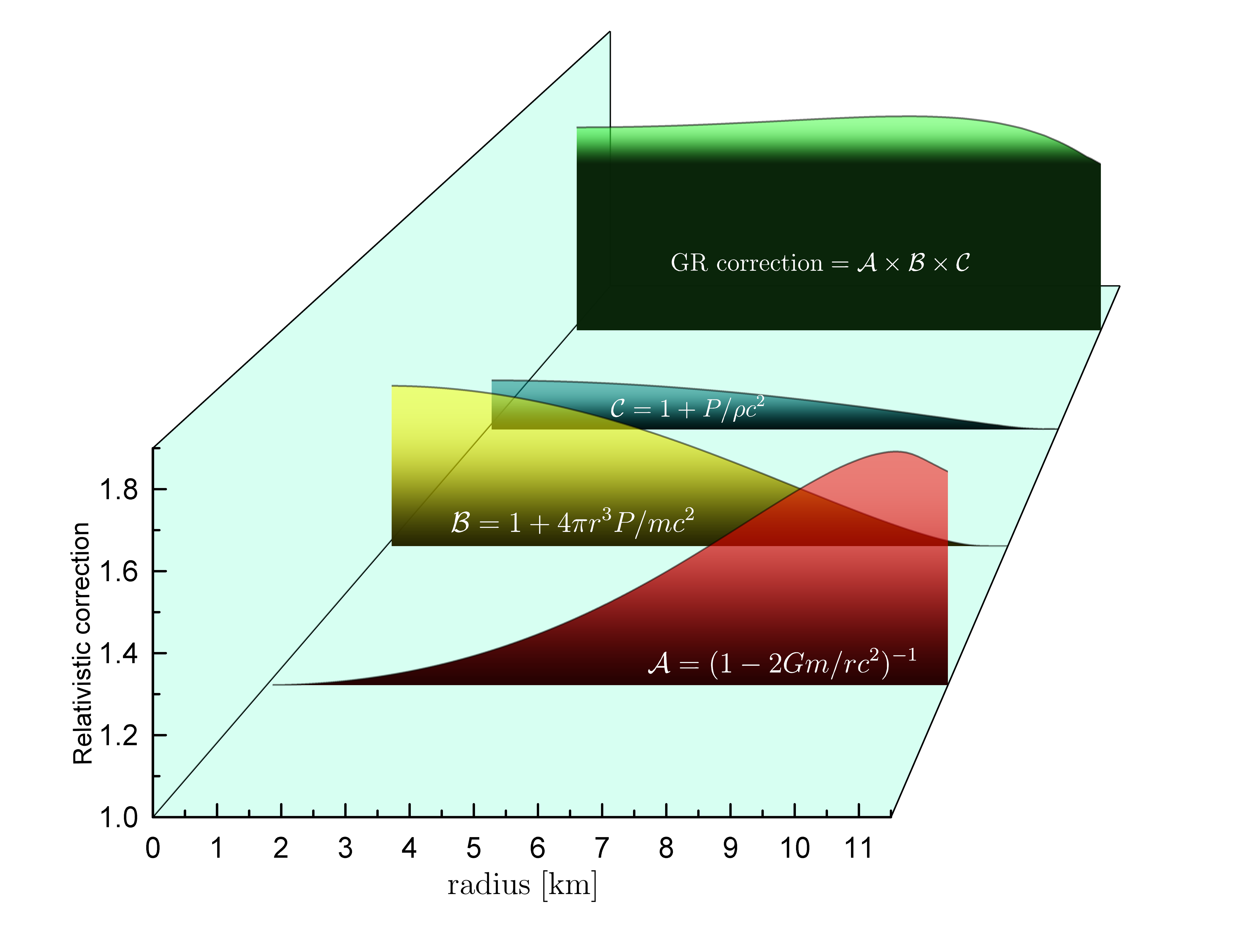}}
\vspace*{0pt}
\caption{Relativistic correction inside a neutron star with the equation of state SLy \cite{douchin_2001} and a central density of $\rho_{c}=1 \times 10^{15} \;\text{g cm}^{-3}$. The star has a mass of $1.40 M_{\odot}$ and a radius of $11.55$ km.  The green-black patch  corresponds to the complete correction term $\mathcal{A\times B \times C}$, the yellow-black patch corresponds to $\mathcal{A}$; the teal-black patch corresponds to $\mathcal{B}$ and the green-black patch correspond to $\mathcal{C}$.
\label{Rel_corr}}
\end{figure}

Fig. \ref{Rel_corr} shows the overall relativistic correction as well as the radial dependency of the relativistic corrections $\mathcal{A} \equiv (1-2Gm/r c^2)^{-1}$, $\mathcal{B} \equiv 1+4\pi r^3 P/m c^2$ and $ \mathcal{C} \equiv 1+P/\rho c^2$. It is clear that no region of the star, even its core, where $m\rightarrow 0$ occurs, has a Newtonian gravity that is accurate. Instead, the overall general relativistic corrections $\mathcal{A ,\;B,\;C}$ always approaches unity. The $\mathcal{B}$ term makes the most significant contribution in the center, although the $\mathcal{C}$ term is also significant. While these components progressively dissipate  towards the surface as $P$ approaches zero, the $\mathcal{A}$ term that disappears in the center takes over and dominates the relativistic correction at the crust.

\section{Uniform-density solution and Buchdahl's limit}
The TOV equations (\ref{tov_m}) and (\ref{tov_p}) must generally be numerically integrated. There are many known analytical solutions, but one of them is especially intriguing since it offers a limiting value of $M/R$ that is applicable to every star in hydrostatic equilibrium.  Normally, a numerical integration is required, however in the simplified situation of a spherically symmetric star with uniform matter density $\rho_{*}$, the TOV equations allow for an analytic solution. This solution, although having a perhaps simplified configuration, can be used to determine the minimal radius of a relativistic star with a given mass. \cite{rezzolla2013relativistic}.

Assuming that the fluid is incompressible results in a simple and rather realistic model of a star where the density $\rho_{*}$, is constant up to the star's surface and vanishes at $r \geq R$;

\begin{eqnarray}
\rho(r) = \begin{cases} \rho_{*},  &  r < R  \\
                  0, & r \geq R.
\end{cases}
\end{eqnarray}

Since $P(r)$ may be calculated from hydrostatic equilibrium, specifying $\rho(r)$ directly replaces the need for an equation of state . Following that, integrating (\ref{tov_m}) to get

\begin{eqnarray} \label{m(r)}
m(r) = \begin{cases} \frac{4}{3}\pi r^{3}\rho_{*},  &  r < R  \\
                  \frac{4}{3}\pi R^{3}\rho_{*}=M, & r \geq R.
\end{cases}
\end{eqnarray}

It is worth noting that the value for the energy density may be set after selecting the star's mass and radius, \textit{i.e.,} $\rho_{*}=3M/(4\pi R^{3})$, but that $M$
and $R$ are not linearly independent. Instead, their ratio is upper limit. This result may be derived by realizing that, although the energy density inside the star is uniform, the pressure is not and is provided by the integral of Eq. (\ref{tov_p}), \textit{ i.e.,}

\begin{eqnarray}\label{int_tov}
     \int_{P_{c}}^{P(r)} \frac{d\bar P}{(\rho_{*} + \bar P)+ (\rho_{*} + 3\bar P)} = \frac{4\pi}{3} \int_{0}^{r} \frac{d\bar r}{1-8\pi \rho \bar r^{2}/3},
\end{eqnarray}
where $P_{c}=P(r=0)$ is the central pressure of the star. Performing these standard integral, one finds that
 \begin{eqnarray}
     \frac{\rho_{*} + 3P}{\rho_{*} +P} = \frac{\rho_{*} +3 P_{c}}{\rho_{*} + P_{c}} \left[ 1-\frac{8\pi \rho_{*} r^{2}}{3} \right]^{1/2}.
 \end{eqnarray}

The left side of the above equation equals unity because the pressure $P$ is zero at the surface of the star, where $r=R$. Thus, we obtain

\begin{eqnarray}
    R= \sqrt{  \frac{3}{8\pi \rho_{*}} \left[ 1-\frac{(\rho_{*}+P_{c})^2}{(\rho_{*}+3P_{c})^{2}}    \right]                }.
\end{eqnarray}

This yields the radius of a uniform density star with a central pressure of $P_{c}$ . We may also rearrange this result and use (\ref{int_tov}) to get a useful equation for central pressure.

As a result, the
    central pressure, $P_{c}$, is given by
\begin{eqnarray}\label{p_c}
    P_{c} = \rho_{*} \left[ \frac{1-(1-2M/R)^{1/2}} {3(1-2M/R)^{1/2}-1}
    \right],
\end{eqnarray}

and

\begin{eqnarray}\label{p(r)}
    P=P(r) = \rho_{*} \left[   \frac{(1-2M r^{2}/R^{3})^{1/2}-  (1-2M/R)^{1/2} }                    
    {3(1-2M/R)^{1/2} -(1-2Mr^{2}/R^{3})^{1/2}}
    \right] \;\;\;\; \text{for} \;\;\; r < R.
\end{eqnarray}

To obtain the complete solution to the problem, it remains to determine the functions $e^{\nu(r)}$ and $e^{\lambda (r)}$ in the metric (\ref{schw_met_chap2}).  From (\ref{metric_pot_e}) and (\ref{m(r)}), we immediately find that

\begin{eqnarray}\label{grr}
g_{rr} = -e^{\lambda} = \begin{cases} \left(1-\frac{2}{r} \frac{4}{3}\pi r^{3}\rho_{*} \right)^{-1},  &  r < R , \\
                  \left( 1-\frac{2M}{r}\right)^{-1}, & r \geq R.
\end{cases}
\end{eqnarray}

    We specifically notice that the following solution agrees with the corresponding formula from the Schwarzschild metric for the exterior solution at the star's surface, where $r = R$. The equation for the function $e^{\nu (r)}$ is given by (\ref{m(r)}), and (\ref{p(r)}). By applying the boundary constraint that $e^{\nu(r)}$ equals the appropriate expression in the Schwarzschild metric at $r = R$, one may fix the integration constant. Accordingly,

\begin{eqnarray}\label{gtt}
\sqrt{g_{tt}} = \begin{cases} \frac{3}{2} \left( 1-\frac{2M}{R}\right)^{1/2} -\frac{1}{2}\left(1-\frac{2Mr^{2}}{R^{3}}\right)^{1/2},  &  r < R , \\
                  \left( 1-\frac{2M}{r}\right)^{1/2}, & r \geq R.
\end{cases}
\end{eqnarray}

The most essential property of the Schwarzschild constant-density solution
described above is that it sets a limitation relating the star's mass $M$
and its (coordinate) radius $R$. To derive this limit, one notes that (\ref{p_c})  implies that $P_{c} \rightarrow \infty$ as $ M/R \rightarrow 4/9$. In other words, if we define the stellar compactness as the dimensionless ratio $\eta := M/R$ then an infinite pressure is required to maintain a star
with compactness bigger than the critical value, $\eta_{\text{crit}} := 4/9\simeq 0.44 $. If the star matter is squeezed to attain a compactness $\eta_{\text{crit}}$  or greater, it can only collapse to form a \textit{blackhole}. The relevance of this conclusion, which is also known as Buchdahl's limit is that while we have obtained it for a constant energy-density star, a theorem by Buchdahl (1959) establishes that it holds for any equation of state, so that the radius of any relativistic star must be bigger
than $R_{\text{crit}}$
\begin{eqnarray}\label{buchdal_chap2}
    R> R_{\text{crit}} := \frac{9}{8}r_{g} = \frac{9}{4} M, \;\;\;\;\;\;\;
    \frac{M}{R} < \frac{4}{9}.
\end{eqnarray}

where $r_{g} := 2M$ is the Schwarzschild radius\footnote{A simple proof of Buchdahl's theorem can be found in Weinberg (1992)\cite{weinberg1972gravitation}}. The condition (\ref{buchdal_chap2}) may manifest as a very limiting one, although this is essentially the case only for compact stars. To fix concepts, the critical radius for the Sun is $\sim 3$ km, but it is as small as $\sim 1 $ cm for the Earth. In practice, most realistic equations of state lead to neutron-star models with compactness $0.1 \leq \eta \leq 0.2 $. Buchdahl's theorem states that (\ref{buchdal_chap2}) is actually valid for \textit{any} equation of state, despite the fact that we have only shown that this constraint holds to an object with constant density \cite{hobson2006general}.

\section{Spacetime curvature inside and outside a compact star}

This section describes how to compute the curvature both within and outside of a NS for various nuclear EoSs.

The energy and momentum of all matter and radiation exist in the cosmos are strongly integrated with the curvature of space-time according to the general theory of relativity. The strength of the curvature is determined by the distortion caused by a massive object in space-time, which is mechanically similar to how a spring works \cite{biswal_2020}. Certain mathematical concepts, such as compactness , Riemann tensor , Ricci scalar, and Kretschmann scalar, have been developed in order to explain this unified theory of gravitation in terms of basic geometric algebra. Ref. \cite{ekcsi_2014} has a more extensive explanation and derivation of these mathematical parameters, which include a wealth of information concerning curvature.

To measure the curvature of space-time, the quantities considered are the Ricci scalar, full contraction of the Ricci tensor, Kretschmann scalar ($\mathcal{K}$) (full contraction of the Riemann tensor), and the full contraction of the Weyl tensor ($\mathcal{W}$). Among these values, $\mathcal{K}$ and $\mathcal{W}$ stand out for measuring the space-time curvature both within and outside the star.

With the boundary conditions $\rho = \rho(r) $, $m=m(r)
$, $P(R) = 0$, and $m(R) = M$, we were able to solve Eqs. (\ref{tov_m}) and (\ref{tov_p}). We used the Skyrme Lyon (SLy) \cite{douchin_2001} and the FPS model by Lorenz et al. \cite{lorenz_1993} as the EoS, with $P=P(\rho)$. We  found that our findings are independent of the EoS that is used.
We also determine the compactness ($\eta$) in terms of $\rho(r)$, $P(r)$, and 
$m(r)$, which is defined according to the following equation

\begin{eqnarray}
    \eta(r)\equiv \frac{2G m(r)}{r c^2},
\end{eqnarray}
the Ricci scalar \cite{ekcsi_2014}
\begin{eqnarray} \label{rici_scal}
    \mathcal{R}(r) = \kappa (\rho c^2-3P), \;\;\;\;\;\; \kappa\equiv \frac{8\pi G}{c^4},
\end{eqnarray}
the full contraction of the Ricci tensor
\begin{eqnarray} \label{rici_ten}
    \mathcal{J}^2 \equiv \mathcal{R}_{\mu \nu} \mathcal{R}^{\mu \nu} = \kappa^2 [(\rho c^2)^2+3P],
\end{eqnarray}
the Kretschmann scalar (the full contraction of the Riemann tensor)
\begin{align}\label{kret_full}
    \mathcal{K}^{2} &\equiv \mathcal{R}^{\mu \nu \alpha \beta}  \mathcal{R}_{\mu \nu \alpha \beta} \nonumber\\
       &= \kappa^2 \left[  3(\rho c^2)^2 +3P^2+2P \rho c^2\right] - \kappa\rho c^2  \frac{16 G m}{r^3 c^2} + \frac{48G^2 m^2}{r^6c^4},
\end{align}

and the full contraction of the Weyl tensor

\begin{eqnarray}\label{weyl_full}
    \mathcal{W}^2= \frac{4}{3} \left( \frac{6 G m}{c^2 r^3} -\kappa \rho c^2\right)^2 .
\end{eqnarray}

inside  the star. Appendix \ref{App:AppendixC} contains the derivations of these curvature scalars for the spherically symmetric metric in GR.

 \begin{figure}[!tbp]
  \centering
  \subfloat[]{\includegraphics[width=0.8\textwidth]{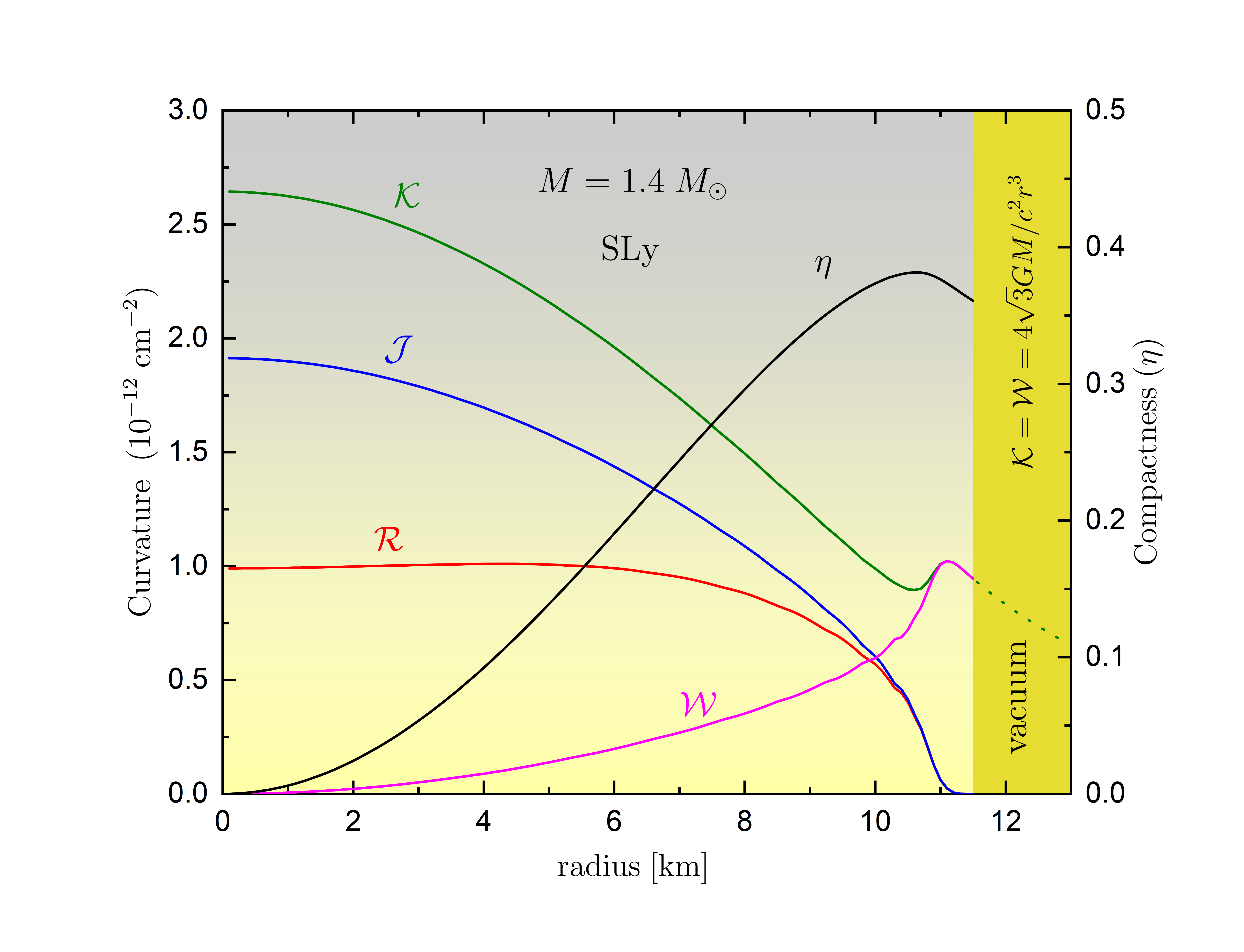}\label{sly_curv}}
  \hfill
  \subfloat[]{\includegraphics[width=0.8\textwidth]{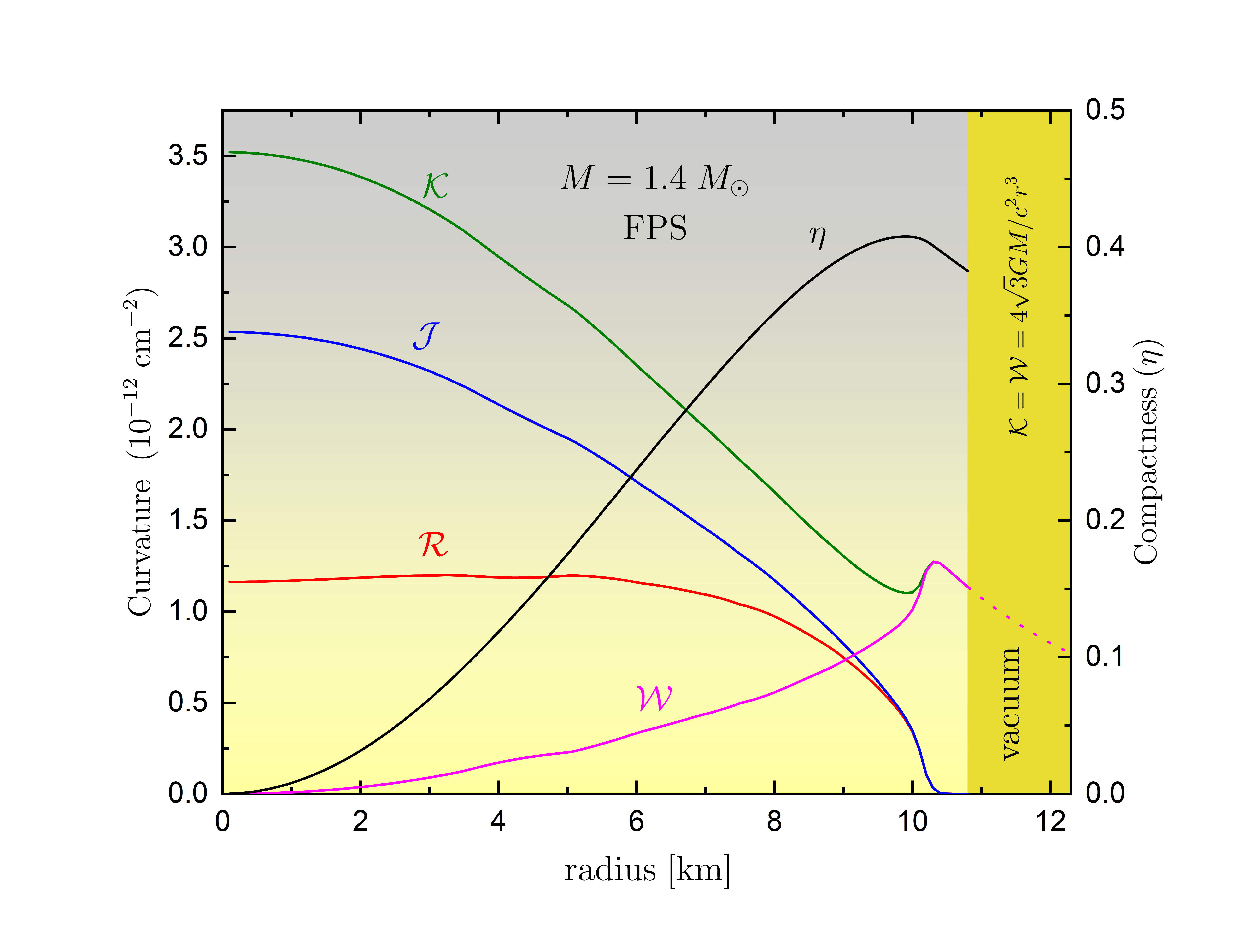}\label{fps_curv}}
  \caption{The neutron star's $\mathcal{R}$, $\mathcal{J}$, $\mathcal{K}$, and $\mathcal{W}$ curvature scalars as well as its compactness $\eta$. The SLy EoS with $M=1.4 M_{\odot}$ and $R=11.55$ km is shown in the upper panel (a), while the FPS EoS with $M=1.4 M_{\odot}$ and $R=10.81$ km in GR is shown in the lower panel (b). The yellow-shaded area represents vacuum, and the dotted lines represent $\mathcal{K}=\mathcal{W}= 4 \sqrt{3} G M/ c^2 r^3$. $\mathcal{K}_c = 2.64 \times 10^{-12} $ cm$^{-2}$ for the SLy and $\mathcal{K}_c =3.52 \times 10^{-12} $ cm$^{-2}$  for the FPS in the center.
  The curvatures and compactness measured in the Solar System experiments, $\mathcal{K}_{\odot} \sim 10^{-28}$ cm$^{-2}$ and $\eta_{\odot} \sim 10^{-5}$, are orders of magnitude less than the curvatures and compactness measured at the crust, $\mathcal{K }\simeq \mathcal{W} \sim 10^{-12}$ cm$^{-2}$ and $\eta \simeq 0.39$.
  }
\end{figure}

It is crucial to see how these curvature scalars are important   as a gauge for neutron star curvature? The Ricci scalar $\mathcal{R}$ and its components, the Ricci tensor $\mathcal{R}_{\mu \nu}$, both vanish outside the star, as is widely known. Since they only operate on $P$,  which vanish at the surface, Eqs. (\ref{rici_scal}) and (\ref{rici_ten}) also indicate the absence of $\mathcal{R}$ and $\mathcal{R}_{\mu \nu}$ in  vacuum\footnote{Due to $r\rightarrow R$, $m \rightarrow M$ at the surface. Since they depends on the $\rho(r)$ and $P(r)$, both of which are zero outside the star, the Ricci tensor and scalar vanish outside the star.}. However, this does not suggest that the Schwarzschild metric's representation of the spacetime of the vacuum is not curved, since the Riemann tensor in a vacuum contains nonvanishing components, such as $\mathcal{R}^1_{ 010} = -2GM/r^3c^2$. Therefore, compared to the Ricci scalar and the components of the Ricci tensor, the non-vanishing Riemann tensor components are better measures of the curvature of spacetime. The square root of the Riemann tensor's full contraction is the Kretschmann scalar. From Eqs. (\ref{kret_full}) and (\ref{weyl_full}), it is clear that the vacuum value for both $\mathcal{K}$ and $\mathcal{W}$ is $\frac{4\sqrt{3}GM}{c^2 R^3}$. So, two appropriate measurements for the curvature inside the star are $\mathcal{K}$ and $\mathcal{W}$. They reach various values within the star, approaches to one another near the crust, and merge outside vacuum \cite{ekcsi_2014,biswal_2020}. The ratio of the curvature at the NS surface $\mathcal{K}(R)$ to the curvature of the Sun $\mathcal{K}_{\odot}$, or surface curvature (SC) =$\mathcal{K}(R)/\mathcal{K}_{\odot}$, is used to describe surface curvature. $\mathcal{K}(R)/\mathcal{K}_{\odot} \approx 10^{14}$, meaning that the NS curvature is $10^{ 14}$ times larger than that of Sun.
\begin{center}
\fbox{\begin{minipage}{25em}
\begin{itemize}
 \item  Neutron star: \\
$ \eta \equiv \frac{2 G M}{R c^2} =0.416 \;M_{1.4}R_{6}^{-1}$  \\ 
$\mathcal{K}\equiv \frac{4 \sqrt{3} G M}{R^{3}c^{2}} =1.18 \times 10^{-12} M_{1.4} R^{-3}_{6}$ cm$^{-2}$ \\
($M_{1.4}=M/1.4M_{\odot}$ and $R_{6}= R/10^6$ cm)  \\
\end{itemize}
\begin{itemize}
    \item The sun : \\
    $ \eta_{\odot} \equiv \frac{2 G M_{\odot}}{R_{\odot} c^2} =4.27\times 10^{-6} $, \\ 
$\mathcal{K_{\odot}}\equiv \frac{4 \sqrt{3} G M_{\odot}}{R_{\odot}^{3}c^{2}} =3.06\times 10^{-27} $ cm$^{-2}$\\
\end{itemize} 
\end{minipage}} 
\end{center}

We show in Fig.\ref{sly_curv} and \ref{fps_curv} the radial variation of the curvature scalars $\mathcal{R}$, $\mathcal{J}$, $\mathcal{K}$, and $\mathcal{W}$ as well as the compactness of two neutron stars, $M =1.4 M_{\odot}$ is taken as the same for both nuclear EoSs SLy \cite{douchin_2001} and FPS \cite{haensel_2004}. We used a central density of $\rho_c = 0.99 \times 10^{15}$ g cm$^{-3}$ to produce a $1.4 M_{\odot}$ star, which is equivalent to a central pressure of $P_c = 1.37 \times 10^{35}$ dyne cm$^{-2}$ for the SLy EoS. For the FPS EoS,  We used $\rho_c = 1.30 \times 10^{15}$ g cm$^{-3}$ to construct a star of $1.4 M_{\odot}$, which  corresponds to $ P_c = 1.08 \times 10$$^{35}$ dyne cm$^{-2}$.
At the centers of the star for the EoSs  SLy and FPS, $\mathcal{K}_c =2.64 \times 10^{-12}$ cm$^{-2}$ and $\mathcal{K}_c =3.52 \times 10^{-12}$ cm$^{-2}$ are observed, respectively. $\mathcal{K}$ drops significantly as it approaches the surface, with the remaining $15$ orders of magnitude more than $\mathcal{K}_{\odot}$. If one considers $\mathcal{K}$ as a gauge of curvature, one easily argues that GR is in a less explored regime than EoS across the whole star. We chose the square root of the full contraction of the Weyl tensor $\mathcal{W}$, which disappears at the center and decreases towards the surface as a parameter of curvature to give the EoS unknown a   ``chance`` to equal that of gravity (at certainly at the inner regions of the star).

These findings show that the gravitational strength inside a neutron star is orders of magnitude more than the gravitational strengths investigated in Solar System studies, emphasizing that the notion of the validity of GR in forecasting neutron star structure is an ``extravagant`` extension \cite{ekcsi_2014}. In contrast, the EoS is just an order of magnitude greater than what is being investigated in nucleon scattering studies. This implies not only that an M-R measurement could very well constrain gravity models much as EoS models, but also that, because EoSs produce very different M-R relations, it would be impossible to investigate deviations from GR using any M-R measurement unless the EoS is fully constrained by Earth-based experiments \cite{cooney_2010,arapouglu_2011,deliduman_2012}.

Near the crust, the compactness and curvature both hit their peak levels. We are able to conclude that the crust of neutron star offers the optimum places in the Universe for finding violations from GR under extreme gravity since the EoS of the crust is reasonably well studied \cite{negele_1973,baym_1971}. Examples of such phenomena that the crust shows include the control of neutron star cooling and the glitch activity of newborn pulsars \cite{Baykal_2005,yakovlev_2004}.
Considering that GR is the least confined physics input in the description of the events by 14 orders of magnitude in curvature and 5 orders of magnitude in compactness, one may anticipate variations from GR to be noticed from, say, pulsar glitch activity. The latest observation \cite{andersson_2012} that the moment of inertia of the crust of neutron stars is insufficient to explain some of the glitch activity seen using the very accurate vortex creep model \cite{alpar_1984,alpar_1993} may indicate a deviation from GR under strong gravity rather than a need for model improvement.

\section{Gravity models for compact stars}

The study of spacetime curvature is the foundation of Einstein's general theory of relativity. Spherical symmetry is often used in  GR to analyze the features and behavior of gravitating objects. Because compact stars have a modest magnetic field, spherical symmetry may be considered \cite{wentzel_1961}. Furthermore, it is considered that spherical symmetry is required for many solutions of Einstein's field equations. Hillebrandt and Steinmetz \cite{hillebrandt_1976} investigated the stability of compact spherically symmetric objects. Das et al. \cite{das_2003} solved partial differential equations with initial value issues to develop a model of spherically symmetric gravitational equations. Herrera et al. \cite{herrera_2008} provided a method for solving Einstein's field equations in a spherically symmetric fashion.

The notion of internal fluid is crucial in understanding compact objects. A variety of fluid configurations, including ideal fluid, isotropic fluid, anisotropic fluid, etc., may be used to investigate various aspects of compact stars. The compact object's tangential and radial pressures are equal in an isotropic fluid. The difference between an anisotropic fluid and a perfect fluid is that an anisotropic fluid has unequal radial and tangential pressures. Due to the existence of a solid core and the occurrence of numerous physical processes in stars, compact objects may also manifest anisotropy in internal fluid pressure \cite{mardan_2021,gedela_2019}.

A different approach to modeling highly dense stars known as the hybrid model or core envelope model was considered by some researchers as a result of the uncertainty regarding the form of the equation of state for a stellar  matter beyond the nuclear regime.  A star's core and envelope are two distinct but contiguous layers. They include various matter distributions and possess various macroscopic physical attributes, such as mass of inertia, a standard measure for density, an assessment of mass, or radius. In the last few decades, researchers have investigated the core envelope models for massive relativistic stars \cite{mardan_2021,durgapal_1969,das_1975,negi_1990,sharma_2002,ramesh_2005,thomas_2005, hansraj_2016,mafa_2016,gedela_2019,mafa_2019}. For the core and envelope regions, Durgarpal and Gehlot \cite{durgapal_1969} developed specific interior solutions with various energy densities. The gravitational redshifts from the core-envelope interface to an object's surface under different situations and their impact on the equation of state were calculated by Das and Narlikar \cite{das_1975}.

The development of core-envelope models for changing physical characteristics, specifically, different EoS for core and envelope parts, was studied by Negi et al.\cite{negi_1990}.  The Vaidya-Tikekar model for a star with a quark phase in its inner layer and a less dense baryonic envelope was demonstrated by Sharma and Mukherjee \cite{sharma_2002}. The core portion of the star was proposed by Tikekar and Thomas \cite{ramesh_2005} for an anisotropic matter structure. According to Ward and Whitworth \cite{ward_2011}, the core region of the star plays a crucial role in the last phase of the formation of a compact star by undergoing fusion, which transforms lighter elements into heavier ones. Additionally, it maintains the star's gravity and pressure balance. The core region of the star becomes larger due to the increased population of heavier components.

A core-envelope star model of PSR J1614-2230 that has a core layer with a quark matter distribution with linear EoS and an envelope layer with a matter distribution with quadratic EoS was recently studied by Takisa et al. \cite{mafa_2019} to include anisotropic fluid. Some of the physical characteristics, including energy density, pressures, anisotropy, radial sound speed, and adiabatic index, were successfully confirmed to be continuous. Similarly, Gedela et al.\cite{gedela_2019} constructed core-envelope models, which estimate the properties of anisotropic compact stars using linear and quadratic EoS for core and envelope regions, respectively. 

In the next chapter, we construct EoSs using GR-based core-envelope models   and describe distinct classes of compact stars.

\chapter{Distinct class of compact stars based on the General Relativistic core-envelope framework}
\label{chap:3}

``\textit{Even the brightest minds aren't able to imagine all the consequences of their research}''                                
\begin{flushright}
-- Serge Haroche
\end{flushright}

\bigskip

Neutron stars are one of the densest objects in the observable universe. It represents state of  matter with highest densities. As such, they are valuable laboratories for  the study of dense matter. Such studies include  interplay between various disciplines like general relativity, high-energy astrophysics, nuclear and particle physics etc \cite{lattimer2004physics, potekhin_2010}. Neutron stars have masses
of about (1$\sim $3 \(\textup{M}_\odot\)). These stars with masses about 1.2 \(\textup{M}_\odot\)  have central densities more than normal nuclear matter density and  radius of the order of 10 km. The average mass density $\rho$ of  the neutron star is approximately $10^{15}$ g cm$^{-3}$, which is about 3 times  the nuclear saturation density $\rho_{n}$$=2.7 \times 10$$^{14}$ g cm$^{-3}$ and at the core $\rho$ $>$ $\rho_{n}$ \cite{glendenning_2012}. The magnetic field of such a compact stars lies  between $10^{8}$ -10 $^{15}$ gauss and possess gravitational field of  $2\times 10^{11}$ cm s$^{-2}$ times stronger than that of earth's gravitational field. The structure of these stars can be considered having an outer and  an inner crust. The envelope (outer crust) matter consists of  atomic nuclei (ions) and electrons. The thickness of envelope is few hundred meters. The inner crust occurs at a density of $4\times 10^{11}$g cm$^{-3}$ which consists of  neutron-rich atomic nuclei, free neutrons, exotic nuclear matter, hyperons etc. The thickness of this crust is typically about few kilometers. The outer crust envelopes the inner crust, which expands from the neutron drip density to a transition density $\rho_{tr}$ $\sim$ $1.0 \times 10^{14}$ g cm$^{-3}$. Furthermore, beyond the transition density one enters the \textit{core}, where all atomic nuclei have been melt down into their components, neutrons and protons. Caused by the high density and Fermi pressure, the core might also contain more massive baryon resonances or possibly a gas of free up, down and  strange quarks. Ultimately, $\pi$ and K mesons condensates may be found there too. 
In view of our inadequate knowledge of the equation of state of matter at extremely high densities ($\rho$ $>$ $\rho_{n}$), it is difficult to have proper elucidation of matter in the form of an equation of state and quantitative calculations for the structure of neutron stars become obscure. A methodical valuation on the structure and properties of neutron stars can be found in \cite{potekhin_2010, haensel_2007,datta_1988}(and references therein). Many theorists have developed theoretical models for the structure of neutron stars which may be made up of various layers including core (inner and outer), crust (inner and outer) in which atomic nuclei are arranged into a crystal and the liquid ocean composed of the coulomb fluid \cite{potekhin_2010}. The central region, i.e., core contains hyperons \cite{balberg_1999,weissenborn_2012} or quark matter \cite{witten_1984}. A detailed analysis of quarks core models are discussed by Bordbar, Bigdell and Yazdizadeh \cite{bordbar_2006}. 
All these dissimilar internal structure lead to different physical equation of state and hence contrasting mass-radius (M-R) relations. It then suggest to look for alternative methods devoid of the types of matter compositions  \par

This chapter is laid out as follows. We begin by introducing   our methodology in  section  \ref{sec:methodology_3}.  
Section \ref{sec:compact_structure} discusses stellar stability of compact stars in the context of two different geometrical EoSs. Other properties such as keplerian frequency, surface gravity, surface gravitational redshift and moment of inertia are also discussed.
Our results and discussions are summarised  in 
section \ref{sec:discussion_chap3}. For simplicity, we restrict ourselves to non-rotating spherically symmetric equilibrium models.

\section{Methodology}
\label{sec:methodology_3}

Alternative method to study compact high-density astrophysical objects is through the space-time metric of the general theory of relativity and solving the relevant Einstein's equations. Such attempts particularly for compact objects like  the neutron star exist\cite{vaidya_1982}. Thus, for the present study we make use of the core-envelope model for neutron stars  based on the geometric approach in the general relativistic framework . The core-envelope model of a neutron star\cite{thomas_2005,gedela_2019,mafa_2016} has different physical properties in envelope and core regions. From this we have considered two different EoS, based on anistropic pressure in core or envelope region. The core-envelope models studied by Thomas, Ratanpal and Vinodkumar\cite{thomas_2005}(TRV model)
have considered anisotropic pressure in the envelope region and isotropic pressure in the core region. While in another case studied by S. Gedela, N. Pant, J. Upreti and R. Pant (SNJR model)\cite{gedela_2019} have taken both  the core and envelope region as anistropic. In both the cases valid solutions of the Einstein's equations were studied in appropriate metrics. The EoSs deduced from such models are then used to compute the neutron star properties. Brief descriptions of these two models \cite{thomas_2005, gedela_2019} are given in the following section.

\subsection{ Core-Envelope Model in the General Relativistic Framework}
\label{sec:rel_space-time}

In general relativity, a model of an isolated star (or other fluid ball, such as boson stars\footnote{A boson star is a hypothetical astrophysical object made up of boson particles (conventional stars are formed from mostly protons, which are fermions, but also consist of Helium-4 nuclei, which comprise bosons).}) typically consists of an external region that is an asymptotically flat vacuum solution and a fluid-filled inner region that is theoretically a perfect fluid solution of the Einstein field equation. These two parts need to be precisely matched across the zero-pressure surface of a spherical world sheet. Compact nonrotating objects' outer regions are described in terms of the Schwarzschild solution.

Our primary focus in this chapter is based on the models belonging to the core-envelope family as discussed by Thomas et al. (TRV) and Gedela et al. (SNJR). We summarize below only the relevant parts of the formalism  adopted for the study of compact objects with appropriate geometric consideration. More details can be found in   \cite{thomas_2005,gedela_2019}.

We outline below the derivation of relativistic equations of stellar structure for static spherically symmetric star objects. The beginning point is the formula for the metric of a stationary, spherically symmetric space-time as discussed in chapter \ref{chap:2},

\begin{eqnarray} \label{schw}
    \mathrm{d}s^{2}= e^{\nu(r)}\mathrm{d}t^{2}-e^{\lambda(r)}\mathrm{d}r^{2}-r^{2}(\mathrm{d}\theta^{2}+\mathrm{sin}^{2}\theta \;\mathrm{d}\phi^{2})
\end{eqnarray}
where $t$ is a time-like coordinate, $r$ is a radial coordinate, $\theta$ and $\phi$ are the polar and azimuthal angles, respectively, while $\nu =\nu(r)$ and $\lambda =\lambda(r)$  are some functions of $r$.
 The right boundary condition for the metric is to match (\ref{schw}) with the Schwarschild exterior metric at the surface of the star. It is implemented as 
 
 \begin{eqnarray} \label{ext_schw}
   \mathrm{d}s^{2} =
     \bigg(1-\frac{2 G M}{R c^{2}}\bigg) \mathrm{d}t^{2}- \bigg(1-\frac{2G M}{R c^{2}}\bigg)^{-1} \mathrm{d}r^{2} 
   - r^{2} \mathrm{d}\theta^{2}- r^{2} \sin^{2} \theta \mathrm{d}\phi^{2} 
\end{eqnarray}
 \begin{eqnarray}
    \nu(r=R)=\ln \bigg(1-\frac{2GM}{R c^{2}}\bigg)
\end{eqnarray}
\begin{eqnarray}
    \lambda(r=R)=-\ln \bigg(1-\frac{2GM}{R c^{2}}\bigg)
\end{eqnarray}
\noindent Here, $R$ and $M$ is the radius and mass of the star.
One may simply analyse this metric and understand the significance of all variables. Let us emphasize that $r$ and $t$ are Schwarzschild coordinates, appropriate for an observer at infinity.

The relationship between geometry of space-time and material properties of
massive bodies is given by the Einstein equations
\begin{eqnarray}\label{EFE}
    \mathcal{R}_{\mu \nu}-\frac{1}{2} \mathcal{R} g_{\mu \nu }= \frac{8\pi G}{c^{4}} T_{\mu \nu}
\end{eqnarray}
\noindent where $\mathcal{R}_{\mu \nu}$ is the Ricci curvature tensor and $\mathcal{R}=\mathcal{R}^{\mu}_{\nu}$ is the scalar curvature, whereas $T_{\mu \nu}$ is the stress-energy tensor. The tensor $\mathcal{R}_{\mu \nu}$ and $\mathcal{R}$ are calculated from the metric tensor $g_{\mu \nu}$ (see Appendix \ref{App:AppendixB}).

It has been solved for the metric given by Eqn (\ref{schw})
for an energy momentum tensor relevant for perfect fluid \cite{thomas_2005,gedela_2019}
\begin{eqnarray}\label{EMT}
    T_{\mu \nu}=(\rho +p)u_{\mu}u_{\nu}-p g_{\mu \nu} +\pi_{\mu \nu}, \;\;\;\;\;\;\; u_{\mu}u^{\nu}=1
\end{eqnarray}
 where  $\pi_{\mu \nu}$ denotes anistropic stress tensor give by
 \begin{eqnarray}
     \pi_{\mu \nu}=\sqrt{3} S \bigg[C_{\mu}C_{\nu}-\frac{1}{3}(u_{\mu}u_{\nu}-g_{\mu \nu})\bigg]
 \end{eqnarray}

where $S$=$S(r)$ is the magnitude of anisotropy stress tensor and $C^{\mu}=(0,  -e^{-\frac{\lambda}{2}},0,0)$,  is a radial vector.

   Thomas et al \cite{thomas_2005} have discussed core-envelope model on pseudo-spheroidal spacetime with core consisting of isotropic distribution of matter and envelope with anisotropic distribution of matter. While anisotropic core-envelope models by assuming linear equation of state in the core and quadratic equation of state in the envelope have been studied by Gedela et. al \cite{gedela_2019}. In the following sub-sections we derive important aspects of these two models.
 
\subsection{The TRV Core-Envelope model}

It has been shown that core and envelope regions consist of different physical features. They have chosen ansatz for a pseudo-spheroidal geometry of spacetime to solve the Einstein's equations. According to their metric, potential for pseudo-spheroidal geometry is expressed as

\begin{eqnarray} \label{met_pot}
    e^{\lambda(r)}=\frac{1+K \frac{r^{2}}{\mathfrak{R}^{2}}}{1+\frac{r^{2}}{\mathfrak{R}^{2}}}
\end{eqnarray}

 \noindent where $K$ and $\mathfrak{R}$ are geometric variables.

 \noindent The energy momentum tensor components (\ref{EMT}) with anisotropic stress tensor $\pi_{\mu \nu}$ has non-vanishing components

\begin{eqnarray}\label{E_tensor}
    T^{0}_{0}=\rho, \,\,\,\,\ T^{1}_{1}=-\bigg(p+\frac{2S}{\sqrt{3}}\bigg),\,\,\,\,\ T^{2}_{2}=T^{3}_{3}=-\bigg(p-\frac{p}{\sqrt{3}}\bigg).
\end{eqnarray}

  \noindent The magnitude of anistropic stress is given by \cite{thomas_2005}
 \begin{eqnarray}
     S=\frac{p_{r}-p_{\perp}}{\sqrt{3}}
 \end{eqnarray}
 The boundary conditions for the core and envelope regions are
\begin{eqnarray}\label{chap3_cond}
   S(r)=0 \; \mbox{for} \;\;0\leq r\leq R_{C} \quad\text{and}\quad 
      S(r)\neq 0\;\;\; \mbox{for}\;  R_{C}\leq r\leq R_{E}
  \end{eqnarray}
   \noindent where $R_{C}$ refers to the core boundary radius and $R_{E}$ corresponds to the envelope boundary radius.  Making use of these conditions with Eqn. (\ref{schw}), (\ref{EFE}), (\ref{met_pot}) and (\ref{E_tensor}), the Einstein field equations give the equations for density and pressure.
   
The Einstein field equation (\ref{EFE})  relating to the metric (\ref{schw}) employing ansatz (\ref{met_pot}) is provided by a combination of three equations:

\begin{eqnarray}
    8\pi\rho = \frac{K-1}{\mathfrak{R}^2} \bigg(3+ K\frac{r^2}{\mathfrak{R}^2}\bigg) \bigg(3+ K\frac{r^2}{\mathfrak{R}^2}\bigg)^{-2},
\end{eqnarray}

\begin{eqnarray}
    8\pi p_{r} = \bigg[\bigg( 1+ \frac{r^2}{\mathfrak{R}^2}\bigg) \frac{\nu^{'}}{r}-\frac{K-1}{\mathfrak{R}^2}\bigg] \bigg(1+K \frac{r^{2}}{\mathfrak{R}^{2}}^{-1}\bigg),
\end{eqnarray}

\begin{eqnarray}\label{ani_eq}
\begin{split}
8\pi \sqrt{3}S 
&=-\bigg(\frac{\nu^{''}}{2}+\frac{\nu^{'2}}{4}-\frac{\nu^{'}}{2r}\bigg) \bigg(1+\frac{r^{2}}{\mathfrak{R}^{2}}\bigg) \bigg(1+K \frac{r^{2}}{\mathfrak{R}^{2}}\bigg)^{-1}+   \\
& \frac{(K-1)}{\mathfrak{R}^{2}}r \bigg(\frac{\nu '}{2}+\frac{1}{r}\bigg)  \bigg(1+K \frac{r^{2}}{\mathfrak{R}^{2}}\bigg)^{-2}+  \\
& \frac{K-1}{\mathfrak{R}^{2}}         \bigg(1+K \frac{r^{2}}{\mathfrak{R}^{2}}\bigg)^{-1}.
\end{split}
\end{eqnarray}

\noindent Accordingly, the density distribution ( core and envelope region ) is expressed as 
\begin{eqnarray}\label{density}
    \rho = \frac{1}{8\pi \mathfrak{R}^{2}}\bigg[ 3+2\frac{r^{2}}{\mathfrak{R}^{2}}\bigg] \bigg[1+2 \frac{r^{2}}{\mathfrak{R}^{2}}\bigg]^{-2}.
\end{eqnarray}

 \noindent where R is a geometrical parameter. Equation (\ref{density}) provides the density distribution in core and envelope region by using boundary condition for $0\leq r\leq R_{C}$ for core and $ R_{C}\leq r\leq R_{E}$ for envelope region.

\noindent The radial and transverse pressure in the envelope region is given by
\begin{eqnarray}\label{chap3_env_pre_r}
    8\pi p_{E}= \frac{C\sqrt{1+\frac{r^{2}}{\mathfrak{R}^{2}}}\big(3+4\frac{r^{2}}{\mathfrak{R}^{2}}\big)+D} 
    {\mathfrak{R}^{2} \big(1+2\frac{r^{2}}{\mathfrak{R}^{2}}\big)^{2}\big(C \sqrt{1+\frac{r^{2}}{\mathfrak{R}^{2}}}+D\big)},
\end{eqnarray}
\begin{eqnarray} \label{chap3_env_pre_t}
    8\pi p_{E_{\bot}}=   8\pi p_{E}
   -\frac{\frac{r^{2}}{\mathfrak{R}^{2}}\big(2-\frac{r^{2}}{\mathfrak{R}^{2}}\big)       }
  {\mathfrak{R}^{2} \big(1+2\frac{r^{2}}{\mathfrak{R}^{2}}\big)^{3}}.
\end{eqnarray}
 and the anisotropy parameter $S$ is obtained as
\begin{eqnarray} \label{chap3_env_ani}
  8\pi\sqrt{3} S= \frac{ \frac{r^{2}}{\mathfrak{R}^{2}}\big(2-\frac{r^{2}}{\mathfrak{R}^{2}}\big)}
  {\mathfrak{R}^{2}\big(1+2 \frac{r^{2}}{\mathfrak{R}^{2}}\big)^{3}}.
 \end{eqnarray}
 The constants $C$ and $D$ are to be obtained by matching the solution with the Schwarzchild exterior spacetime metric :
 \begin{eqnarray}
      \mathrm{d}s^{2}= \bigg(1-\frac{2m}{r}\bigg)\mathrm{d}t^{2}-\bigg(1-\frac{2m}{r}\bigg)^{-1}\mathrm{d}r^{2}-r^{2}(\mathrm{d}\theta^{2}+\mathrm{sin}^{2}\theta \;\mathrm{d}\phi^{2})
 \end{eqnarray}
 all over the boundary $r=R$ of the star, where $p_{r}(R)=0$.
 The consistency of metric potentials and pressure and along radial direction across $r=R$ suggests the following correlations :
 \begin{eqnarray}
     e^{\nu(R)}= \frac{1+\frac{R^{2}}{\mathfrak{R}^{2}}} {1+ 2\frac{R^{2}}{\mathfrak{R}^{2}}} = 1-\frac{2m}{R},
 \end{eqnarray}
 
\begin{figure}[h]
\centerline{\includegraphics[scale=0.5]{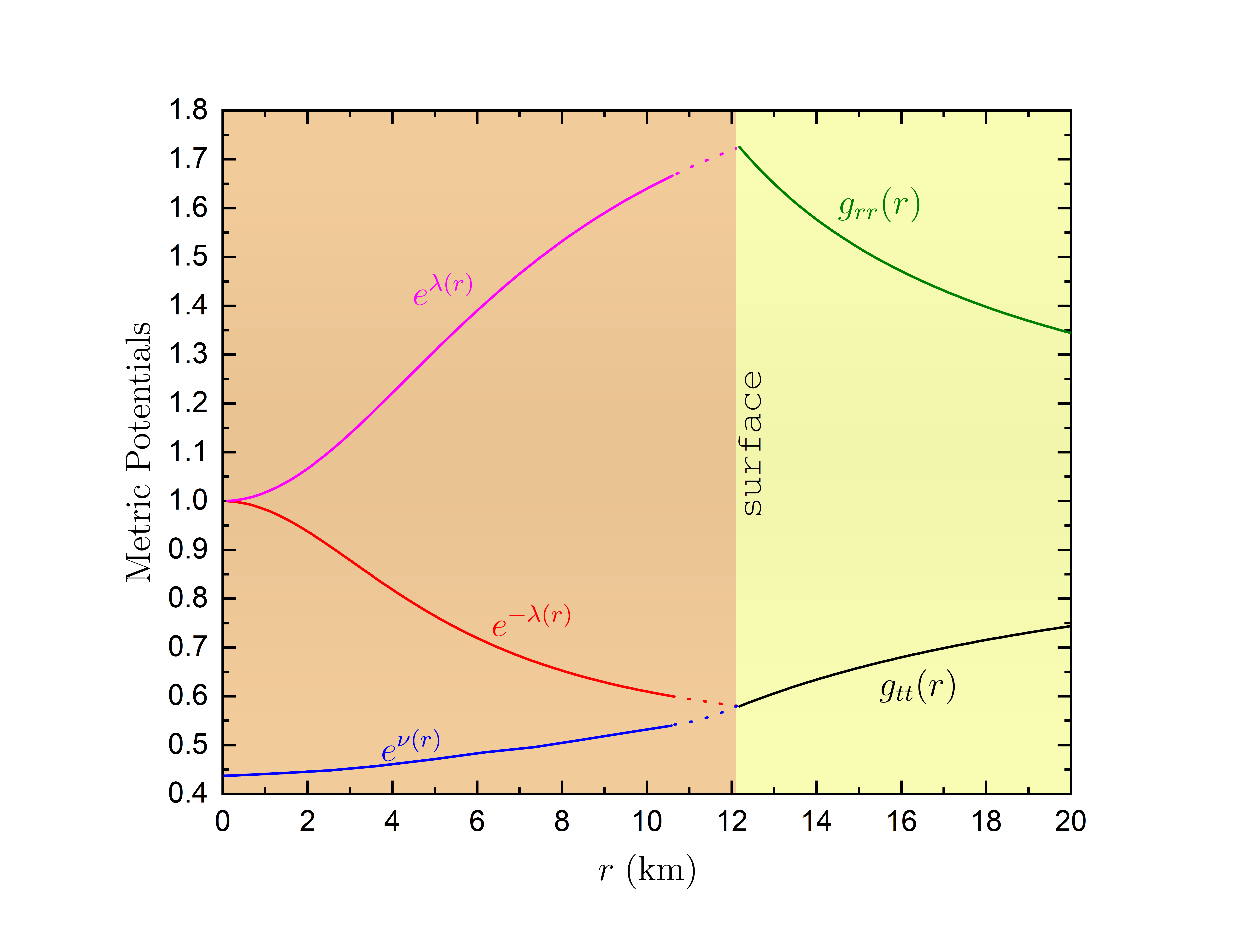}}
\vspace*{0pt}
\caption{ Metric function of static superdense stars as function of radius in units of km. The metric function $e^{-\lambda}$ is flat near the center and reaches a minimum near the surface of the star, where it joins the exterior Schwarschild metric, For a particular choice of $\lambda=0.07$. 
 \label{met_pot_trv}}
\end{figure}

 Fig.\ref{met_pot_trv} depicts the radial profile of the temporal and spatial metric potentials, $e^{\nu(r)}$ and $e^{-\lambda(r)}$, respectively. Both metric potentials are clearly finite at their centres and maintain consistent at all points. The redshift factor $e^{\nu(r)}$ steadily increases from the center of the star towards the asymptotic region. The metric function $e^{-\lambda(r)}$ is flat near the center and reaches a minimum near the surface of the star, where it joins the redshift factor. The metric potentials $e^{-\lambda(r)}$, $e^{\nu(r)}$  are continuous and well behave in the core and the envelope regions.

\begin{eqnarray}
C=-\frac{1}{2}\bigg( 1+2\frac{R^{2}}{\mathfrak{R}^{2}}\bigg)^{-\frac{7}{4}},
\end{eqnarray}
\begin{eqnarray}
    D=\frac{1}{2}  \sqrt{1+\frac{R^{2}}{\mathfrak{R}^{2}}} \bigg(3+4 \frac{R^{2}}{\mathfrak{R}^{2}}\bigg) \bigg(1+2 \frac{R^{2}}{\mathfrak{R}^{2}}\bigg)^{-\frac{7}{4}}.
\end{eqnarray}

\begin{figure}[H]
\centerline{\includegraphics[scale=0.50]{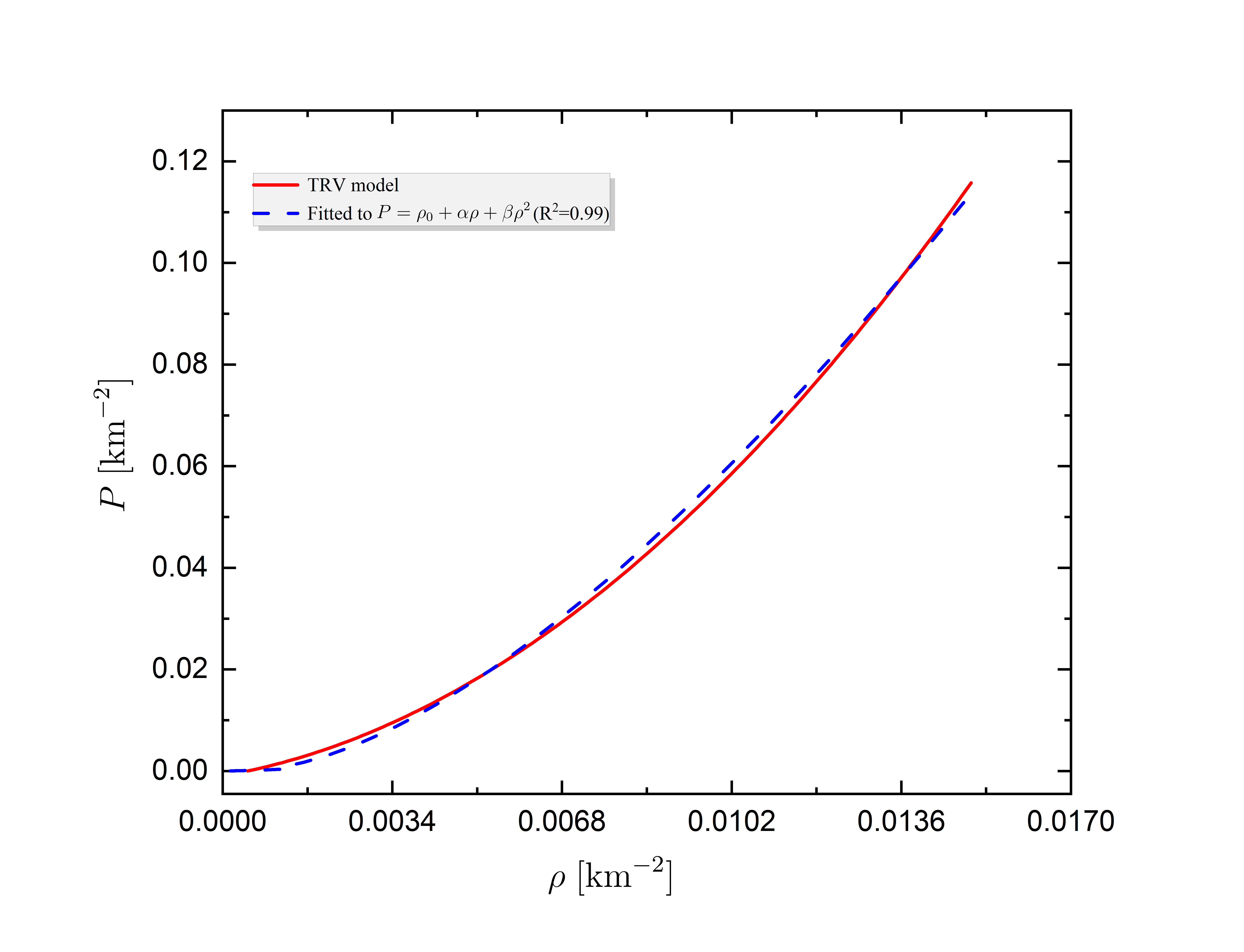}}
\vspace*{0pt}
\caption{The radial pressure and density are given by Thomas et al.\cite{thomas_2005}.(given in units of  km$^{-2}$), is plotted with solid curve. The dashed curve corresponds to the fitted curve with $\alpha= 406$, $\beta=1.69$ and $\rho_{0}=-9.30\times 10^{-4}$. For a density variation ($\lambda=0.01$).
 \label{trv_eos}}
\end{figure}

\noindent The radial pressure in the core  region is given by
\begin{equation}\label{chap3_rad_pre}
8\pi  p_{C}=  \frac{A\sqrt{1+\frac{r^{2}}{\mathfrak{R}^{2}}}+B \bigg[\sqrt{1+\frac{r^{2}}{\mathfrak{R}^{2}}}L(r)+\frac{1}{\sqrt{2}}\sqrt{1+2\frac{r^{2}}{\mathfrak{R}^{2}}}\bigg]}{ \mathfrak{R}^{2} \big(1+2\frac{r^{2}}  { \mathfrak{R}^{2}}\big) \bigg [A +\sqrt{1+2\frac{r^{2}}{\mathfrak{R}^{2}}}+B \bigg(\sqrt{1+\frac{r^{2}}{\mathfrak{R}^{2}}}  L(r) -\frac{1}{\sqrt{2}} \sqrt{1+ 2 \frac{r^{2}}{\mathfrak{R}^{2}}}\bigg) \bigg]}
\end{equation}
  where
\begin{eqnarray*}
    L(r)= \ln\left(\sqrt{2}\sqrt{1+\frac{r^{2}}{\mathfrak{R}^{2}}} +\sqrt{1+2 \frac{r^{2}}{\mathfrak{R}^{2}}}\right).
\end{eqnarray*}  
where $A$ and $B$ are given by
\begin{equation}
    A=\frac{[5\sqrt{5}- 3\sqrt{2} (\sqrt{3} L(R_{c})-\sqrt 2.5)   ] C+ \frac{1}{\sqrt{3}} [5 \sqrt{5}+2\sqrt{2}(\sqrt{3} L(R_{c})-\sqrt{2.5}     ]      D       }
    {5\frac{5}{4}},
\end{equation}
\begin{eqnarray}
    B= \frac{ \sqrt{2} } {5\frac{5}{4}} [3\sqrt{3}C-2D].
\end{eqnarray}
 Equation (\ref{density}) implies that the matter density at the center is explicitly related with geometrical variable $R$ as 
\begin{alignat}{2}
 \mathfrak{R}=\sqrt{\frac{3\lambda}{8\pi \rho(R)}} &\quad , \:  \lambda=\frac{\rho(R)}{\rho(0)}= \frac{1+\frac{2 R^{2}}{3 \mathfrak{R}^{2}}}
  {\left(1+2\frac{R^{2}}{\mathfrak{R}^{2}}\right)^{2}} 
\end{alignat}
In Fig.\ref{trv_eos}, we plot pressure vs density obtained from the TRV model ( solid line in  Fig. \ref{trv_eos}) for density variation parameter $\lambda=0.01$. It is found that the pressure-density curve can be fitted to a quadratic form (dashed curve of Fig. \ref{trv_eos}) as
\begin{eqnarray}
   p=\rho_{0}+\alpha\rho+\beta\rho^{2}
\end{eqnarray}
where  $\rho_{0}=-9.30\times 10^{-4}$, $\alpha = 406 $  and $\beta=1.69$ are the fitted parameters. It can be shown that the model reveals a quadratic equation of state for different choices of the density variation parameter $\lambda$.

\subsubsubsection{Physical Plausibility}

\begin{figure}[H]
\centerline{\includegraphics[scale=0.45]{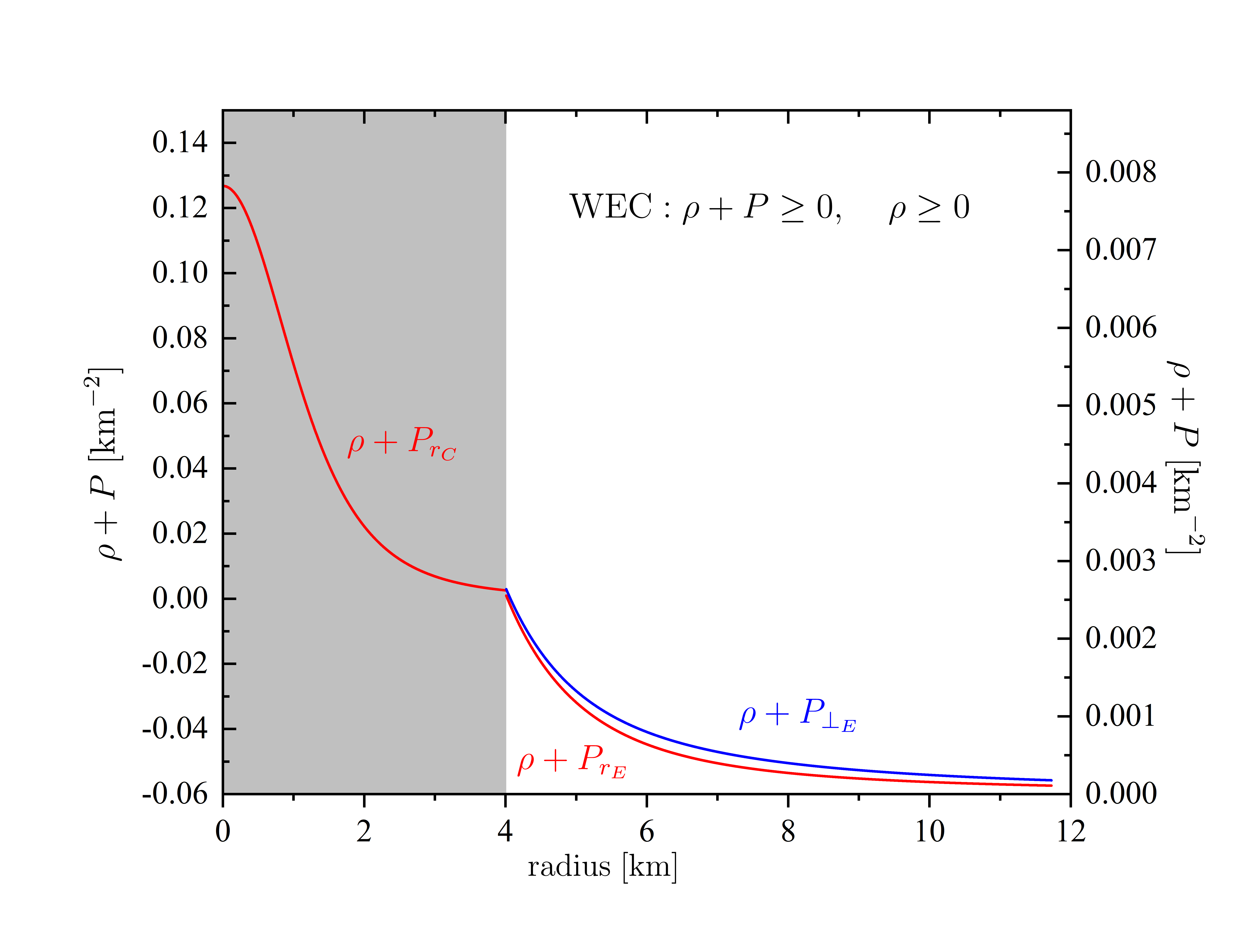}}
\vspace*{0pt}
\caption{Variation of  weak energy
conditions ($\rho+P \geq 0$, $\rho \geq 0$) with radial
coordinate. For a density variation ($\lambda=0.01$).
 \label{WEC}}
\end{figure}

\begin{figure}[H]
\centerline{\includegraphics[scale=0.45]{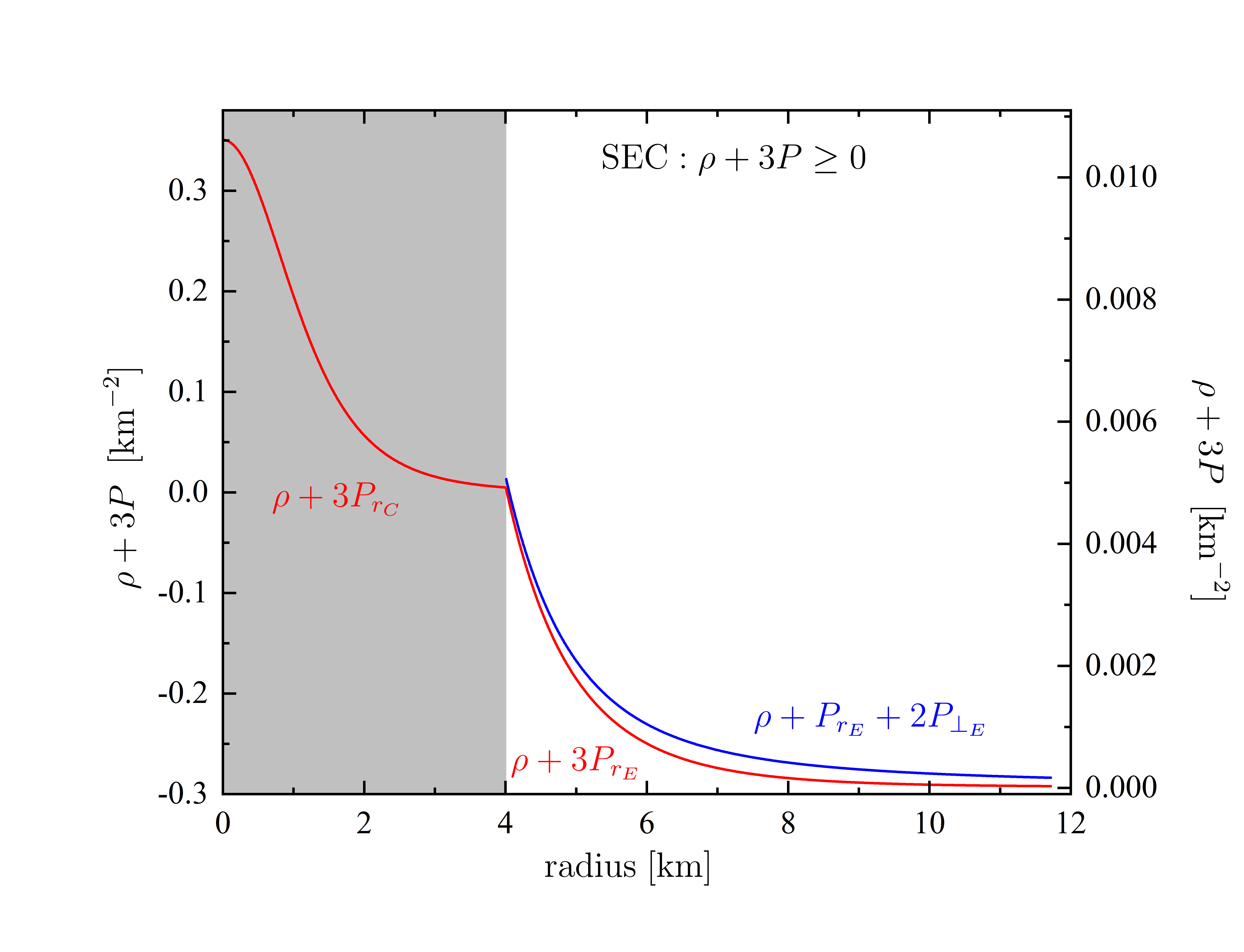}}
\vspace*{0pt}
\caption{Variation of  strong energy
conditions ($\rho+P_{r}+2 P_{\bot} \geq 0$) with radial
coordinate. For a density variation ($\lambda=0.01$).
 \label{SEC}}
\end{figure}

For a physically stable configuration, the core and envelope of the stars should satisfy the following inequalities concurrently (which are known as energy conditions \cite{maurya_2019}):

\begin{enumerate}[label=(\roman*)]
    \item Null Energy Condition (NEC) $\rho + P_{r} \geq 0$ 
    \item Weak Energy Condition (WEC) $\rho + P_{r} \geq 0$, $\rho \geq 0$ (WEC$_{r}$)
    and $\rho + P_{\bot} \geq 0$, $\rho \geq 0$ (WEC$_{\bot}$) and 
    \item Strong Energy Condition (SEC) $\rho + P_{r}+2 P_{\bot} \geq 0$ . 
\end{enumerate}
From the figures (\ref{WEC}) and (\ref{SEC}) it is clearly visible that the variation  of energy conditions with 
$r$ of the core and the envelope of the star is continuous at the interface and satisfying realistic conditions.

\subsection{The SNJR core-envelope model}
In the second case of core-envelope anisotropic model of Gedela et al.\cite{gedela_2019}, two distinct EoSs for core and envelope region are proposed.

The Einstein's field equations (EFEs) (with $G = c = 1$) considering the matter within the fluid sphere anisotropic is represented as
\begin{eqnarray}\label{EFE_2}
 - 8\pi  T^{\mu}_{\nu} =  \mathcal{R}^{\mu}_{\nu}-\frac{1}{2} \mathcal{R} g^{\mu}_{\nu}
\end{eqnarray}
where
\begin{eqnarray} \label{ETS_2}
    T^{\mu}_{\nu}= [(p_{\bot}+\rho)v^{\mu}v_{\nu}=p_{\bot}g^{\mu}_{\nu}+(p_{r}-p_{\bot})\chi^{\mu}\chi_{\nu}]
\end{eqnarray}

is the energy-momentum tensor, $\mathcal{R}^{\mu}_{\nu}$ is the Ricci tensor, $\mathcal{R}$ stands for scalar curvature, $p_{r}$, and $p_{\bot}$ are the energy density, radial pressure measured in the spacelike vector's direction, and transverse pressure measured in the perpendicular direction to $p_{r}$, respectively.

  The following system of equation is produced by the EFEs for the geometry and matter taken into consideration by the line element (\ref{schw}) and energy momentum tensor (\ref{EFE_2}).

\begin{eqnarray} \label{density_snjr}
    8 \pi \rho = \frac{(1-e^{-\lambda})}{r^{2}}+ \frac{\dot{\lambda} e^{-\lambda}}{r},
\end{eqnarray}
\begin{eqnarray}\label{pr_snjr}
    8\pi p_{r}= \frac{\dot{\nu} e^{-\lambda}}{r} + \frac{(1-e^{-\lambda})}{r^{2}},
\end{eqnarray}
\begin{eqnarray}\label{pt_snjr}
    8\pi p_{\bot} =\frac{e^{-\lambda}}{4} \bigg( 2 \ddot{\nu}+\dot{\nu}^{2}-\dot{\nu}\dot{\lambda} +\frac{2\dot{\nu}}{r}-\frac{2\dot{\lambda}}{r} \bigg),
\end{eqnarray}
where overhead dots denotes the derivatives with respect to the radial coordinate $r$.

Equations (\ref{pr_snjr}) and (\ref{pt_snjr}) allow us to calculate the measure of anisotropy ($\Delta$) as defined by

\begin{eqnarray} \label{anisotropy_snjr}
\begin{split}
\Delta & = 8\pi(p_{\bot}-p_{r}) \\
 & = e^{-\lambda} \bigg[  \frac{\ddot{\nu}}{2} -\frac{\dot{\lambda}\dot{\nu}}{4} +  \frac{\dot{\nu}^{2}}{4}- \frac{\dot{\nu}+\dot{\lambda}}{2r}+ \frac{e^{\lambda}-1}{r^{2}}
 \bigg].
\end{split}
\end{eqnarray}

The system of Eqs. (\ref{density_snjr} - \ref{anisotropy_snjr}) is transformed as a result of the following transformations: $x = r^2$, $z(x) = e^{-\lambda(r)}$, and $y(x) = e^{\nu(r)}$
\begin{eqnarray}\label{diff_den_snjr}
    8\pi \rho = \frac{1-z}{x}-2 z',
\end{eqnarray}
\begin{eqnarray}\label{diff_pres_snjr}
    8\pi p_{r}=2 z \frac{y'}{y}-\frac{1-z}{x},
\end{eqnarray}
\begin{eqnarray}
    8 \pi p_{\bot} = z \bigg[ 
    \bigg( \frac{2y''}{y}-\frac{y'^{2}}{y^{2}} \bigg)x+\frac{2y'}{y}\bigg] +
    \bigg(  1+x\frac{y'}{y}
    \bigg),
\end{eqnarray}
\begin{eqnarray} \label{diff_ani_snjr}
    8\pi \Delta = z \bigg( \frac{2y''}{y}-\frac{y'^{2}}{y^{2}} \bigg)x + z'\bigg(  1+x\frac{y'}{y}
    \bigg) + \frac{1-z}{x}
\end{eqnarray}

where ($'$) and ($''$) stand for $x$'s first and second derivatives, respectively.

In  a core-envelope model,  space-time geometry is divided into distinct
regions. These regions are formed of the core ($0\leq r \leq R_{C}$),
the envelope ($R_{C}\leq r \leq R_{E}$),  and the ( $R_{E} > r$ ) exterior. The three regions' matching line elements could well be considered  as
 \begin{itemize}
     \item The core region ($0\leq r \leq R_{C}$) :\\
      \begin{eqnarray}
   \mathrm{d}s^2\mid_{C} =  e^{\nu_{C}(r)}dt^{2}-e^{\lambda_{C}(r)}\mathrm{d}r^{2}-r^{2}(\mathrm{d}\theta^{2}+sin^{2}\theta  \mathrm{d}\phi^{2})
    \end{eqnarray}
 \end{itemize}
 \begin{itemize}
     \item The envelope region ($R_{C}\leq r \leq R_{E}$) : \\
     \begin{eqnarray}
     \mathrm{d}s^2\mid_{E} =  e^{\nu_{E}(r)}\mathrm{d}t^{2}-e^{\lambda_{E}(r)}\mathrm{d}r^{2}-r^{2}(\mathrm{d}\theta^{2}+sin^{2}\theta  \mathrm{d}\phi^{2})
\end{eqnarray}
 \end{itemize}
 \begin{itemize}
     \item  The Schwarzschild exterior metric solution given in Eq.(\ref{ext_schw}).
 \end{itemize}

(i) For core region ($0\leq r \leq R_{C}$), Gedela et al. \cite{gedela_2019} have derived a metric potential that satisfies the linear EoS  where in
\begin{eqnarray} \label{met_pot_snjr_linear}
    z=e^{-\lambda_{C}} =\frac{a x}{(b x +1)^{2}}+1,
\end{eqnarray}
and
\begin{eqnarray}\label{eos_liner_snjr}
    p_{r_{C}}= \alpha \rho -\beta ,
\end{eqnarray}
where $a$, $b$, $\alpha$ and $\beta$ are constants. Substituting $z$ value from Eq. (\ref{met_pot_snjr_linear}) in Eq. (\ref{diff_ani_snjr}) and utilizing Eqs. (\ref{diff_den_snjr}, \ref{diff_pres_snjr}, \ref{met_pot_snjr_linear}) and Eq. (\ref{eos_liner_snjr}), we get the following differential equation:

\begin{eqnarray} \label{diff_linear_snjr}
    \frac{y'(x)}{y(x)}+ \frac{a(3\alpha+(1-\alpha)bx+1)+8\pi \beta (bx+1)^3}               {2(bx+1)(ax+(bx+1)^2)}=0.
\end{eqnarray}

Eq. (\ref{diff_linear_snjr}) is integrated, and the result is

\begin{eqnarray} 
\begin{split}
y & = e^{\nu_{C}} \\
 & = c_{1}(bx+1)^{2\alpha} e^{p_{2}-4\pi \beta x} p_{1}^{\frac{2\pi a \beta}{b^{2}}-\alpha},
\end{split}
\end{eqnarray}
where
\begin{equation}\nonumber
    p_{1}= a x+ (bx+1)^{2},\\
    p_{2}=\frac{\sqrt{a}(4\pi \beta (a+2b)+(\alpha+1)b^{2})\tanh^{-1} \big(\frac{a+2b (bx+1)}{ \sqrt{a} \sqrt{a+4b} }\big)} 
    {b^{2}\sqrt{a+4b}},
\end{equation}
and $c_{1}$ is an integration constant.  In view of Eqs. (\ref{met_pot_snjr_linear}) and (\ref{diff_linear_snjr}) the system of Eqn. (\ref{diff_den_snjr})-(\ref{diff_ani_snjr}) becomes

For the core region $(0\leq r \leq R_{C})$, a linear EoS as given below is used\cite{gedela_2019}.
\begin{eqnarray}\label{c_eos_num_snjr}
    p_{C}=(0.170)\rho-(7.833 \times 10^{-5})
\end{eqnarray}
 \noindent The numerical values appeared in equation (\ref{c_eos_num_snjr}) are the same as given in\cite{gedela_2019}. The expressions of density and pressure for core region are given by
\begin{eqnarray}
    \rho_{C}= \frac{a(b r^{2}-3)}{8\pi (b r^{2}+1)^{3}}
\end{eqnarray}
\begin{eqnarray}
    p_{C}= \frac{a\alpha(b r^{2}-3)}{8\pi (br^{2}+1)^{3}}-\beta
\end{eqnarray}
\noindent where $a$, $b$, $\alpha$ and $\beta$ are constants whose numerical values are $-0.00735$ km$^{-2}$, $0.0038$ km$^{-2}$, $0.1707$ km$^{-2}$ and $0.7833 \times 10^{-5}$ km$^{-2}$, respectively.

(ii) For envelope region $(R_C\leq r \leq R_{E})$, they have considered quadratic EoS in the form
\begin{eqnarray} \label{E_eos_snjr}
    p_{E}=\kappa\rho^{2}-\gamma
\end{eqnarray}
\noindent
where  $\kappa$ ($=108$) and $\gamma$ (=$1.088 \times$ 10$^{-5}$) are free parameters. The numerical values of these constants have been fixed based on the observed properties of the Vela X-1 pulsar ( mass of the pulsar is estimated to be at least $1.88 \pm 0.13 M_{\odot}$ ) \cite{gedela_2019}.

\begin{figure}
\centerline{\includegraphics[scale=0.5]{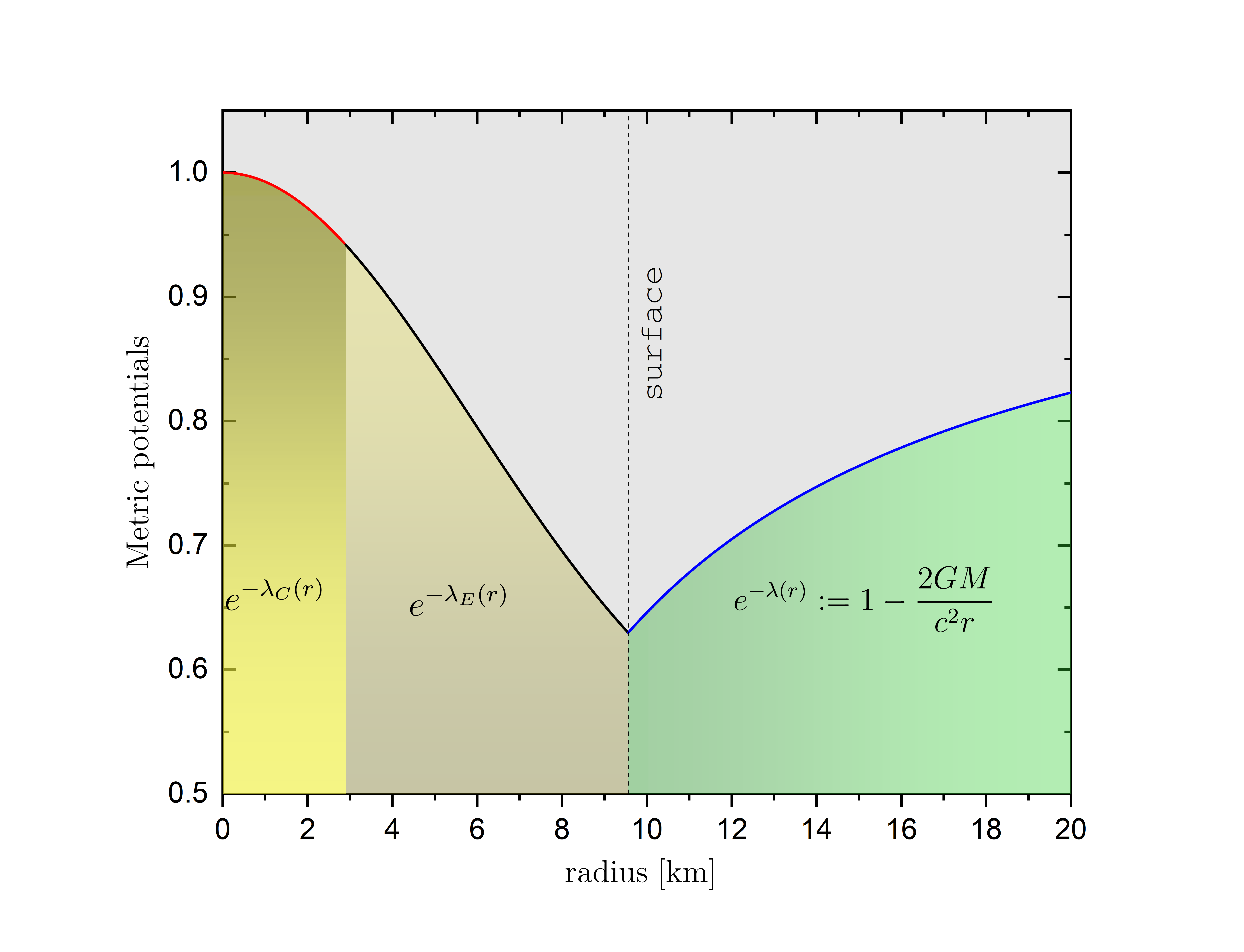}}
\vspace*{0pt}
\caption{ Metric function of static superdense stars as function of radius in units of km. The metric function $e^{-\lambda}$ (as given in Eq. (\ref{met_pot_snjr_linear})) is flat near the center and reaches a minimum near the surface of the star, where it joins the exterior Schwarschild metric (\ref{ext_schw}). 
 \label{met_pot_snjr}}
\end{figure}

The metric potentials of the core region should match smoothly with the gravitational potentials of the envelope region and also the gravitational potentials of the envelope region should link smoothly across the boundary  with the Schwarzschild exterior metric. The profiles of the two metric functions (interior and exterior) are shown in figure \ref{met_pot_snjr}
for a typical superdense star.

\begin{figure}[H]
\centerline{\includegraphics[scale=0.50]{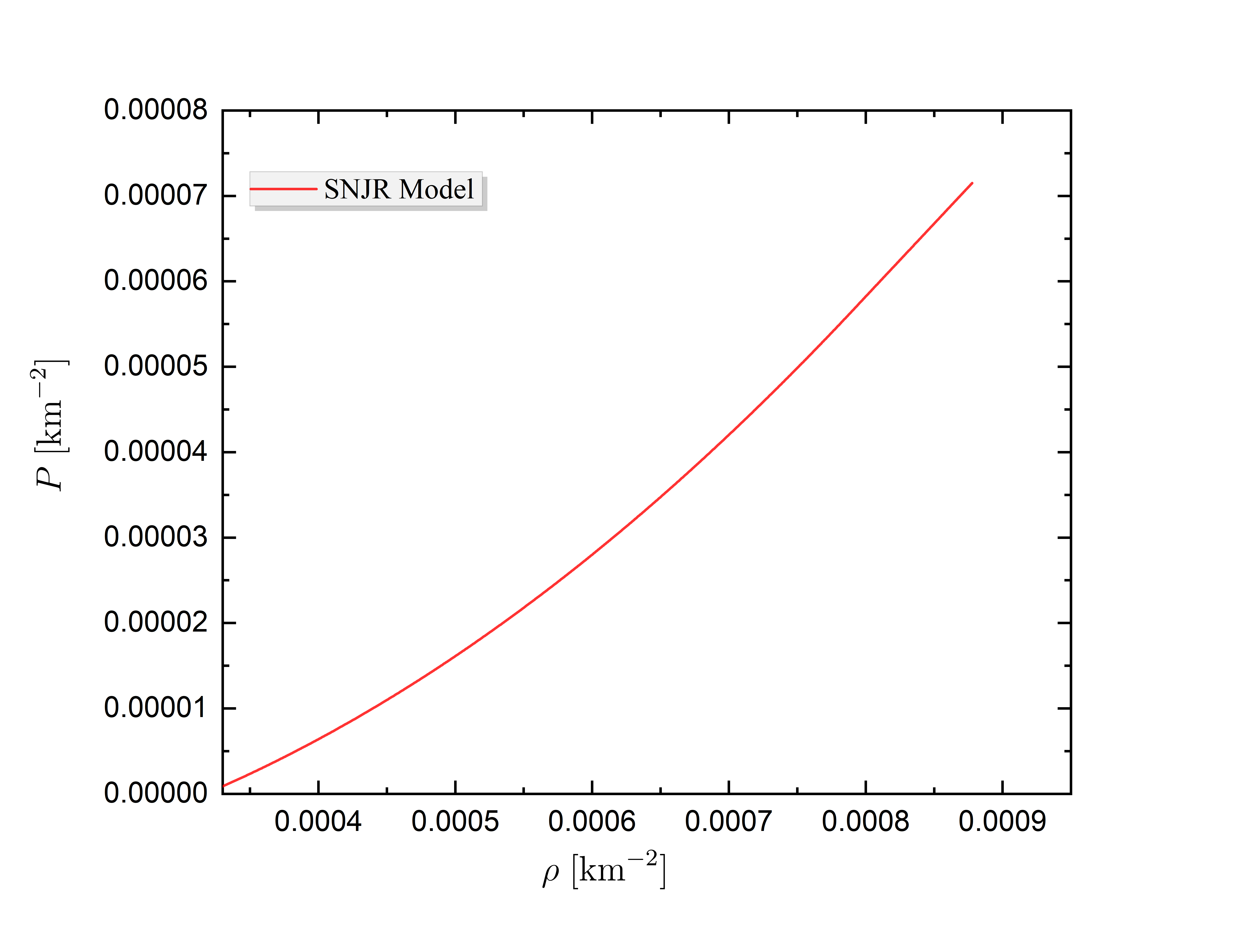}}
\vspace*{0pt}
\caption{  Variation of a pressure $P$ in  (km$^{-2}$) with respect to a density $\rho$ in (km$^{-2}$) . Figure based on Eqn. (\ref{c_eos_num_snjr}) and (\ref{E_eos_snjr}) with $\kappa=$$108$ km$^{-2}$ and $\gamma =$$1.088 \times$ 10$^{-5}$ km$^{-2}$.
 \label{snjr_eos}}
\end{figure}

In Fig. \ref{snjr_eos}, we plot pressure against density obtained from the SNJR model.

 \noindent An important feature of both of these core-envelope models (TRV and SNJR) is that they have the stable equilibrium under hydrostatic configuration. Theoretical study of the relativistic core-envelope model using  paraboloidal spacetime by Ratanpal and Sharma \cite{sharma2013relativistic} have shown that paraboloidal geometry also admit quadratic equation of state. Other EoSs that we have considered in the present work for comparison include those considering different physical compositions of nuclear matter reported by \cite{alford2005hybrid, akmal1998equation,potekhin2013analytical,engvik1995asymmetric,douchin2001unified,wiringa1988equation,zdunik2000strange}. The different models used in this study are listed in Table~\ref{ta1}.

\noindent Further, the density and pressure profile in the envelope region are given by\cite{gedela_2019} 
   \begin{eqnarray}
      \rho_{E}= \frac{a(br-3)}{8\pi (br-1)^{3}}
  \end{eqnarray}
\begin{eqnarray}
    p_{E}=\frac{a^{2} \kappa(b r^{2}-3)^{3}}{64\pi^2 (br^{2}+1)^{6}}-\gamma
\end{eqnarray}

\begin{figure}[H] 
\centerline{\includegraphics[scale=0.5]{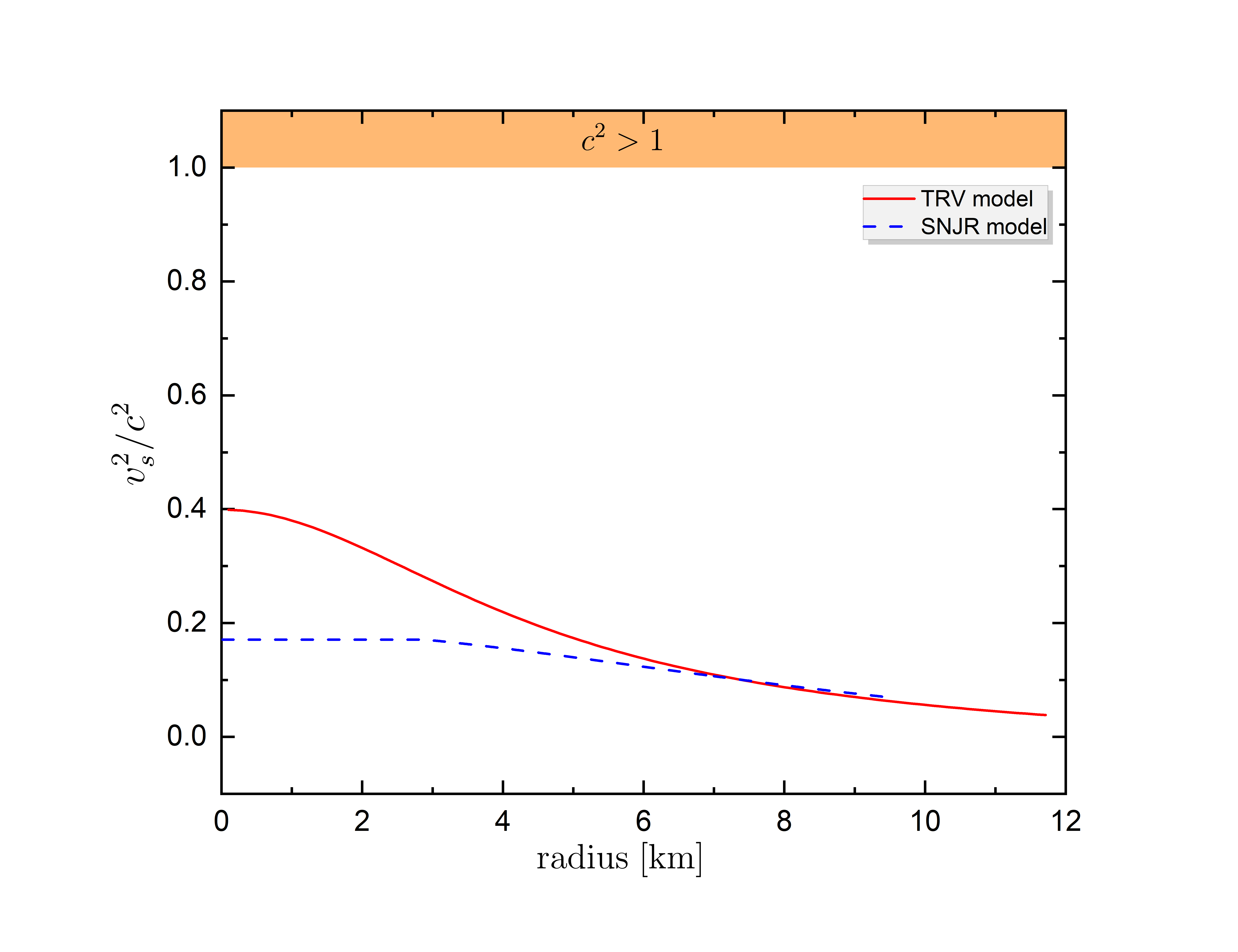}}
\caption{Velocity of sound, $v^{2}_{s}$, in unit of the speed of light, $c$, as a function of radius calculated for the TRV (Red solid line) equation of state (for $\lambda=0.01$) and SNJR (Blue dashed line) equation of state.
 \label{sound_trv_snjr} }
\end{figure}

  \noindent  A very crucial feature of the equation of state is the causal limit( a sound signal cannot propagate faster than the speed of light, $\nu^{2}_{s}=dp/d\rho\leq c^{2}$). In both cases based on the geometrical models (TRV and SNJR) the causality condition is satisfied. In particular, Thomas et al.\cite{thomas_2005} have studied the causality limit for different density variables. The computed  speed of sound ($\nu_{s}^2$) versus radius is shown in Fig.\ref{sound_trv_snjr}. Both the cases, clearly indicate the validity of causality condition.

\section{Compact Star Structure : Static Equilibrium configurations}\label{sec:compact_structure}
It is vital to explore static and spherical symmetrical gravity sources in general relativity, especially when it comes to internal structure of compact objects. For simplicity , we consider only nonrotating, spherically symmetric stars. The geometry inside the star is described by the familiar TOV equation, which is valid for a perfect fluid \cite{oppenheimer_1939}. The equation of state is all that is required to solve the TOV equations. For static, spherically symmetric stars in hydrostatic equilibrium,
the TOV equations may be written as a pair of first-order differential equations. The calculation of neutron star structure is obtained by
numerically integrating the Tolman-Oppenheimer-Volkoff equation \cite{oppenheimer_1939}
\begin{eqnarray}
\frac{dP}{dr}=- \frac{G m(r)\rho(r)}{r^{2}}\frac{\big(1+\frac{P(r)}{\rho(r) c^{2}}\big)\big[1+\frac{4\pi r^{2}P(r)}{m(r)c^{2}}\big]}{1-\frac{2Gm(r)}{r c^{2}}},
\label{tov_press}
\end{eqnarray}

\begin{eqnarray}\label{tov_mass}
\frac{d m(r)}{dr}=4\pi r^{2}\rho(r).    
\end{eqnarray}
Here $P$ is the radial pressure, $\rho$ is the mass density, $r$ is the radial distance measured from the center, and $m(r)$ is the enclosed mass from
the center $r = 0$ where $P=P_{c}$ and $\rho=\rho_{c}$ to a radial distance $r$.  In the present work we have fixed the central density for both geometrical models at $\rho_{c}= 1.34\times 10^{15}$g cm$^{-3}$. Equations (\ref{tov_press}) and (\ref{tov_mass}) are  integrated numerically to determine the global structure (e.g. radius and mass) of a neutron star.
\noindent To begin with, the density close to the center of the compact star is assumed to be homogeneous, with the density $\rho =\rho_{c}$, the radius $r=0.1$ cm and m(0.1 cm)= $4\pi \rho_{c} r^{3} /3 $. Equations (\ref{tov_press}) and (\ref{tov_mass}) are integrated numerically from $r=0.1$ cm to the boundary of the star, where the pressure falls to zero  ($P(R)=0$ ). The total mass of the star is then given by  $M= m(R)$.\\
  \noindent Using the data files provided by $\ddot{\text{O}}$zel et al. \cite{ozel2016masses}, we have re-ploted pressure-density profile  corresponds to all the model EoS's  listed in Table~\ref{ta1} along with geometric EoS's of TRV and SNJR. It can be seen that  these EoS's distinctly differ from each other.

\begin{table}[h]
\begin{center}
\resizebox{\columnwidth}{!}{
\begin{tabular}{ c c c c }
\hline
\bf{Label}    &\bf{EoS}  &\bf{Composition and model} &\bf{References} \\[1ex]
\hline\hline
    
1 & ALF1 & nuclear plus quark matter (MIT Bag Model)  &  Alford et al. (2005)\cite{alford2005hybrid} \\ [1.2ex]
\hline
    
2 & APR &  \makecell{$n p e \mu$, variational theory, Nijmegen NN \\ plus Urbana NNN potential}  & Akmal et al. (1998)\cite{akmal1998equation} \\ [1.2ex] 
\hline
 
3 &\hphantom{0} BKS19 & \makecell{cold catalyzed nuclear matter \\analytical  unified EoSs} & Potekhin et al. (2013)\cite{potekhin2013analytical}\\ [1.2ex]
\hline
 
4 & ENG &\makecell{Dirac-Brueckner HF\\asymmetric nuclear matter }  & Engvik et al. (1996)\cite{engvik1995asymmetric}\\ [1.2ex]   
\hline   

5 & SLy  & \makecell{potential method,  n p e $\mu $ \\effective nucleon energy functional  }  & \makecell{Douchin and \\ Haensel et al. (2001)\cite{douchin2001unified}} \\[1.2ex]
\hline

6& WWF1 & \makecell{variational method\\ dense nucleon matter}  &  Wiringa et al. (1988)\cite{wiringa1988equation}\\ [1.2ex]
\hline

7 & SQM1 & \makecell{MIT Bag Model \\ (Strange quark matter)} & Zdunik  (2000)\cite{zdunik2000strange}\\[1.2ex]
\hline

8 & TRV &\makecell{core   :   isotropic fluid distribution \\envelope : anistropic fluid distribution } & Thomas et al. (2005)\cite{thomas_2005}\\ [1.2ex]
\hline

9 & SNJR & \makecell{core  :  linear equation of state \\envelope : quadratic equation of state} & Gedela et al. (2019)\cite{gedela_2019}\\ [1.2ex]
\hline \hline

\end{tabular}
}
\captionsetup{
	justification=raggedright,
	singlelinecheck=false
}
\caption{Nuclear and Geometrical equations of state used  for the construction of models of general relativistic static neutron stars}
\label{ta1}
\end{center}
\end{table}

\begin{figure}[hb!]
\centerline{\includegraphics[scale=0.5]{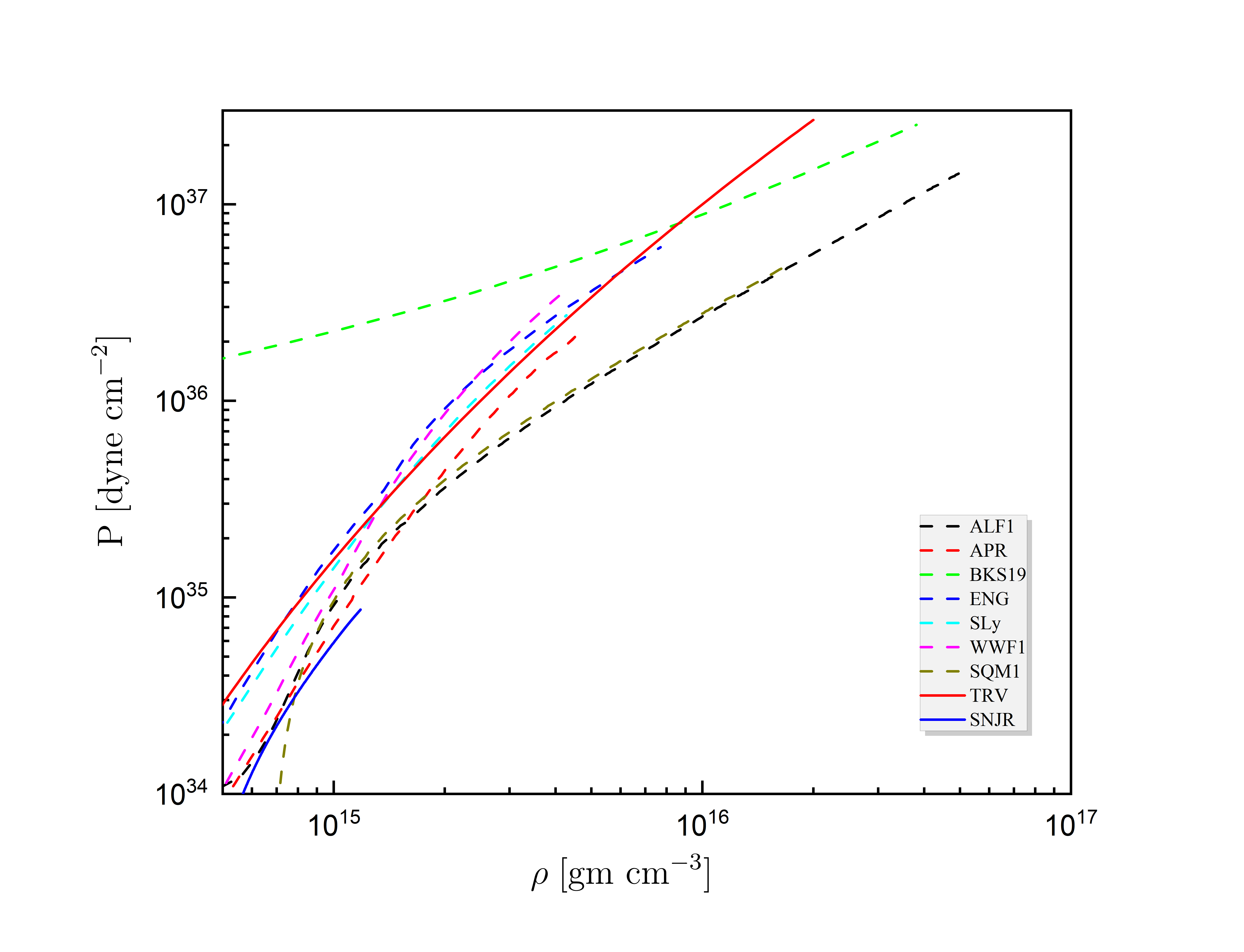}}
\vspace*{1pt}
\caption{Geometrical EoS TRV( red solid-line) and SNJR ( blue solid-line) compared to the selected nuclear EoS's ( ALF1 (black dash-line), APR (red dash-line), BKS19 (green dash-line), ENG (blue dash-line), SLy (cyan dash-line), WWF1 (magenta dash-line) and SQM1 (dark yellow dash-line). Details of these EoSs are listed in Table~\ref{ta1}.
 \label{f4}}
\end{figure}

\subsection{Mass, Central Density, and Radius of Neutron Star} \label{mr_section}
In this work, we have considered two general relativity inspired equations of state and  compared with seven different nuclear equations of states as listed in Table~\ref{ta1}. The pressure-density profile of selected EoSs are plotted in Fig.\ref{f4}.

\begin{figure}[h!]
\centerline{\includegraphics[scale=0.5]{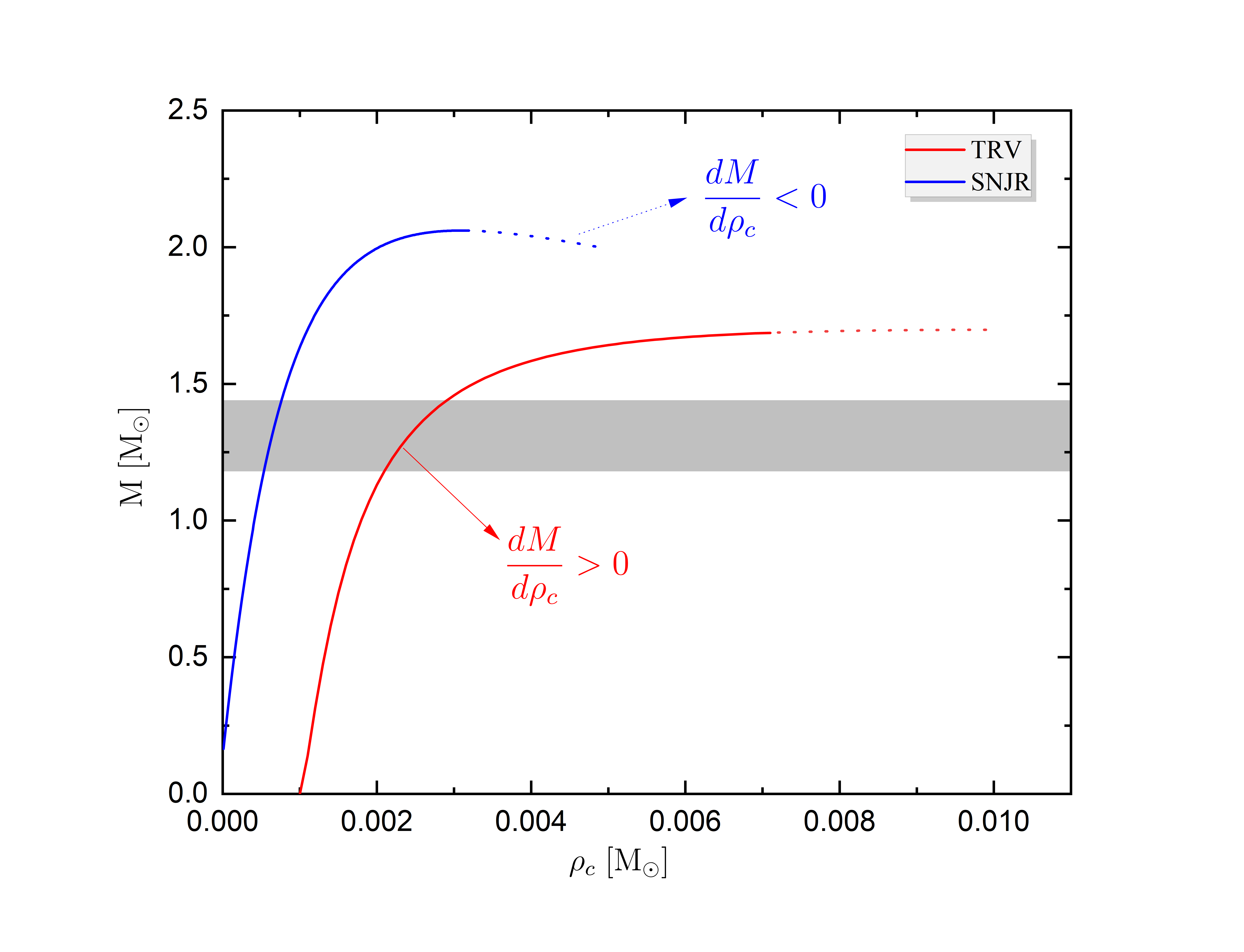}}
\vspace*{1pt}
\caption{ Mass as a function of central density (km$^{-2}$, in geometrical unit, see Appendix \ref{App:AppendixA}) for two geometrical EoS (TRV and SNJR). Configuration to the right of the maxima (dotted segments) are unstable with respect to small radial perturbations. The shaded band is the range of precisely measured masses of binary radio pulsars \cite{lattimer2012nuclear}
 \label{rhoc_mass}}
\end{figure}

The central density of the maximum allowable mass configuration is the maximum one which  can be reached within static neutron stars. Models with $\rho_{c} > \rho_{c}(M_{\text{max}})$ have $d M/d\rho_{c} < 0$. They are therefore unstable with respect to small radial perturbations and collapse into black holes. The  maximum central density of the static stable \footnote{One can show that in this case the stellar model is \textit{stable} if its mass $M$ increases with growing central density ($dM/d\rho_{c}>0$). This so called \textit{static stability criterion} \cite{harrison1965gravitation,jakov1971relativistic} is widely used in the literature. This condition is \textit{necessary} but not sufficient. The opposite inequality $\mathrm{d}M/\mathrm{d}\rho_{c}<0$ always implies instability of stellar models.
} neutron stars is, for the geometrical TRV EoS, $0.0070 \; \text{km}^{-2}$ ($=9.57\times 10^{15} \;\text{g cm}^{-3}$). The corresponding maximum values of stellar mass is $M_{\text{max}}= 1.68 M_{\odot}$. For the SNJR EoS, the permissible maximum mass configuration is $M_{\text{max}}=2.06 M_{\odot}$, for $\rho_{c}=0.0031 \;\text{km}^{-2}$ ($=4.29\times 10^{15} \;\text{g cm}^{-3}$). The stable branch of both geometrical EoS chosen in this study is shown by a solid line in Fig.\ref{rhoc_mass}, while the unstable branch is represented by a dotted line.
In Table \ref{ta2}, we present the maximum star masses obtained from the maximum stable value of the central density.

\noindent The composition and model used for all these equation of state and their respective bibliographic references are also listed in Table~\ref{ta1}.   Making use of these  equations of state, we obtained  the mass-radius relationship for a compact star. 

\begin{figure}[h!]
\centerline{\includegraphics[scale=0.5]{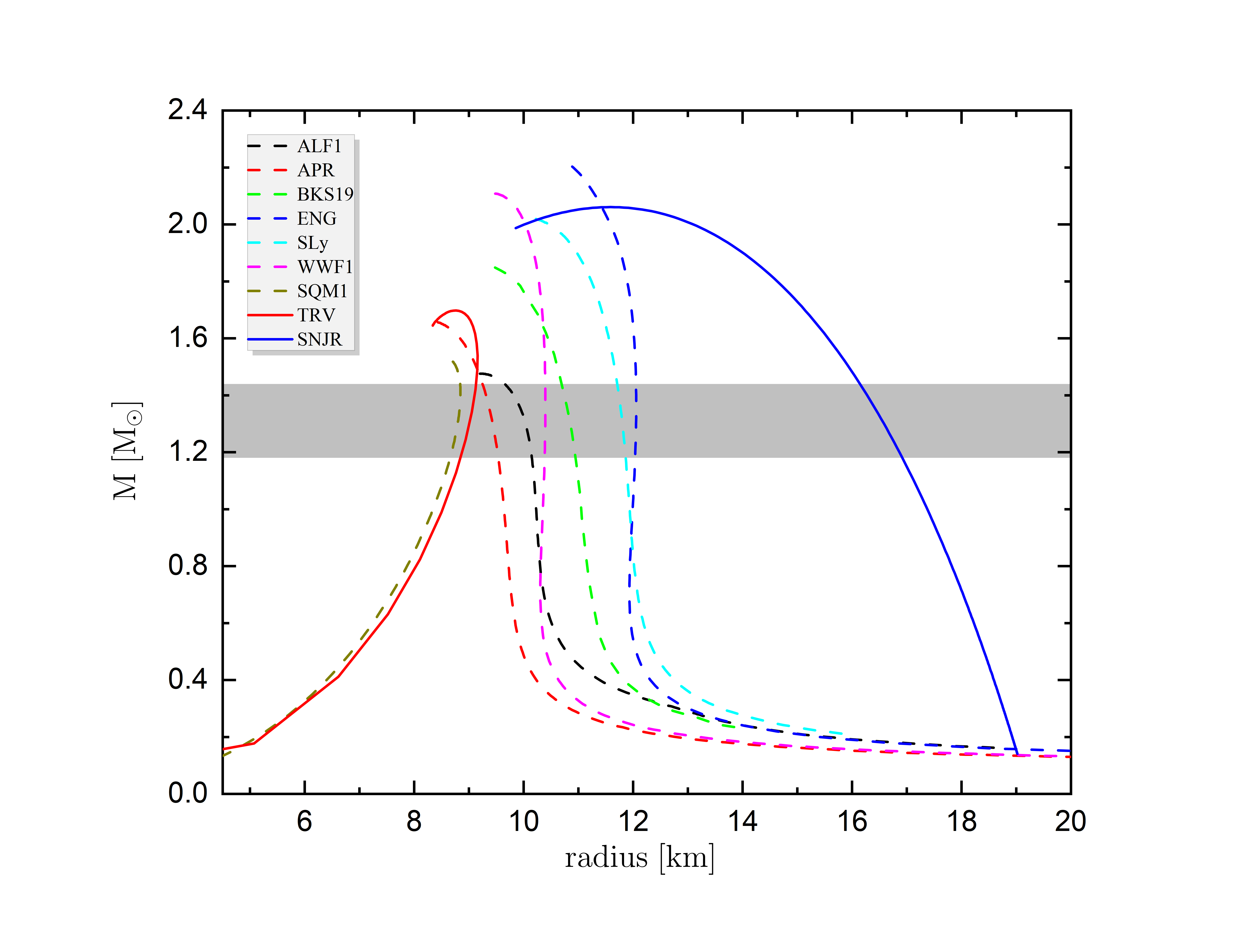}}
\vspace*{1pt}
\caption{  Neutron star mass as a function of radii for pure nuclear matter EoSs vs. geometrical EoSs. Where the dashed lines all describe the nuclear EoS and the solid lines the geometrical EoS.The labels are explained in Table~\ref{ta1}. In which the TRV EoS is calculated 
at $\lambda=0.01$, for the M-R relation shown. The maximum stable mass at corresponding maximum central density are represented in Table \ref{ta2}, for all the EoSs.  The horizontal grey band is the range of precisely measured masses of binary radio pulsars \cite{lattimer2012nuclear}
 \label{f5}}
\end{figure}

\noindent The mass-radius relations obtained with the help of nuclear equations of state of different compositions are compared with the geometrical equations of state and are plotted in Fig.~\ref{f5}. These plots reiterate the fact that  nuclear and geometrical equations of state  manifest three distinct types of compact stars. The first one corresponds to the two  cases represented by the models 7 and 8 of Table~\ref{ta1} , the second one corresponds to the models (1 to 6) largely represented by the nuclear matter EoSs and third type corresponds to the model (9) represented by the geometric model (SNJR). In all the three cases the maximum masses correspond to stable structure varies from 1.4 to 2.3 \(\textup{M}_\odot\), while the radius at their maximum masses lie 8 - 9 kms in the case of the first category, 9 - 12 kms in the cases of (second category) and  beyond 12 km in the case of the third category. The central density at maximum mass obtained here for the stable configurations are listed in Table~\ref{ta2}.

\noindent The M-R diagram obtained from the two geometrically deduced models behave differently. We found that TRV equation of state resulted into the mass-radius curve similar to the one obtained for strange quark matter stars (SQM1, label-7) \cite{li2010strange}. The monotonically increasing mass with radius ($M \propto R^{3}$) is expected for the class of ultra compact objects which are self-bound \cite{weber2012structure}. The surface density of strange star is roughly fourteen orders of magnitude larger than the surface density of normal neutron stars \cite{weber2012structure}. The TRV model gives a stable configuration in the same orders of magnitude, with the surface density,  $\rho_{s}$ $\approx$ $2 \times 10^{14}$ g cm$^{-3}$. Thus, it is an appropriate geometrical model for the study of  ultra compact stars having exotic matter composition.

\noindent The isotropic fluid distribution in the core part of the TRV model\cite{thomas_2005} is justified if the core matter distribution is of quarks or strange matter, governed by MIT bag model. Further, the envelope with anisotropic fluid distribution can be viewed as due to hadronization to baryonic matter. Thus the TRV model prediction fit well  with that of the strange quark matter stars with its maximum mass, ($M_{max}=1.69$ $M_{\odot}$) and radius, 8.76 km. The SNJR model that predicts the third category in which EoS has  the linear behaviour inside the core and quadratic behaviour at the envelope has resulted into the M-R diagram  different from all other cases. Its M-R curve is broader as compared to all other cases studied here. Recent observations of binary neutron-star mergers (GW170817) have reported an estimation for the radius of the  neutron star in the range 10.6 to 11.5 kilometers\cite{capano2020stringent}. 
 The goal of the next subsection is to constrains this mass-radius configuration with gravitational wave observations and  other astrophysical events.

\subsection{Observational constraints}
\subsubsubsection*{Gravitational wave observations}

For decades, scientists have hoped that measurements of gravitational waves (GWs) and related electromagnetic radiation from mergers of binaries composed of compact stars would place more stringent constraints on the properties of matter's equation of state  at densities higher and much higher than nuclear density. Since August 2017, when the first detection of gravitational radiation interpreted as the coalescence of a binary-neutron-star (BNS) merger was recorded as GW170817 \cite{abbott2017gw170817}, such anticipation has become a reality. There have been multiple follow-up observations of electromagnetic radiation \cite{abbott2017multi,abbott2017estimating} from what are assumed to be the merger's material ejecta (initiated by the GW detection).

The GW170817 \cite{abbott2017gw170817} event data from LIGO and Virgo were utilized by the LIGO/Virgo Collaboration (LVC) to cast the first gravitational wave constraints on the EoS \cite{abbott2018gw170817}, which we now discuss. The GW170817 event comprised of gravitational waves released in the inspiral, late inspiral and merger of a compact binary consisting of compact objects with masses ($m_{1},m_{2}$) $\sim$ ($1.5, 1.3$ $M_{\odot}$) \cite{abbott2017gw170817}.

The LVC employed two procedures. In the first approach, which  term the ``universal relations approach'' they employed roughly EoS-insensitive relations to execute EoS inferences. These are universal connections between the two tidal deformabilities of compact objects in a binary (the so-called binary Love relations) \cite{yagi2016binary,yagi2016approximate}, as well as between the tidal deformability of any one of the compact objects and their compactness (the so called Love-C relations) \cite{yagi2016approximate}.

The second LVC approach, which  refer to as the ``spectral EoS approach'' used a specific model of the EoS to carry out parameter estimation. The formula $ v^{2}_{s}= dp/d\rho =\Gamma(\rho)p/(\rho+p)$, where $p$ and $\rho$ represent pressure and  density, respectively, and $\Gamma$ is the adiabatic index for a polytrope, is used to construct the phenomenological spectral EoS \cite{tan2022extreme}.

There are advantages and downsides to both approaches. The universal relations method has the virtue that no functional form for the EoS has to be specified, but since the universal relations are not accurate, one must minimize over the remaining EoS sensitivity \cite{tan2022extreme}. The spectral EoS technique, on the other hand, has the benefit of immediately inferring a posterior on the EoS, allowing all EoS-dependent parameters to be easily estimated. Specifying a functional form for the EoS, on the other hand, has its own difficulties since it could bias nuclear physics conclusions in circumstances when Nature's EoS cannot be accurately represented by the spectrum EoS model.
However, the posterior distributions in the mass-radius plane obtained by these two approaches are consistent, with the spectral EoS posteriors being tighter than the universal relations \cite{raithel2019constraints,tan2020neutron}.

\subsubsubsection*{Other Astrophysical observed events :}

We consider here, the initial NICER  constraints on the EoS were given by the NICER team using x-ray data from PSR J0030 0451. PSR J0030 0451 is an isolated, old ( roughly 7.8 Gyr), millisecond pulsar with a period of approximately 4.87 ms \cite{riley2019nicer,miller2019psr}. We  refer to these approaches as the ``Amsterdam analysis'' (AM) and the ``Illinois-Maryland analysis'' (IL/MD) since they were used in the collaboration. The modeling of the hot spots, the exploration of the parameter space (e.g., Amsterdam employed nested sampling, Illinois-Maryland used Markov- Chain Monte Carlo sampling), and the priors used vary between the two approaches.
Although the discrepancies are statistically compatible with one another \cite{riley2019nicer,miller2019psr}, they resulted in distinct confidence regions in the mass-radius plane. The employment of MultiNest may have resulted in nonconvergent results with live-point number, according to a recent comparison report \cite{riley2021nicer}, which offers a possible reason for the difference between the IL/MD and the AM posteriors in the mass-radius plane.

The millisecond pulsar PSR J0740-6620, with a period of roughly 2.9 ms, in a binary system with a white dwarf has also lately been the subject of investigation by the NICER team \cite{riley2021nicer,miller2021radius}. The companion allowed pulsar radio measurements to determine that the pulsar's mass was around $2.14 \;\text{M}_{\odot}$. The NICER team was able to determine the background for this pulsar using x-ray data from XMM (X-ray Multi-Mirror) Newton. This data set was subjected to two analyses, as earlier, and two posteriors in the mass-radius plane were produced that are statistically compatible with one another \cite{riley2021nicer,miller2021radius}.

\begin{figure}[h!]
\centerline{\includegraphics[scale=0.5]{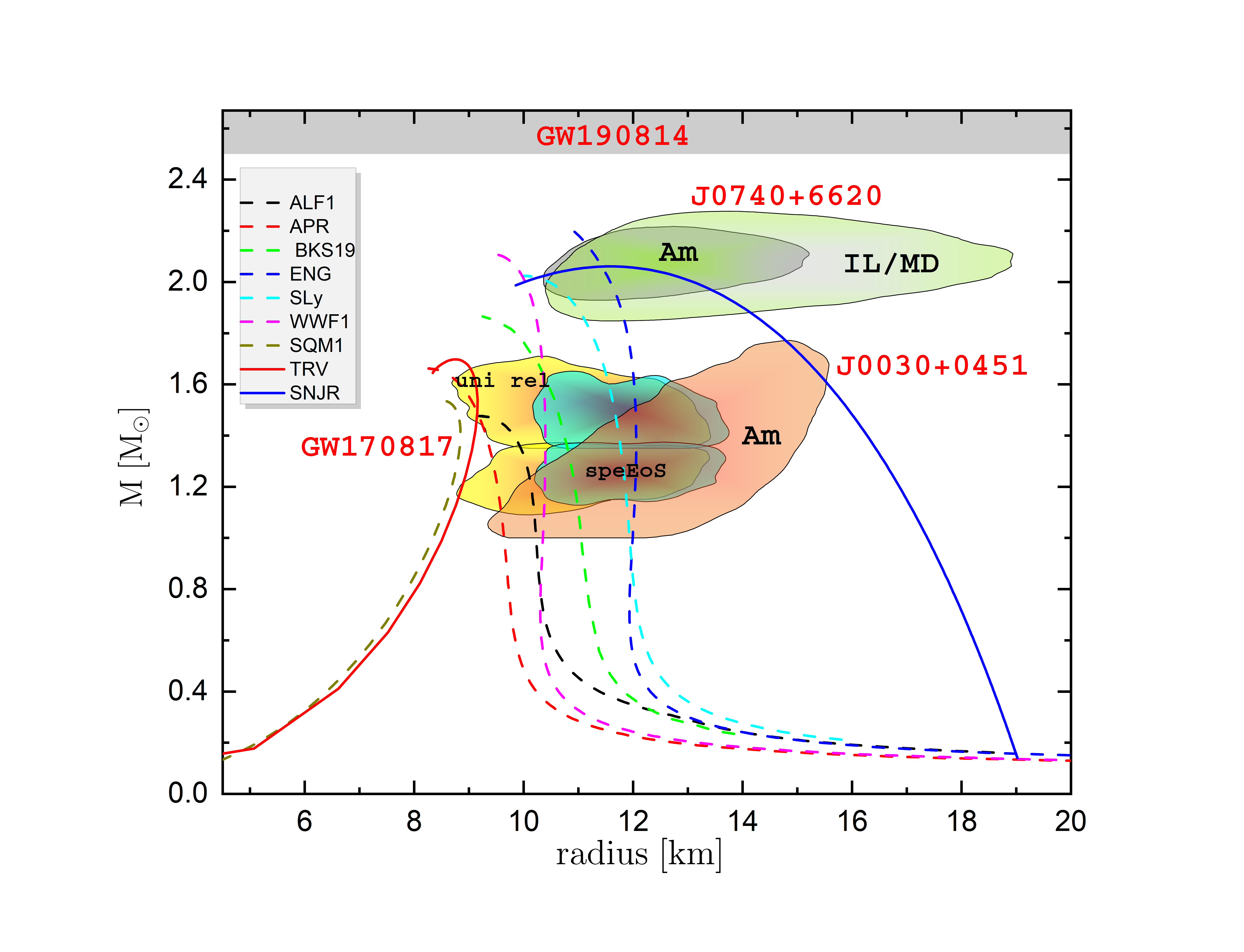}}
\caption{ Observational constraints on the theoretical neutron star mass-radius plane from LIGO/Virgo and NICER data. The yellow and cyan regions correspond to 90\%  confidence regions, obtained from the GW170817 event using the universal relations and spectral EoS approaches, respectively. The orange and green regions correspond to 90\% confidence regions obtained from NICER data on PSR J0030+451 and J0740+6620, using the Illinois-Maryland and the Amsterdam analysis, respectively. Finally, the horizontal region is 90\% confidence region for the mass of lighter object in GW 190814. Selected nuclear and geometrical EoSs are shown by dotted and solid lines in the plot.
 \label{mr_gw}}
\end{figure}

Fig. \ref{mr_gw} shows the mass-radius curves for the selected nuclear and geometrical EoSs (as discussed in \ref{mr_section}) with various gravitational waves and astrophysical observational constraints.  Gravitational wave and pulsar observations show a nuclear EoS that is largely consistent with the computed neutron star mass-radius. The  radius of  these Neutron stars are about $10-12$ km, it is  consistent with the posteriors of GW170817. In the case of two geometrical EoS deduced through the general relativity framework ( not based on nuclear interactions), they behave differently compared to a normal neutron star (nuclear EoSs). The TRV EoS  shows mass-radius plane similar to self-bound compact stars (e.g., quarks stars), which is not gravitational bound as per their radius branch, but it  shows that this different radius curve  belongs to self-bound nature of the star. In constrains with observational data, To some extent the posterior of GW170817 containment of the TRV mass-radius plane. The mass-radius plane from the SNJR EoS corresponds to a large extent with the posterior of the pulsar J0740+6620 and in addition this geometrical EoS gives a radius of a broad range and its mass is also greater than $2 M_{\odot}$. Still these observational constraints are not sufficient for comparison, more sensitive observational methods are still being conducted.

Furthermore, since the physics of neutron star core is still unknown, TRV EoS might be useful to described such  ultra-compact stars. Various scientists believe that there are quark-like matter in the centers of neutron stars, and many EoSs constructed of strange and quark-like matter have been studied. Nonetheless, unless exact information based on observations is obtained, as well as knowledge on the actual materials of compact stars such as neutron stars, all of these models will be regarded as assumptions.

\subsection{Keplerian frequency (rotation frequency of neutron star) } \label{sec_kepler_fre}

It is important to first present some essential information before describing the process used to construct the diagrams below: first, what is the Kepler
frequency, followed by a few observational constraints.

Kepler frequency is defined as the frequency at which a rotating mass would shed matter at its equator. In actuality, the limit of rotation for the star itself is the angular velocity of a particle in a stable circular orbit at the equator of the star. For conciseness, frequency is used instead of angular velocity. It is dependent on the star's internal structure in a complex manner through the interior metric and the frame dragging frequency.

Antoniadis et al. measured the most massive NS observed (PSR J0348 + 0432) in 2013  \cite{antoniadis2013massive}, with a mass of $2.01 \pm 0.04 M_{\odot}$. This pulsar has a
rotation frequency of $46 Hz$, thus deviations from spherical symmetry are deemed
 negligible. This signifies that for every static NS, this number is the minimum limit for the maximum stable mass predicted by any mass-radius relation \cite{cipolletta2015fast}. A recent investigation on a massive millisecond pulsar (MSP J0740 + 6620) yielded a mass value of $2.14^{+0.10}_{-0.09}$, with a 68.3\% coefficient of determination \cite{cromartie2020relativistic}.

The (PSR J1748 2446ad) pulsar has the highest measured rotational frequency of $716 \; Hz$. Given that several NS models anticipate rotational rates of up to $2000 \; Hz$, this is really a fairly weak limitation.

 The Kepler frequency expresses the balance of centrifugal and gravitational force on a particle on equatorial plane at the surface of a star. It is expressed as

\begin{eqnarray} \label{kepler_eqn}
    \Omega_{C}=\sqrt{\frac{M}{R^{3}}} \; ,
\end{eqnarray}
where the subscript $C$ denotes classical symmetry of the centrifugal and gravitational forces, which is the Newtonian expression for the Kepler angular velocity. This equation do to not hold in General Relativity, but as it turn out, it holds to very good accuracy if the right side is multiplied by a prefactor$(C)$ \cite{haensel2009keplerian}. It has been shown by J. M. Lattimer, et al\cite{lattimer2004physics}., Haensel et al.\cite{haensel2009keplerian} and B. Haskell et al.\cite{haskell2018fundamental} that the numerical value of the Keplerian frequency, namely the maximum rotational frequency of a neutron star accounting for the effects of general relativity, deformation , and independent on the EoS, can be well fitted from the simple formula 

\begin{eqnarray} \label{kep_haskell}
      \Omega_{K} \approx C \bigg(\frac{M}{\text{\(M_\odot\)
}}\bigg)^{1/2}\bigg(\frac{10 \: \text{km}}{R}\bigg)^{3/2} \;  \text{Hz}\: ,
\end{eqnarray}2009
 providing the neutron star mass is not very close to the maximum stable value, $M$ and $R$ are the mass and the radius of the nonrotating star respectively. The constant $C$ of Eq. (\ref{kep_haskell}) are given by  B. Haskell et al.\cite{haskell2018fundamental}. For the self bound compact stars, it is given as $1.15$ kHz and for other gravitationally bound neutron stars it is given as  $1.08$ kHz.
 
 \noindent The deduction of $\Omega_{K}$ generally requires the calculation of rotating general relativistic configurations. Nevertheless , Haensel et al. (2009) have shown  to a good degree of accuracy that the mass-shedding frequency $\Omega_{K, max}$ can be determined by the EoS-independent empirical formula  as given in Eq. (\ref{kep_haskell}). On the other hand, it allows to determine $\Omega_{K}$ using the mass and radius  of the nonrotating star.

\begin{figure}[h!]
\centerline{\includegraphics[scale=0.5]{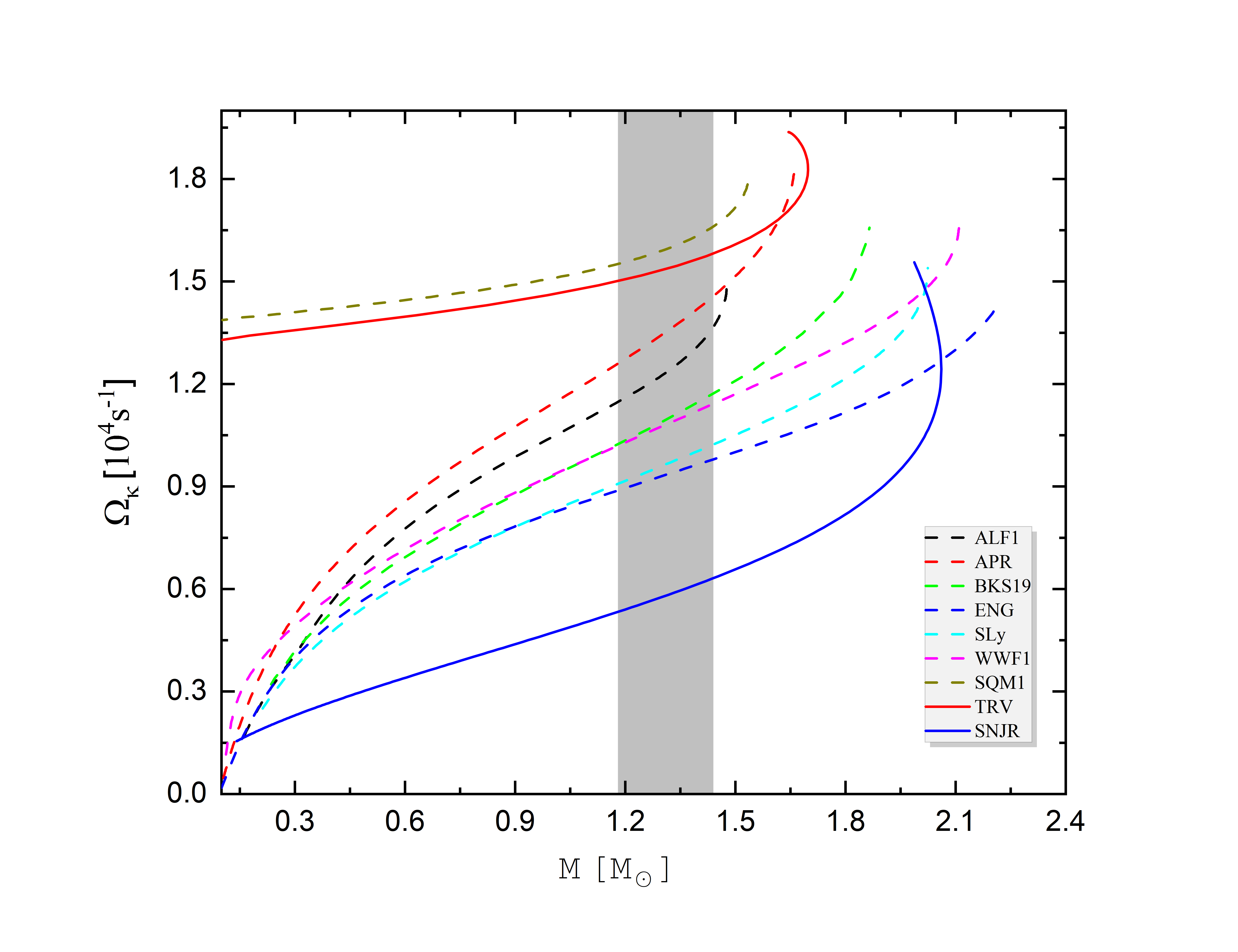}}
\caption{ Kepler frequency, $\Omega_{k}$, as a function of neutron star mass using the two different classes of EoS. The vertical band shows the range of precisely measured masses of binary pulsars \cite{lattimer2012nuclear})
 \label{f6}}
\end{figure}

\noindent The calculated Keplerian frequency based on the mass-radius relations obtained using all the nine equations of state  are shown in   Fig.~\ref{f6}.  Here we found that Keplerian frequency
corresponds to TRV and SQM1 are similar with higher values of $\Omega_{K}$ ( 14-18 kHz). While other cases $\Omega_{K}$ varies from 2 kHz to 18 kHz. The results of Keplerian frequency for the maximum mass of stable stars are shown in Table~\ref{ta2} for all the nine models.

\subsection{Moment of inertia versus $M$}
By evaluating a neutron star's moment of inertia, we may learn important things about the star's interior that are now out of our sight or its circumference.

Observer in an inertial frame of reference (where a frame is given in \cite{glendenning_2012})
near a spinning star will feel revolution about its core. A frame of reference is a set of coordinates that can be used to calculate the locations and velocities of objects in that frame; inertial implies that the frames are not accelerating toward one another. It would spin more quickly as it goes nearer to the star. The interior structure of revolving star is significantly impacted by the rotation of local inertial frames. As a result,
the Lense-Thirring effect, also known as the dragging of local inertial frames, describes the motion of revolving stars in spacetime. The centrifugal forces acting upon a fluid element of the star, depend not
on $\Omega$ (where $\Omega$ is the angular velocity of the star) but on the difference between the frequencies $\Omega$ and $\omega(r)$(the angular velocity of the local inertial frames), $\bar  \omega\equiv \Omega - \omega(r)$ \cite{hartle1967slowly}.

\begin{figure}[h!]
\centerline{\includegraphics[scale=0.7]{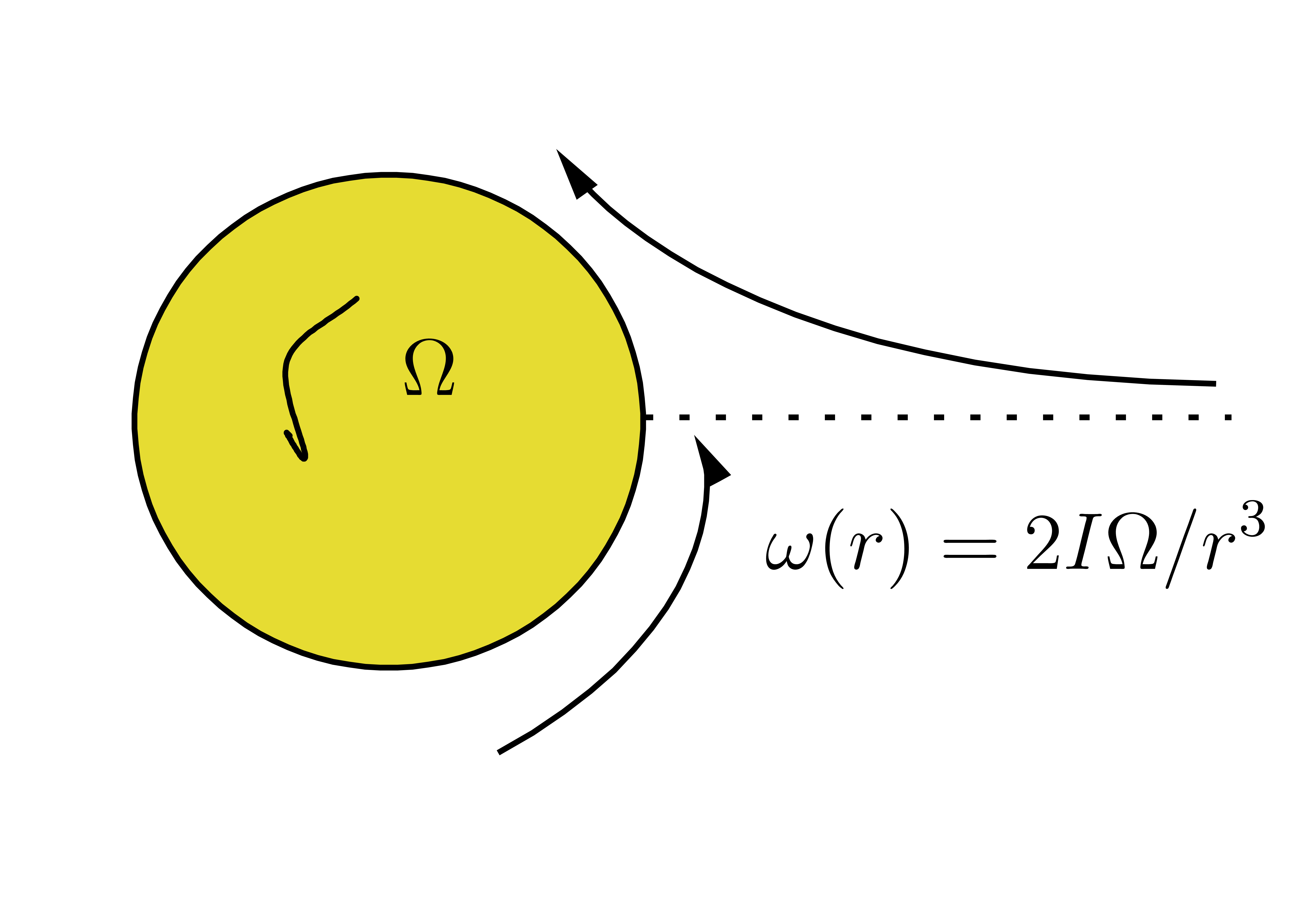}}
\caption{ Schematic of the trajectory of a particle dropped from infinity in the equatorial plane of a rotating star.\label{mitoy}}
\end{figure}

We only discuss characteristics based on static NS in this study, therefore we neglect the slowly rotating NS approximation \cite{hartle1967slowly}. There are five mass-radius based analytical equations for computing the moment of inertia :
\begin{itemize}
      \item Ravenhall and Pethik (1993) \\
       Ravenhall and Pethik \cite{ravenhall1994neutron} presented the following equations
      \begin{eqnarray}
         I \simeq 0.21 \frac{M R^{2}}{1-2GM/Rc^{2}}
      \end{eqnarray}
    \item Lattimer and Prakash (2000) \\
  In addition, Lattimer and Prakash employed three analytical approaches in 2000 \cite{lattimer2001neutron} to
 approximate the moment of inertia of a stationary neutron star using Einstein's equations. The Schwarzschild interior solution for an incompressible fluid was the first solution they explored \cite{schwarzschild1999gravitational}. Buchdahl(1967) \cite{buchdahl1967general} developed the second solution. The third solution is the one proposed by Tolman (1939)\cite{tolman1939static}  and it corresponds  to the situation where the mass-energy density varies quadratically : $\rho = \rho_{c} [1-(r/R)^{2}]$.
The three approximations for the moment of inertia produced by the analytic
solutions above by Lattimer and Prakash (with the abbreviations they suggested -
$I_{Inc}$ for the approximation produced from the interior Schwarzschild solution, $I_{Buch}$
from Buchadahl's solution and $I_{T\; VII}$ from Tolman's), are:

\begin{eqnarray}
    I_{Inc}\simeq 2 M R^{2} (1-0.87 \eta - 0.3 \eta^{2} )^{-1}/5
\end{eqnarray}
where $\eta$ is the compactness of a neutron star : $\eta = \frac{G M}{R c^{2}}$
 \begin{eqnarray}
     I_{Buch} \simeq M R^{2} \bigg(\frac{2}{3}-\frac{4}{\pi ^{2}}\bigg) (1-1.81 \eta + 0.47 \eta^{2})^{-1}
 \end{eqnarray}
 \begin{eqnarray}
     I_{T\; VII}\simeq 2 M R^{2} (1-1.1\eta -0.6\eta^{2})^{-1}/7
 \end{eqnarray}
\item Lattimer and Schutz (2005) \\
With a 10 \% accuracy, Lattimer and Schutz \cite{lattimer2005constraining} approximated the moment of inertia of the first star in the radio binary pulsar system PSR J0737-3039. The approximation they found, valid for NSs with $M> M_{\odot}$, is 
\begin{eqnarray}
    I \simeq 0.237 M R^{2} \bigg[ 1+1.42 \frac{M}{M_{\odot}}\frac{\text{km}}{R}+90 
 \bigg( \frac{M}{M_{\odot}}\frac{\text{km}}{R} \bigg)^{4} \bigg]
\end{eqnarray}
 which further written as 
 \begin{eqnarray}
     I \simeq (0.237 \pm 0.008 + 0.674 \eta +4.48 \eta^{4})
 \end{eqnarray}
\end{itemize}

Here we use the approximation of Lattimer and Schutz to compute moment of inertia as it is more appropriate for stars with masses greater than $M_{\odot}$ than the other approximations.
Furthermore, all the EoSs  taken in this work provided masses more than $M_{\odot}$, so this approximation can be considered reasonable. 
\begin{figure}[h!]
\centerline{\includegraphics[scale=0.5]{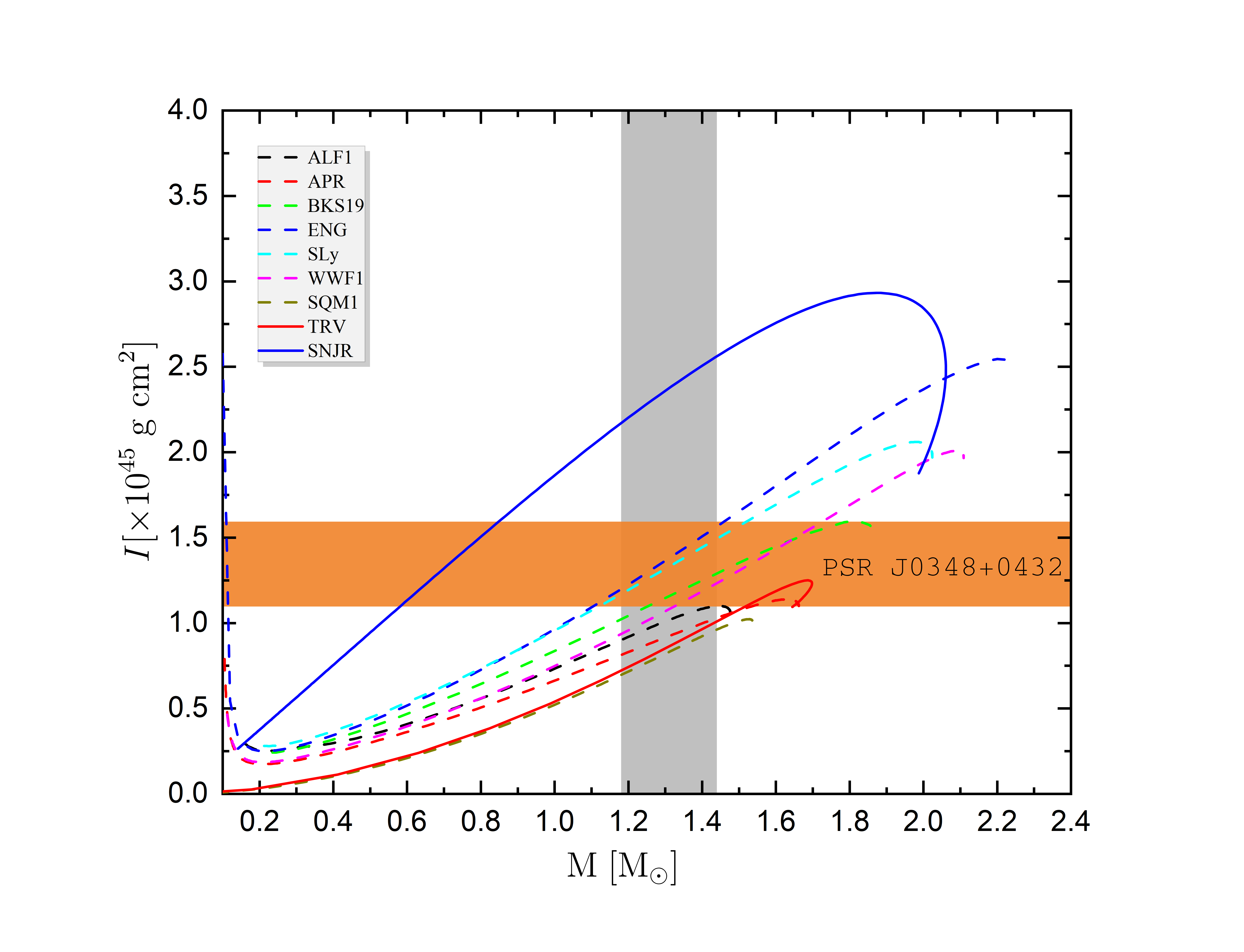}}
\caption{ Moment of inertia $I$ of a non-rotating neutron star versus stellar mass $M$ for several nuclear and geometrical EoSs of dense matter. The vertical band shows the range of precisely measured masses of binary pulsars \cite{lattimer2012nuclear}. The horizontal orange band is the range of estimated $I$ ($=1.09 \sim 1.59 \times 10^{45}$ g cm$^{2}$) of the observed PSR J0348+0432 \cite{zhao2014surface}.\label{mi}}
\end{figure}

Fig. \ref{mi} is presents the moment of inertia against stellar mass for all the selected EoSs. EoSs composed of hadronic matter give roughly a moment of inertia range about$1.0-2.5 \times 10^{45}\; \text{g\; cm}^{2}$ as indicated by the dashed lines.
  At given $M$, the value of $I$ for the nuclear EoS (BKS19, ENG, SLy and WWF1) are significantly higher than those deduced using stiff TRV EoS. Nevertheless, even though SNJR is a soft EoS, it gives a higher $I$ value than all other nuclear and geometrical EoS at maximum stable mass. The significant reason could be that we computed $I$ as a function of mass, and if we look at their M-R configuration (as shown in Fig.\ref{f5}), the mass value from this EoS is higher (except for ENG and WWF1) than other EoSs.
In addition, the computed maximum values of $I_{\text{max}}$ at $M_{\text{max}}$ are tabulated in Table \ref{ta2}.

\subsection{ Surface Gravity}
The surface gravity of neutron stars denoted by $g_{s}$ (i.e., the acceleration due to gravity as measured on the surface),  is another  important parameter for the study of neutron star atmospheres      \cite{gudmundsson1983structure}. The upper bound of the surface gravity for neutron  stars with various baryonic EoSs is studied by Bejger et al. (2004) \cite{bejger2004surface}. The surface gravity of neutron star is  many orders of magnitude larger than that of other stars; it is $\sim$ 10$^{12}$ times stronger than gravity at the Earth surface, and $10^{5}$ times larger than that of the white dwarfs.

The expression for $g_{s}$ is given by \cite{bejger2004surface}:

\begin{eqnarray}
    g_{s}= \frac{G M}{R^2 \sqrt{1-x_{GR}}} 
\end{eqnarray}

\begin{figure}[h!]
\centerline{\includegraphics[scale=0.5]{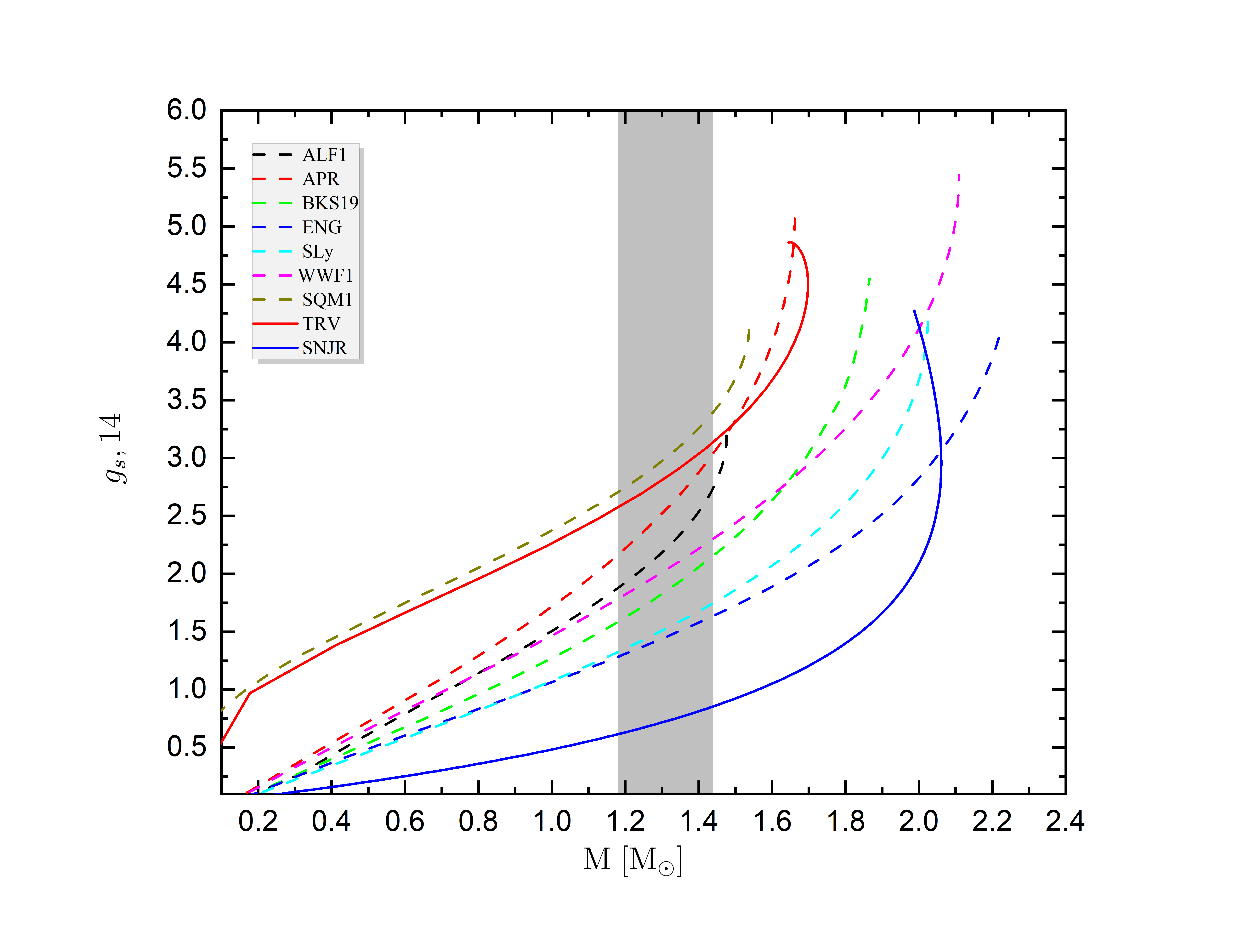}}
\caption{ Plots of $g_{s, 14}$   versus gravitational mass $M$. Surface gravity in the units of 10$^{14}$ cm s$^{-2}$. The vertical band shows the range of precisely measured masses of binary pulsars.\label{f7}}

\end{figure}

\noindent Here, $x_{GR}  =2 G M/R c^{2}=r_{g}/R$, where $r_{g}$ is the Schwarzschild radius.
The importance of relativistic effects for a neutron star mass $M$ and radius $R$ is characterized by the {\it compactness parameter} $r_{g}/R$.  Usually for a neutron star with $M=1.4$ \(\textup{M}_\odot\) and Radius is about 10 km, surface gravity becomes ($g_{s}$) = 2.43 $\times$ 10$^{14}$ cm s$^{-2}$. In consequence it is suitable to measure $g_{s}$ in units of 10$^{14}$ cm s$^{-2}$ and is represented as $g_{s, 14}$ $\equiv$  $g_{s}/(10^{14}$ cm s$^{-2}$). The computed surface gravity, $g_{s, 14}$ for all the cases studied here are shown in  Fig.~\ref{f7} against mass expressed in \(\textup{M}_\odot\). The numerical values of $g_{s, 14}$ correspond to maximum stable mass of the star are also listed in Table~\ref{ta2}.

\noindent  It is found that for $M=1.4$ \(\textup{M}_\odot\), $g_{s, 14}$ ranges from 1.43 to 2.8 and for $M\approx 2.0$ \(\textup{M}_\odot\) the surface gravity lies between 1.88 to 4.38 . The nuclear EoSs (labeled : 1 and 6) with an exotic quarks phase have relatively low $g_{s, \text{max}}$. A similar situation occurs for the SNJR EoS that gives lowest value of surface gravity. The only reason SNJR EoS have low surface gravity is that they have a greater radius compared to other EoSs. The TRV EoS (labeled 8 ) yields $g_{s, \text{max}}$ similar to BKS19 and SLy EoSs. Their values of $g_{s,14 (max)}$ range from 4.10 to 4.60.  
\subsection{ Gravitational Redshift of Compact Star}
In general relativity the ratio of the emitted wavelength $\lambda_{e}$ at the surface of a nonrotating neutron star to the observed wavelength $\lambda_{0}$ received at radial coordinate $r$, is given by $\lambda_{e}/\lambda_{0}=[g_{tt}(R)/g_{tt}(r)]^{1/2}$ \cite{hobson2006general}.  From this the definition of gravitational redshift, $z\equiv (\lambda_{0}-\lambda{e})/ \lambda{e}$ from the surface of
the neutron star as measured by a distant observer $( {g_{tt}(r)}\to -1)$ is given by \cite{straumann2012general}
\begin{eqnarray} \label{grav_redshift_eqn}
    z=\mid - g_{tt}(R)\mid ^{-1/2}-1 =\bigg(1-\frac{2 G M}{R c^{2}}\bigg)^{-1/2}-1
\end{eqnarray}
We compute the limit of the redshift from  the surface of a neutron star using Eqn. (\ref{grav_redshift_eqn}) where $g_{tt}=-e^{\lambda(r)}=-(1-2 G M / c^{2}R)$ is the metric components \cite{oppenheimer_1939}. For a given EoSs the  maximum value $z_{\text{surf}}^{\text {max}}$ increase with increase of $M_{\text{max}}$.  Neutron stars of $M\geq $ \(M_\odot\) are expected to have sizable $z_\text{surf}$. The computed values of $z_\text{surf}$ for all the cases studied here  are listed in Table~\ref{ta2}. The computed values of $z_\text{surf}$ are found to lie between $0.3$ to $0.7$.

For a given EoSs, the maximum value $z_{\text{surf}}^{\text {max}}$  of stable neutron stars is reached at $M = M_{\text{max}}$. On average,   $z_{\text{surf}}^{\text {max}}$ increases with the growth of $M_{\text{max}}$.  However, this is not a strict rule because $z_{\text{surf}}^{\text {max}}$ depends on both $M_{\text{max}}$ and $R_{\text{max}}$ . Taking a softer EoS lowers the value of $M_{\text{max}}$. In particular, $z_{\text{surf}}^{\text {max}} < 0.4$ for the softest ALF1 EoS. In general, the EoSs which maximizes $z_{\text{surf}}^{\text {max}}$ has to be relatively soft in the outer neutron-star layers (to reduce $R$) but stiff in the stellar core (to increase $M_{\text{max}}$). Modern EoSs of the $npe\mu$ matter (APR, BKS19, ENG, SLy and WWF1) are of this type and give $z_{\text{surf}}^{\text {max}} \simeq 0.6$. The presence of hyperonic cores lowers the the value of $z_{\text{surf}}^{\text {max}}$ (e.g., SQM1 and TRV).

\begin{figure}[H]
\centerline{\includegraphics[scale=0.5]{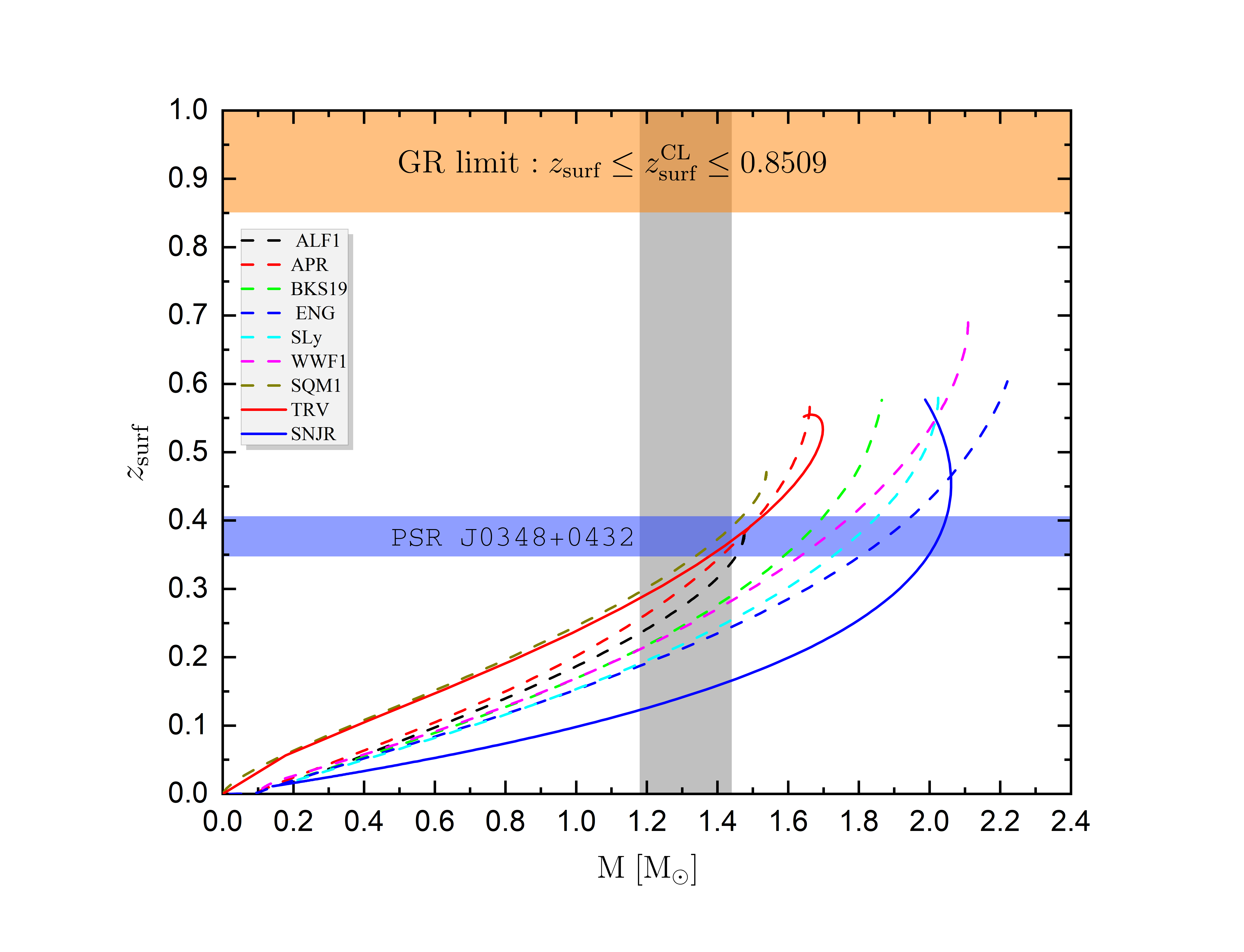}}
\caption{ Surface gravitational redshift $z_{\text{surf}}$ versus gravitational mass $M$ for several EoSs of dense matter.  The solid saffron band  above the horizontal line is prohibited for subluminal EoSs ($v_{s} \leq c$). The solid vertical band is the range of precisely measured masses of double neutron star binaries \cite{lattimer2012nuclear}. The blue horizontal band represent the computed gravitational redshift from the observed PSR J0348-0432, in the range of $z_{\text{surf}}^{\text {max}} \;\; 0.34 -0.40$ \cite{zhao2014surface}.
 \label{f8}}
\end{figure}

\noindent From the listed values of $x_{GR}$ in Table~\ref{ta2}, we found that all the models studied here satisfy the Buchdahl inequality\cite{buchdahl1959general}, $R\geq(9/8)r_{g} = (9/4)G M/c^{2}$   which is stricter than the Schwarzschild bound. A consequence of this is that the gravitational redshift should satisfy $z\leq 2$. The precise upper bound on the surface redshift for neutron star is $z_{\text{surf}}=0.8519$ for subluminal EoSs \cite{gondek1995effects}. In the present study, we found that $z_\text{surf}$ for all the nine cases computed here lie much below the upper bound for the gravitational redshift.

\section{Results and Discussion}
\label{sec:discussion_chap3}

We have computed several properties of a compact star, using nuclear and geometrically deduced equations of state. We have used geometrical equations of state from the core envelope model, that describes different properties of the physics in the core and envelope region.  Many similarities and dissimilarities are observed from the properties computed based on the geometrical EoSs and the nuclear EoSs.

\begin{table}[htb]
\begin{center}
\resizebox{\columnwidth}{!}{   

\begin{tabular}{c c c c c c c c c} 
\hline
 Label &
\makecell{$ M _{max}$ \\$(M_\odot)$} & 
\makecell{$R_{max}$ \\(km)} &
 \makecell{$\rho_{c}$\\($10^{15}$ gm cm$^{-3}$)}&
\makecell{$\Omega_{k}$ \\ ($10^{4}$ s$^{-2}$) }& 
$z_{\text{surf}}$  &
\makecell{$g_{s, 14}$\\(cm s$^{-2}$)} &
$x_{GR}$& 
\makecell{{$I$}\\ ($\times 10^{45}\text{g cm}^{2}$)} \\[1ex]
\hline\hline

1 & 1.47 &  9.21& 3.34& 1.49 & 0.38 & 3.17& 0.47 &1.05\\[1.2ex]
\hline

2 & 1.66 & 8.37 &4.51& 1.85 & 0.57 & 4.89 &0.58 &1.10 \\ [1.2ex]
\hline
3 & 1.86 &  9.25 &2.69& 1.66 & 0.58 & 4.59 &0.59 &1.55\\  [1.2ex]
\hline
4 & 2.22 & 10.76  &2.91& 1.44 & 0.60 & 4.07 &0.60 &2.53\\ [1.2ex]
\hline
5& 2.02 & 10.08  &2.85& 1.53 & 0.57& 4.14 &0.58 &1.99\\  [1.2ex]
\hline
6 & 2.10 &  9.17 &2.93& 1.67 & 0.69 & 5.36 & 0.65 &2.00\\ [1.2ex] 
\hline
7 & 1.53 &  8.48 &3.27& 1.82 & 0.46 & 4.17 &0.53 &1.01\\ [1.2ex] 
\hline
8 & 1.68 &  8.76 &9.57& 1.82 & 0.55 & 4.26 & 0.57 &1.25\\ [1.2ex]
\hline
9 & 2.06 &  11.58 &4.29&1.24 & 0.46 & 1.98 & 0.56 &2.65\\ [1.2ex] 
\hline \hline
\end{tabular} 
}
\captionsetup{
	justification=raggedright,
	singlelinecheck=false
}
\caption{Calculated properties of nonrotating neutron star models}
\label{ta2}
\end{center}
\end{table}

\noindent In Table~\ref{ta2} we have listed computed properties like maximum mass ($M_{\text{max}}$), stellar radius ($R_\text{{max}}$) correspond to the maximum mass , central density ($\rho_{c}$), Keplerian frequency ($\Omega_{k}$) correspond to the maximum mass for the stable structure of the star, gravitational redshift ($z_{\text{surf}}$), surface gravity ($g_{s}$) , compactness parameter ($x_{\text{GR}}$) and moment of inertia ($I$) of a compact star with all the different types of EoSs. We compared all these properties with the properties computed using the two (TRV and SNJR) geometrically deduced equations of state. The properties obtained from TRV equation of state are in good agreement with the properties obtained from other nuclear matter based models. While the parameters obtained using the SNJR model are quite different from others except for $z_\text{surf}$ and ($x_{\text{GR}}$). The central density that yields the maximum mass of $\approx$ 2 \(\textup{M}_\odot\)  in the case of SNJR is very low and the radius is about 12 km. It is also reflected in the low values of the surface gravity. It is noted  that mass-radius configuration as shown in Fig.~\ref{f5} obtained  from geometrical models will be pertinent for divergent class of compact stars. Particularly,  the pseudo-spheroidal spacetime of TRV model seemed to describe the ultra dense compact stars like the strange self-bound stars. The spacetime geometry adopted for the SNJR model represents low density neutron  like star\cite{gedela_2019}, where radius lie between $12\leq R \leq 20$ kms.

\noindent The mass-radius diagram  in  Fig.~\ref{f5} clearly classify the nature of compact stars in three categories :(i) highly compact self-bound stars represented by the TRV Model and SQM1 model with exotic matter compositions (ii) the normal neutron stars with nuclear matter EoS and (iii) the ultra soft compact stars represented by the SNJR geometrical model. At the end,  we are able to identify a correspondence between the geometric description with the structure of the matter distribution in compact objects like a strange star. To summarize, we have been able to classify neutron  stars in three  distinct types each one having different internal structures. We hope that TRV model for compact neutron like stars will be useful for the study of superdense self-bound stars having exotic matter compositions.

Relying on these structural properties, we employed TRV EoS and the anisotropic TOV equation to calculate properties such as anisotropic mass-radius configuration, gravitational energy discrepancy, and moment of inertia based on various anisotropy magnitudes, which are discussed in the next chapter.  
\chapter{Cracking stability of  anisotropic self-bound compact stars }
\label{chap:4}

``\textit{There's no sense in being precise when you don't even know what you're talking about.}''                                
\begin{flushright}
--John von Neumann
\end{flushright}

\bigskip

 
 Properties of  compact stars are one  the most sought-after topics in the field of astroparticle physics in recent times \cite{itoh1970hydrostatic, collins1975superdense, de2011self,horvath2021modeling}. Stability studies, starquakes, wobbling of stars and the processes such as giant gamma-ray burst etc are few examples. They release thousands  times more energy than supernovae. Their distinctive $\gamma$-ray emission lasts from fraction of seconds to  few minutes.  A typical burst releases as much energy as the Sun would in its entire 10-billion-year lifespan \cite{graham2013metal,gendre2013ultra}. Several satellite missions (e.g. HETE-2, FERMI, INTEGRAL, SWIFT, RXTE, ULYSSES) have been launched since the first observation of GRB and recorded flares of GRBs and provided useful data related to the origin of GRBs. The genesis and astrophysical implications of all these  processes are vehemently debated. The energy released in such processes are  estimated to be $\sim 10^{44-47}$ erg \cite{kumar2015physics,nakar2007short,berger2014short}. While the reliable nature of the sources is absolutely fascinating and not yet entirely understood, more spectacular events definitely related with them challenge the researchers' inquisitiveness. Many researchers have suggested models for $\gamma$-ray superflares, which are based on compact objects (e.g. Neutron stars, Black holes binary mergers). The model based on anisotropic fluid for the highly compact stars (i.e. solid quarks)  studied by  Xu et al. \cite{xu2006superflares} have proposed an alternative model from the conventional one (i.e. the magnetar model), that SGR superflares could be caused by substantial quakes in  quark stars.  Xu et al. \cite{xu2006superflares} addressed the prospect of gamma-ray bursts  resulting from the phase transition of a neutron star to a quark star, with a possible energy release of  the order of $10^{52}$ erg. The degradation of the superstrong field ($B \geq 10^{14}$ gauss) in magnetars might fracture the crust and create bursts of high gamma-ray emission \cite{thompson1995soft,thompson1996soft}. Gravitational, magnetic, and superfluid forces could all exert stress on the crust of a neutron star that is evolving. Fracture of the star's crust as a result of these forces might have an impact on the star's spin dynamics and result in high-energy bursts \cite{franco2000quaking}. 

The source of the released energy is one of the most essential and fundamental problems (inferred isotropic equivalent values are given in Table \ref{tabgrb:1}, Column 3.). There are  few options for free energy sources that are worth examining for comparison and future research.
One such attempt is related to the high-density phase transition at the core-evelope interface that can cause instabilities. Mechanical readjustment (crust-cracking) of the star crust can occur, depending on the parameters and the dynamics in which matter undergoes a phase transition which lead to potential  release of   significant amount of energy \cite{marranghello2002phase}. Based on such estimates, the amount of  energy released is limited to around $\Delta E = W (\Delta R / R)$, where $W$ signifies the star's original binding energy and $\Delta R/R$ signifies the radius's fractional change. It is estimated that energy  of the order $10^{53}$ erg could be released in such processes. It has been recently improved to include an exotic solid quark phase which constitutes the bulk of the interior of the neutron star, which is believed to be a possible source of free energy \cite{xu2003solid,peng2008pulsar,zhou2004quakes,xu2006superflares}. The burning of neutron matter to strange matter may be a source of free energy. For example, if the energy per baryon number  of three-flavor quark matter is smaller than that of squeezed hadronic matter, and if other astronomical conditions are satisfied then it induces conversion of a neutron star to a strange star shortly after its formation \cite{berezhiani2002gamma,ouyed2002quark}. However, it is conceivable for a neutron star to evolve into a quark star, particularly a strange quark star. Quark stars are hypothesised extreme compact objects made of quark matter. They are widely considered as the actual ground state of compact baryon stars due to their high density. The presence of strange quark matter raises a significant condition for the selection of the EoS of quark stars: the EoS must incorporate the component provided by strange quark matter in order for the EoS to be considered credible. Strange quark matter has a lower pressure and energy than conventional quark matter with the same quark number density, permitting to be a little more stable.

\begin{table}
\begin{center}
\begin{tabular}{lll}
\hline\noalign{\smallskip}
Object & Date & $E_{\gamma}$ (erg)  \\
\noalign{\smallskip}\hline\noalign{\smallskip}
 GRB 070714b& 2007 Jul 14 & $1.2\times 10^{51}$   \\
 GRB  051221a  & 2015 Dec 21 & $(1.2-1.9)\times 10^{49}$  \\
SGR 0526-66& 1979 Mar 5 & $6\times 10^{44}$  \\
SGR 1900+14 & 1998 Aug 27 & $2 \times 10^{44}$  \\
SGR 1806-20 & 2004 Dec 27 & $3.5\times 10^{46}$  \\
\noalign{\smallskip}\hline
\end{tabular}
\caption{Superflares from GRBs \cite{graham2009grb,soderberg2006afterglow} and SGRs \cite{horvath2005energetics}}
\label{tabgrb:1}
\end{center}
\end{table}

In relativistic astrophysics, the stability of a compact star structure is a major concern. Chandrasekhar \cite{chandrasekhar_1964} investigated radial perturbation for isotropic fluid spheres in general relativity and proposed the pulsation equation to evaluate fluid sphere stability. Generalizing this method for use with anisotropic matter distributions was accomplished by Dev and Gleiser \cite{dev2003anisotropic}. Cracking, also known as overturning, is a notion that was first presented by  Herrera \cite{herrera1992cracking} and  Di Prisco et al.\cite{di1994tidal}, and it proposes a different approach ($-1\leq v_{s\bot}^{2}-v_{s r}^{2}\leq 1$, where $v_{s\bot}$ and $v_{s r}$ are tangential and radial speed of sound respectively)  determining whether or not an anisotropic matter distribution is stable in general relativity.  
The fundamental assumption of this approach is that the fluid constituents on each side of the cracking point experience a relative acceleration towards each other. It was developed with the purpose of describing the behavior of a fluid distribution  away from its equilibrium state. Later on, Chan et al. \cite{chan1993dynamical} showed that even very minor departures from local isotropy might contribute significantly for its structural stability. In conjunction with this, they found that changes in density by themselves do not throw the system out of equilibrium when it is configured with the anisotropic matter. Such deviations can only be induced by perturbations that affect both the density and the local anisotropy \cite{di1994tidal,di1997cracking}.

Increasing theoretical studies suggest that fascinating physical events may cause local anisotropy, such as uneven radial and tangential stresses $p_{r}\neq p_{\bot}$ ( see  \cite{herrera1997local,mak2003anisotropic}, and references therein). In the framework of General Relativity, Lema\^itre \cite{lemaitre1933univers} noted that local anisotropy can relax the upper limits on the maximum surface gravitational potential. Since the early works of  Bowers and Liang \cite{bowers1974anisotropic} its significance in General Relativity has been investigated.   Hillebrandt and Steinmetz  \cite{hillebrandt1976anisotropic} studied the stability of completely relativistic anisotropic neutron star models and found stability criteria comparable to that of isotropic models. Subsequently, Chan et al.  \cite{chan1993dynamical} evaluated the influence of  the local anisotropy in the emergence of dynamical instabilities. They observed that  moderate anisotropies might have a significant impact on the system's evolution.

 Many studies have been reported on core-envelope anisotropic relativistic objects in recent times \cite{sagar_2022,pant_2020three}. In which compact stars are divided into two distinct regions in accordance with the structure of various physical considerations  \cite{thomas_2005,mafa_2016,gedela_2019,mardan_2021,pant_2020three}. Any model for an anisotropic compact object is ineffective if it is unstable against variations of its physical variables. Distinct degrees of stability or instability lead to different patterns of development in the collapse of self-gravitating objects. In the present study, we shall investigate the effect of fluctuations in density and local anisotropy have on the potential cracking of local and non-local anisotropic matter configurations within the  general relativity framework. In particular, we  concentrate on how these disturbances cause phenomena that are conducive for starquakes. Here, we incorporate Einstein's equations that determine cracking, overturning, expansion, or collapse. Certain occurrences might radically change the system's evolution. If a configuration does not really crack (or overturn), it is indeed potentially stable (not definitely stable), since further perturbations might trigger expansion or collapse.  Abreu et al. \cite{abreu2007sound}, Gonzalez et al. \cite{gonzalez2015cracking}, and Ratanpal \cite{ratanpal2020cracking}
proved that the regions for which $-1\leq v_{s\bot}^{2}-v_{sr}^{2}\leq 0$ are potentially stable and the regions for which $0\leq v_{s\bot}^{2}-v_{sr}^{2}\leq 1$ are potentially unstable , where $v_{s\bot}^{2}=\frac{\partial p_\bot}{\partial \rho}$ and $v_{sr}^{2}=\frac{\partial p_{r}}{\partial \rho}$.


Thus, compact stars are astrophysical laboratories of many extreme  physics. The core and envelope of a highly compact stars are made up of distinct physical materials, according to the current knowledge of the strong interaction processes leading to a phase transition to quark matter, the study becomes extremely difficult due to the non-perturbative aspects of quantum chromodynamics (QCD)\cite{annala2020evidence}. An alternative treatment to such a self bound system is being studied recently \cite{khunt2021distinct} based on geometrical approach wherein the core-envelope model has been employed. The model with isotropic core and anistropic envelope naturally supports quake formation.
Following such studies  \cite{xu2006superflares} and  \cite{shu2017gamma}, the prospects of  a progenitor for   starquakes  from a self bound compact stars are being addressed in this study. 
The core-envelope models studied by Thomas, Ratanpal and Vinodkumar (TRV model) have considered anisotropic pressure in the envelope region and isotropic pressure in the core region \cite{thomas_2005}.  Physically such a scenario is  possible with a core containing  pure quark phase and the envelope accumulating quark-hadron mixed phase. The comprehensive investigation of the TRV model is described in our previous work \cite{thomas_2005,khunt2021distinct}, the structural properties like mass-radius relation, gravitational red-shift, Keplerian frequency, and surface gravity are studied using the TRV model. Their structural properties show that this model is important for the study of highly compact self-bound stars. A density perturbation at the core-envelope interface can lead to an event like starquake. The rearrangement of the star's mass distribution caused by a starquake affects the star's moment of inertia, resulting in precession and polar drifting \cite{link1998starquake}. As a result, unusual spin behaviour might be an indicator of the occurrence of a stellar quake. Evidence of crust cracking may already exist in few isolated pulsars, showing that the process is underway.

\section{Anisotropic matter configuration in general relativity}
\label{sec:2_Ch4}

Using equations (\ref{full_EFS_comp2}) and (\ref{full_EFS_comp3}), or correspondingly the conservation law $T^{\mu}_{\nu ;\mu}=0$, it is now convenient to transform the above equations into a form where the hydrodynamical properties of the system are more evident and that reduces to the TOV equations for systems with isotropic pressure, i.e.,

\begin{eqnarray}
     e^{-\lambda}=1-\frac{2 m(r)}{r},
\end{eqnarray}
\begin{eqnarray}\label{nuu}
    \nu^{'}=\frac{2m(r)+8\pi r^{3}P_{r}}{r(r-2m(r))},
\end{eqnarray}
\begin{eqnarray}
    P^{'}_{r}=-(\rho +P_{r})\frac{\nu^{'}}{2} +\frac{2}{r}(P_{\bot}-P_{r}),
    \label{rpre}
\end{eqnarray}
where the \textit{mass function} $m(r)$ is defined by 
\begin{eqnarray} \label{mass}
m(r)=4\pi \int_{0}^{r} \rho ~ \bar r^2 d\bar r.
\end{eqnarray}
Moreover, it corresponds to the mass inside  a sphere of  radius $r$ as perceived by a distant observer.
 
 Combining (\ref{nuu}) and (\ref{rpre}), we obtain
 
 \begin{eqnarray}
     P^{'}_{r}=-(\rho+P_{r})\frac{m(r)+4\pi r^{3} P_{r}}{r(r-2m(r))}+\frac{2{\Delta}}{r}P_{r}.
     \label{hydro}
 \end{eqnarray}
 which is the generalized Tolmann-Oppenheimer-Volkoff (TOV) equation. It can be seen that Eq. (\ref{hydro}) readily reduces to the standard TOV when $P_{r}=P_{\bot}$. Where $P_{\bot}=(1+\Delta)P_{r}$ is introduced here to account for the anisotropy in the envelope. We will choose $\Delta$ values based on the magnitude of Eq.(\ref{chap3_env_ani}) for $\lambda = 0.07$.  Interestingly, Eq.~(\ref{hydro}) which recalls the Newtonian hydrostatic-equilibrium equation and where the final term is obviously zero in the case of isotropic pressures, i.e. $P_{\bot}=P_{r}$.\\

\section{Anisotropy profile}
\begin{figure}[h!]
\centerline{\includegraphics[scale=0.5]{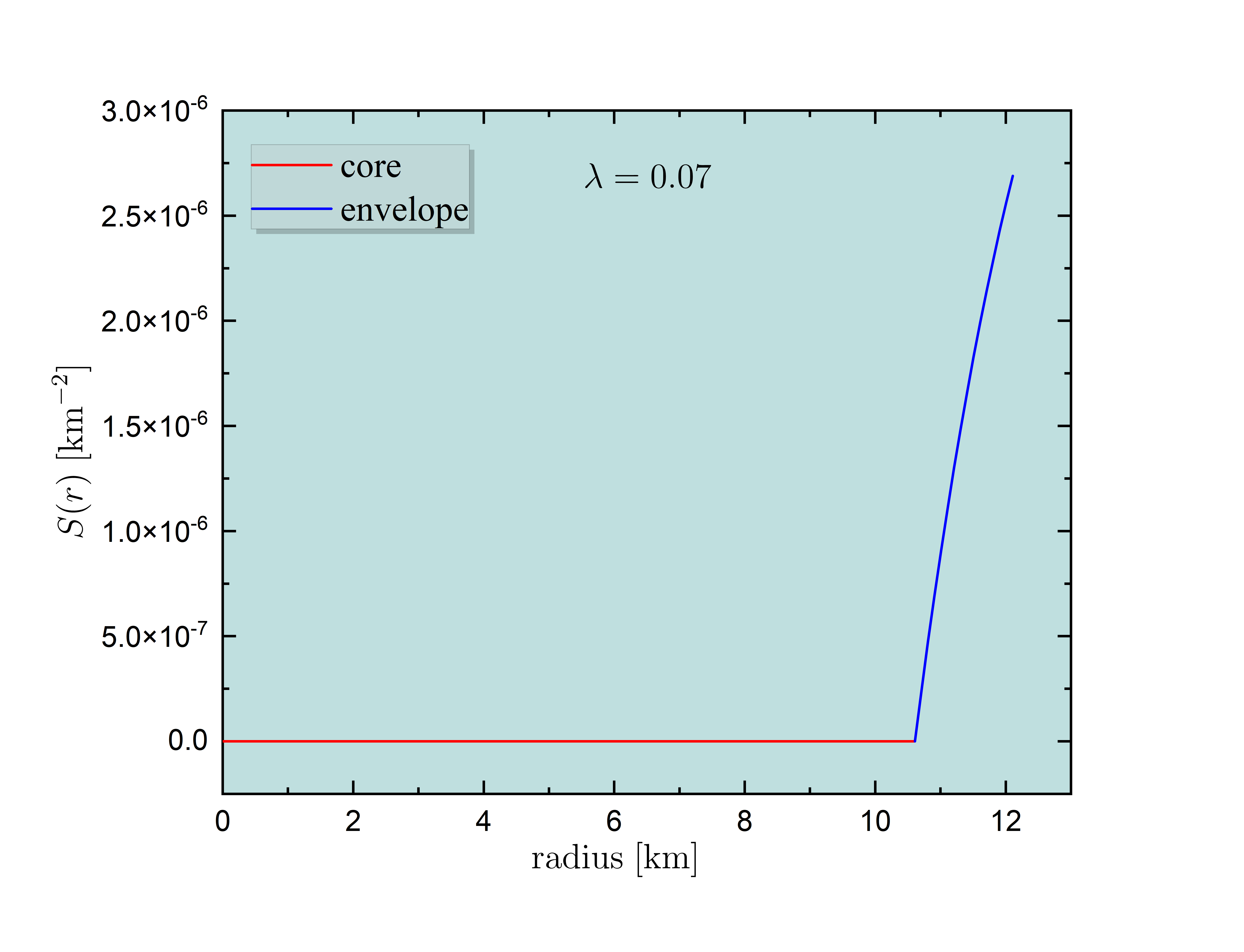}}
\caption{Variation of an anisotropy $S$ in km$^{-2}$ with respect to a radius of the star within a range of boundary condition as given in Eq.~(\ref{chap3_cond}). For a density variation $\lambda=0.07$}
\label{chap4_fig_1}
\end{figure}

\begin{figure}[h!]
\centerline{\includegraphics[scale=0.5]{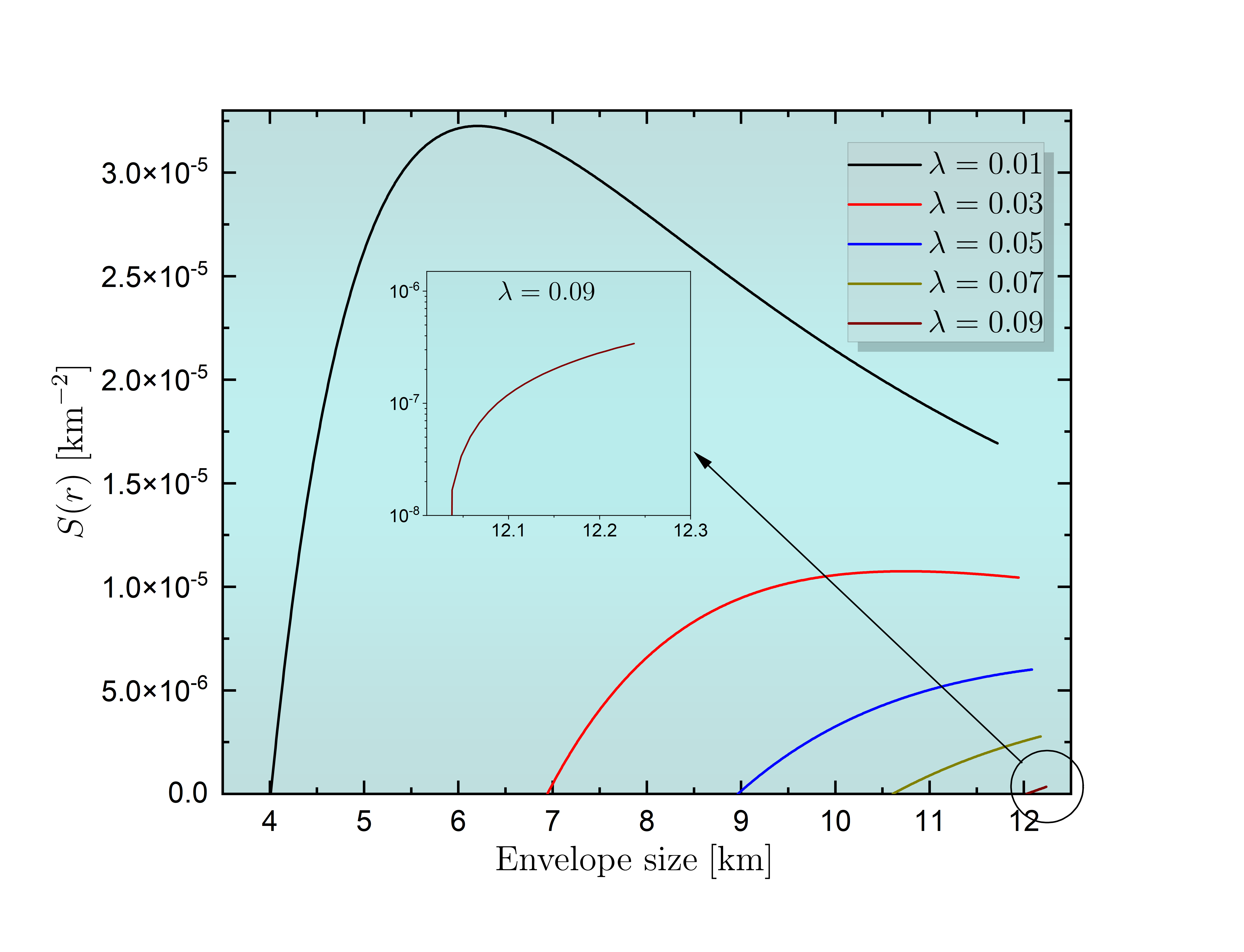}}
\caption{Variation of an anistropy $S$ in km$^{-2}$ with respect to a envelope size of the star. For a density variation of  $\;\; \lambda =0.01, 0.03, 0.05, 0.07$  and, $0.09$.}
\label{chap4_fig_2}
\end{figure}

The radial and tangential pressures differ due to local pressure anisotropy ($p_{r}\neq p_{\perp}$). In a conventional isotropic Tolman- Oppenheimer- Volkoff  equation, the difference between radial and tangential pressure generates an additional force. The discussion on anisotropic pressure and the physical conditions for anisotropic stars can be found in Refs. \cite{sulaksono2015anisotropic, setiawan2019anisotropic}, and further details can be seen in an anisotropic pressure review paper \cite{herrera1997local}. In the present work,
we  examine quantitatively, the changes on the properties of the compact star due to  the local anisotropic pressure. In order to adopt the model suggested in \cite{thomas_2005}, we have computed anisotropic magnitude for the core-envelope regions using the conditions (\ref{chap3_cond}) with Eqs. (\ref{chap3_env_pre_r}), (\ref{chap3_env_pre_t}), (\ref{chap3_env_ani}) and (\ref{chap3_rad_pre}). For a density parameter, $(\lambda=0.07)$, defined as the ratio of surface density to the central density, the distributed anisotropy for the core and the envelope regions is shown Fig. \ref{chap4_fig_1}.
For the core region, the anisotropy $(S)$ is zero according to the TRV model \cite{thomas_2005}, while in the envelope region anisotropy is non-zero. The anisotropy magnitude inside the envelope region computed using the Eq.~(\ref{chap3_env_ani}) and their result is represented as blue solid curve in  Fig. \ref{chap4_fig_1}. It can be noticed that the anisotropy is an increasing function of $r$. For 10.61 km $\leq r$ $\leq$ 12.17 km, the anisotropy varies in the range  $[2.45 \times 10^{-8}, 2.77\times 10^{-6}]$. It is  important to know its influence on the stability of the envelope region. The main reason for choosing a thin crust is to study processes like cracking or overturning in the medium of anisotropy. In Fig. \ref{chap4_fig_2}, we also show the anisotropy profile for various density variation values.

 \subsection{ Physical plausibility}
The interior solution should fulfill certain fundamental physical criteria. Some of the physical acceptability conditions for anisotropic materials have been described in \cite{herrera1997local,thomas_2005,mak2003anisotropic} as

\begin{enumerate}[label=(\roman*)]
\item radial pressure $p_{r}$, tangential pressure $p_{\bot}$ and density $\rho$  should be positive everywhere within the configuration;
\item gradients for radial pressure and density should be negative, 
\begin{equation*}
  \frac{\partial p_{r}}{\partial r} \leq 0   \;\;\; \text{and} \;\;\;  \frac{\partial \rho}{\partial r} \leq 0
\end{equation*}

\item Sound speed at the interface should be continuous and should satisfy the causality condition at the core of a compact star model, as well as decreasing monotonically outwards.
\begin{equation*}
    \frac{\partial p_{r}}{\partial \rho} \leq 1   \;\;\; \text{and} \;\;\;  \frac{\partial p_{\bot}}{\partial \rho} \leq 1
\end{equation*}

\item  The core and the envelope for the star should satisfy the energy conditions besides being
continuous at the interface.
\end{enumerate}

\section{The geometrical  EoS of extreme dense matter and stellar stability }
\label{sec:3}

In this work, we use geometrical EoSs which are based on our previous work \cite{khunt2021distinct} to study the canonical properties of compact stars. In which, the pressure and density profile are computed from the Eqs.~(\ref{density}), (\ref{chap3_rad_pre}) and (\ref{chap3_env_pre_r}) for different density parameters ($\lambda=0.01$ and $\lambda =0.07$). The best fit for the pressure-density curve is found to be in the quadratic form \cite{khunt2021distinct}
\begin{eqnarray}
   p=\gamma+\alpha\rho+\beta\rho^{2}
\end{eqnarray}
where  $\gamma$, $\alpha  $  and $\beta$ are the fitted parameters. For two different density variation parameters and their fitted values of EoSs are listed in Table {\ref{tab:2}}. Based on such an equation of state, the mass-radius relation has been reported in our earlier study \cite{khunt2021distinct}. The TRV model  based on the pseudo-spheroidal geometry, can support different choices of $\lambda$. Thomas et al. \cite{thomas_2005} shows that $\lambda$ can varies from $0.01$ to $0.09$ for this particular model of superdense stars. Looking for a thin crust situation, we  present our results for the choice of $\lambda=0.07$.  The results of core radius  and envelope size ($R_{E}=R-R_{c}$)  for two different values of $\lambda$  are given in Table {\ref{tab:2}}.

\bigskip

\begin{table}[htb]
\begin{center}
{\begin{tabular}{cccccc}
\hline\noalign{\smallskip}
$\lambda$ & $\gamma$ & $\alpha$  & $\beta$ & \makecell{$R_{C}$ \\(km)}  & \makecell{\text{Envelope size} $R_{E}$ \\(km)}  \\
\noalign{\smallskip}\hline\noalign{\smallskip}
 0.01& $-9.30 \times 10^{-4}$ & 1.69  &   406 & 4.00  & 7.72\\
 0.07 & $-2.09\times 10^{-4}$ & $0.77$  &$1291.49$ & 10.60 &1.57 \\
\noalign{\smallskip}\hline
\end{tabular}}
\caption{Values of constants that generate EoSs, core and envelope radii ($R_{C}$, $R_{E}$) for two different values of density variation parameters}
\label{tab:2}
\end{center}
\end{table}

\begin{figure}[h]
\centerline{\includegraphics[scale=0.50]{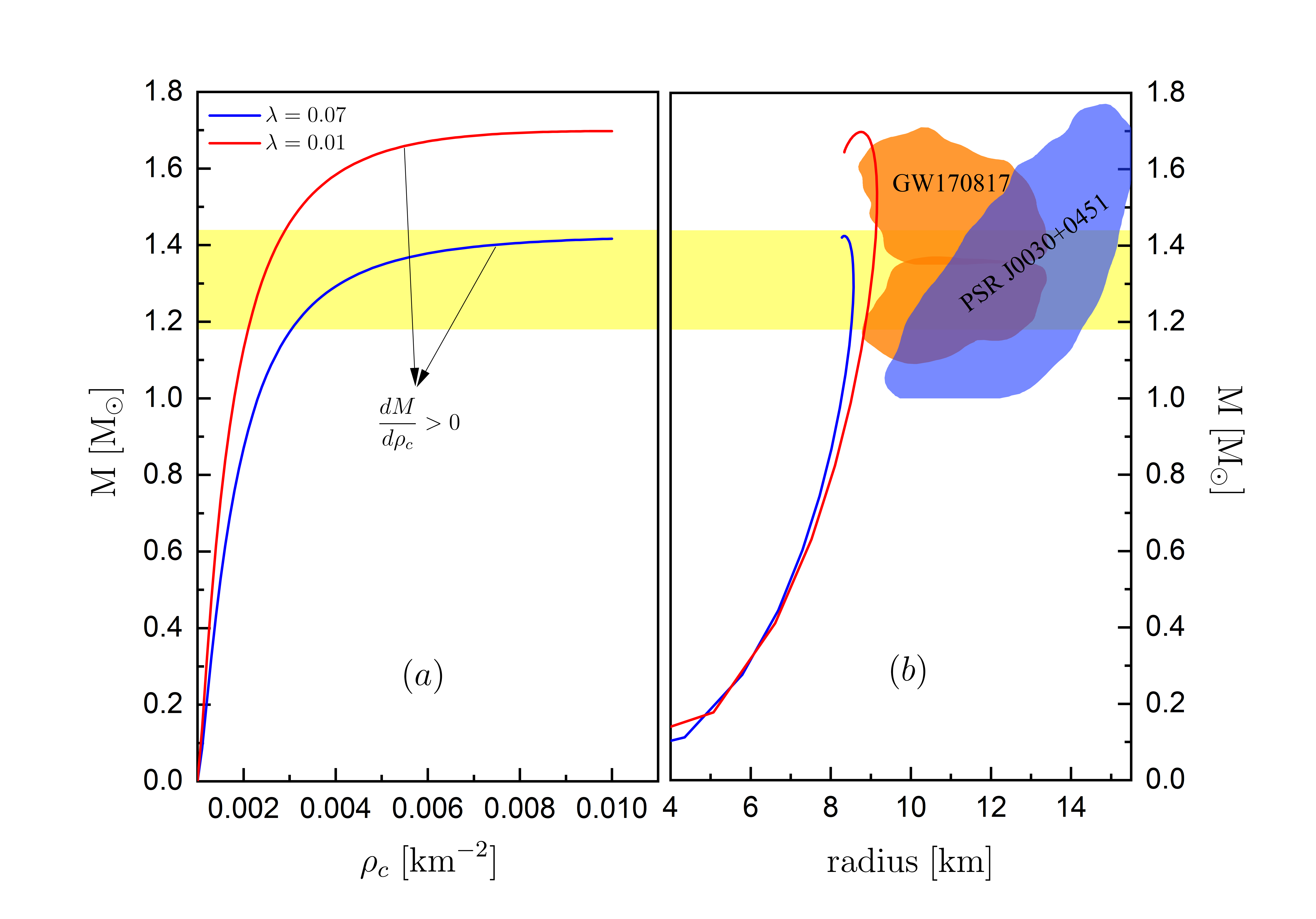}}
\caption{ Mass-central  density relationship  (panel (a)) and Mass-radius relationship (panel (b)) for the two geometrical EoSs models considered in these plots for comparison. The hatched (Yellow color) region in panel (a) and (b) represents the range of precisely measured masses of binary radio pulsar \cite{lattimer2012nuclear}. The  orange regions are the mass-radius constraints from the GW170817 event \cite{abbott_2018}. The blue region represent the pulsar, is the NICER estimations of PSR J0030$+$0451 \cite{miller2019psr}.}
\label{fig:mr_ch4}       
\end{figure}

In Fig. \ref{fig:mr_ch4}, we show the Mass-Radius diagram with the TRV EoS considered in this work.  The density profile at a stellar structure's centre, $r = 0$, which determines how stable \footnote{Not all branches of sequence $M=M(\rho_{c})$ are stable. This can be unstable by means of radial oscillations. Degenerate stars with $dM/d\rho_{c}<0$ are found to be unstable and will finally collapse towards Neutron stars, or Black holes.} it is against its own gravitational attraction. At an equilibrium radius, the confined mass will be larger as higher the centre density. The mass of a self - gravitating compact star increases with central density, in the stable region, in accordance with the Harrison-Zel'dovich-Novikov criterion \cite{zeldovich_1971,harrison1_965,saklany_2023}.   The static stability condition necessitates that $dM/d\rho_{c}>0$ for all points within the confining region, where $r < R$. As can be seen from Fig.\ref{fig:mr_ch4} (panel (a)), the mass of the model star $M(_{\rho_c})$ increases monotonically with $dM/d\rho_{c}>0$, resulting to a stable stellar structure as the central density $\rho_{c}$ increases.  As a consequence, we find by  solving the TOV equation (Eq.(\ref{hydro}) with $\Delta=0$) for a given EoS a maximum possible mass  corresponding to the central maximal density. Mass-radius decreases with decreasing density variation parameter $\lambda$, the ratio of surface density to the central density. The maximum mass at maximum central density is about $1.68 M_{\odot} $ and the corresponding  stellar radius 8.76 km for $\lambda=0.01$, while for $\lambda=0.07$ the stellar mass is about $1.42M_{\odot}$ and radius 8.33 km. The TRV model of superdense stars is  very useful for classification of different class of compact stars. For stars with $M \geq \text{\(M_\odot\)}$, the radius changes very little with $M$, $R \simeq 8-10$ km, comparable to neutron stars with a moderately stiff EoS. However, at smaller $M$ the radius of bare strange star behaves in an entirely different manner. It falls monotonically as $M$ decreases, with $R \propto M^{1/3} $ for $M \leq 0.5\text{\(M_\odot\)}$ as expected from the fact that low-mass strange stars can be described by the Newtonian theory, which gives $M \simeq \frac{4\pi}{3} \rho R^{3}$ . The decrease of $R$ with decreasing $M$ is unique feature of strange star and quark stars or compact star built of abnormal matter generally referred as self-bound stars \cite{de_2011}. It's important to note that the mass-radius constraint from the gravitational wave event GW170817 represented by the orange clouded region in Fig. \ref{fig:mr_ch4} corresponding to the heavier and lighter neutron star, respectively \cite{abbott_2018}. In this figure,  the blue clouded region show the NICER (Neutron Star Interior Composition Explorer) mass-radius measurements on  PSR J0030+0451  \cite{miller2019psr,riley2019nicer}. Besides, the yellow horizontal band show the range of precisely measured masses of binary radio pulsar \cite{lattimer2012nuclear}. Because we have the results of various EoSs in GR, we use the geometrically deduced EoS to compute the mass-radius relation for different density variation parameters, $\lambda$ that matches  with the observational data. Strictly speaking we investigated the mass-radius relation of neutron star in a general relativistic framework for the class of highly compact self-bound stars (e.g,. quark stars).

After calculating an equilibrium stellar model we analyze its stability as stable equilibrium stars of astrophysical relevance. Hereafter we briefly address the stability with respect to density perturbations. For simplicity, we restrict ourselves to non-rotating spherically symmetric equilibrium models.

\section{Cracking  Stability}
\label{sec:3_chap4}

 The sound speed is a microscopic quantity of immense importance, particularly the potential upper bounds that might be obtained, because it impacts the EoS. One approach to quantify dense materials is by the velocity of sound, which is given by $v_{s}^{2}= dp/d\rho$, where $p$ is the pressure and the $\rho$ matter density. Causality demands an absolute restriction on $v_{s}\leq 1$, and thermodynamic stability ensures that $v_{s}^2 \geq 0$ \cite{lattimer2001neutron}. The fundamental assumption concerning the upper bound of the velocity of sound is that it satisfies causality ($v_{s}^{2}= dp/d\rho \leq c^2$), i.e. the speed of sound cannot reach the speed of light, which is the significant barrier imposed on the speed of sound  by general principles according to Zel'dovich et al. \cite{YBZ}. Moreover,  \cite{bedaque2015sound} showed that the possibility of neutron stars with masses of around two solar masses, together with knowledge of the EoS of hadronic matter at low densities, is incompatible with the constraint $c / \sqrt{3}$. The speed of sound and its impact on tidal deformability have also been analyzed in refs \cite{reed2020large,moustakidis2017bounds,van2017upper,ma2019sound}.

When examining arbitrary and independent density and anisotropy perturbations (as in all earlier cracking investigations \cite{herrera1992cracking,di1994tidal,di1997cracking,abreu2007sound,ratanpal2020cracking}), there are few physical criteria to determine the magnitude (absolute and/or relative) of the perturbation, i.e., relatively small (or huge) the perturbations should be.  In addition, these earlier works have only considered continuous perturbations in their studies. It is possible that variable perturbations have the potential to be more effective in inducing cracking within a specific matter configuration. However, we are looking for physical characteristics that can be tested to see whether cracks are forming. A cracking-like scenario could be produced by perturbations of varying orders of magnitude (and relative size $\delta \Delta / \delta \rho$).
\begin{eqnarray}\label{pertub}
    \frac{\delta \Delta}{\delta \rho} \sim \frac{\delta (p_{\bot}-p_{r})}{\delta \rho} \sim \frac{\delta p_{\bot}}{\delta  \rho}-\frac{\delta p_{r}}{\delta \rho} \sim v_{s \bot}^{2} - v_{s r}^{2}
\end{eqnarray}
$v_{s \bot}^{2}$ and $ v_{s r}^{2}$ stand for the radial and tangential sound speeds, respectively \cite{abreu2007sound}.

This fundamental idea  employed in reviewing Herrera's technique to identify possibly unstable anisotropic matter structures is based on the concept of cracking. Now, by addressing the sound speeds and calculating (\ref{pertub}), we  not only have a more precise notion of the relative order of magnitude of the perturbations ($\delta \Delta$ and $\delta \rho$) but also on what regions are more likely to be potentially unstable inside a matter configuration. There are two possible stable and two possible unstable regions: those where the gradient of anisotropy with respect to radial variable $r$ is larger than or equal to zero, and those where it is negative. It is clear that because $0\leq v_{sr}^{2} \leq 1$ and  $0\leq v_{s \bot}^{2} \leq 1$, we have $\mid v_{s \bot}^2-v_{s r}^2 \mid \leq 1$ \cite{abreu2007sound}. Hence,

\begin{align}
    -1 \leq v_{s\bot}^{2}-v_{s r}^2 \leq 1 \Rightarrow
        \begin{cases}
           -1 \leq v_{s\bot}^{2}-v_{s r}^2 \leq 0 & \text{potentially stable} \\
         0 < v_{s\bot}^{2}-v_{s r}^2 \leq 1 & \text{potentially unstable}
        \end{cases}
        \label{condition:1}
   \end{align}

\begin{figure}[h]
\centerline{\includegraphics[scale=0.5]{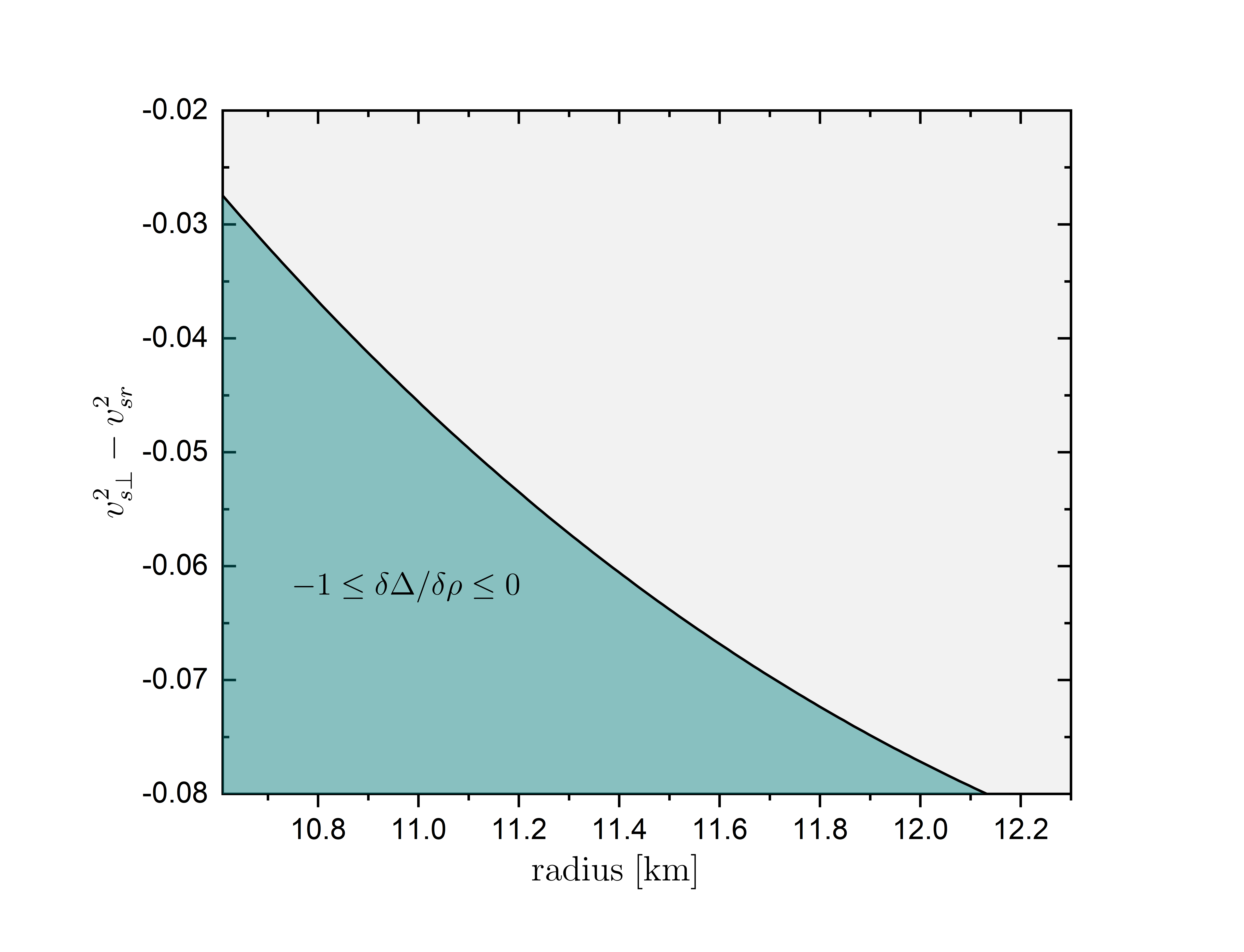}}
\caption{ Variation of the difference of radial and tangential sound speeds for the anisotropic envelope configuration. } 
\label{fig:pert}       
\end{figure}

Accordingly, we may now examine possible stable/unstable regions inside anisotropic models based on the difference of the propagation of sound within the matter configuration. Those regions where $v_{sr}^2 \geq v_{s \bot}^2$ will be  unstable. On the other hand, if $v_{s r}^{2}\leq v_{s \bot}^{2}$ everywhere within a matter distribution, no cracking will occur. For physically plausible models, the size of perturbations in the anisotropy should always be lower than those in density, i.e. $ \mid v_{s\bot}^{2} - v_{sr}^2 \mid $ $\leq $ $ 1 $ $\Rightarrow $ $\mid \delta \Delta \mid $ $\leq$  $\mid \delta \rho \mid$. When $\delta \Delta /\delta \rho > 0$, such perturbations lead to the possibility of unstable configuration. Profile for the $\delta \Delta / \delta \rho $ is displayed in Fig.\ref{fig:pert} for the anisotropic envelope region. For the TRV  core-envelope model under consideration , the perturbation relation $\delta \Delta / \delta \rho \equiv v_{s\bot}^{2}-v_{sr}^{2} $ satisfies the physical limit as given by  Eq.(\ref{condition:1}).

\begin{figure}[!tbp]
  \centering
  \subfloat[]{\includegraphics[width=0.9\textwidth]{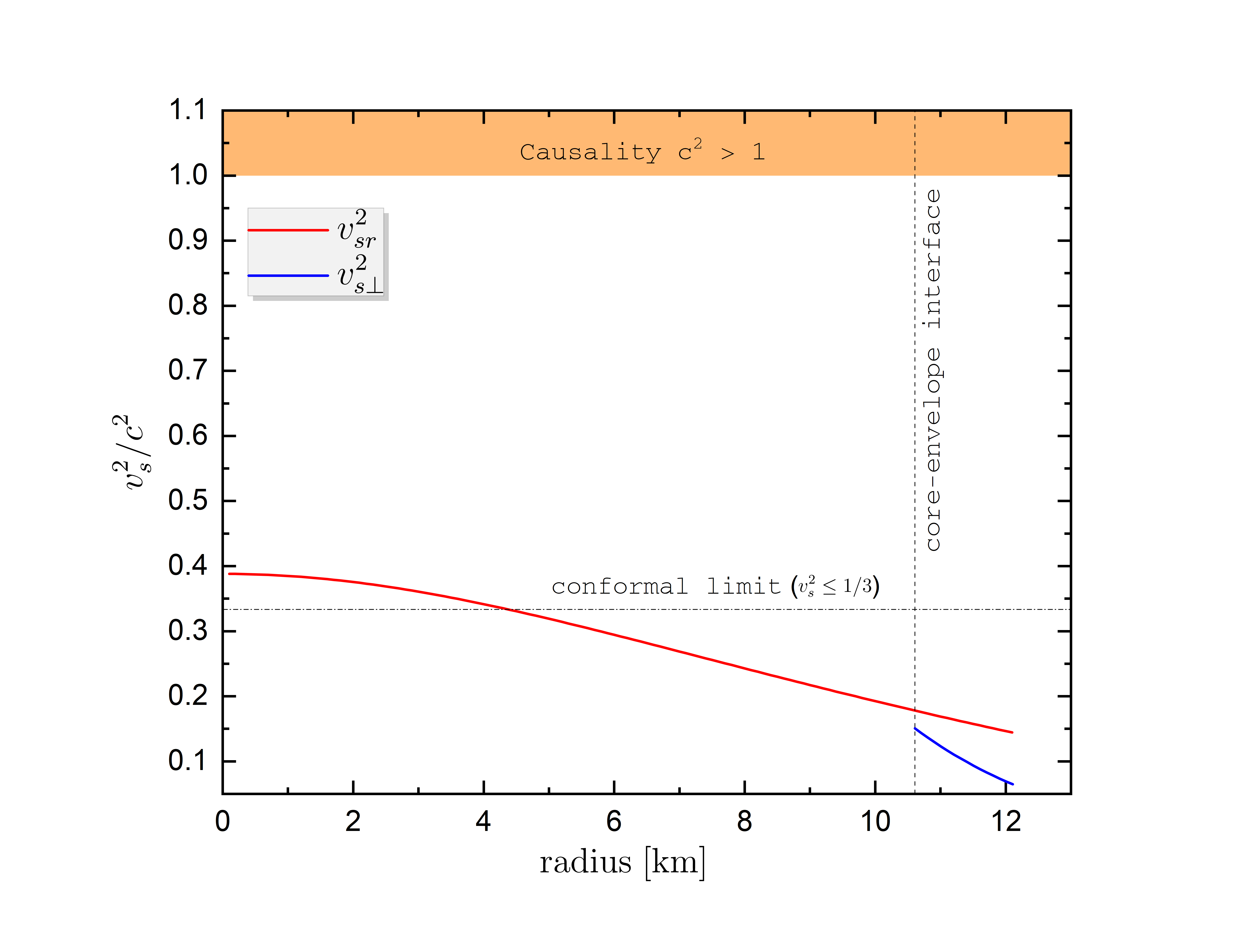}\label{fig:vs_r}}
  \hfill
  \subfloat[]{\includegraphics[width=0.9\textwidth]{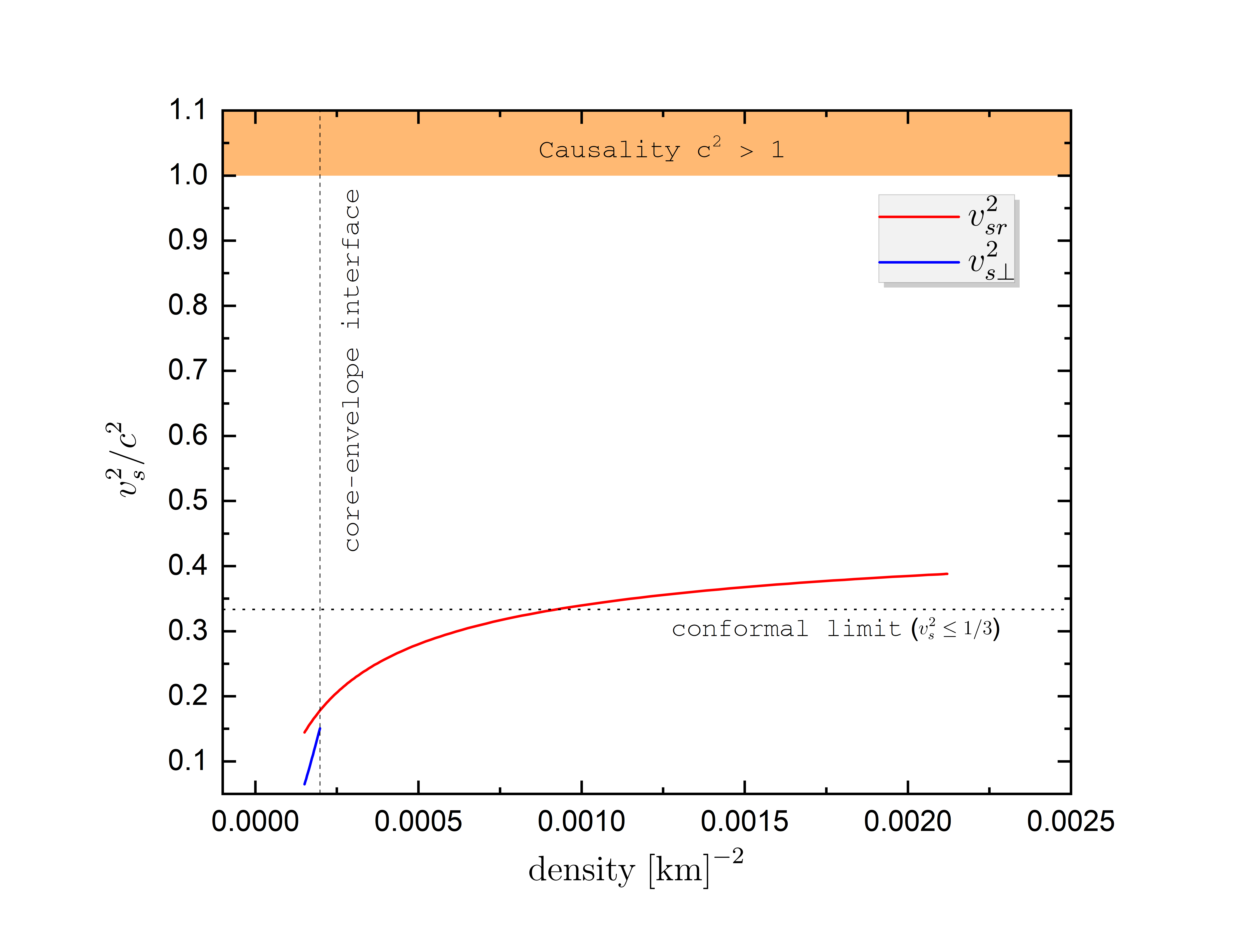}\label{fig:vs_rho}}
  \caption{panel (a): Speed of sound $v_{s}^2$ as a function radial distance for the TRV EoS. panel (b): Speed of sound $v_{s}^2$ versus density. The vertical dotted line shows the core-envelope interface and the horizontal dotted line represents for the conformal limit ($v_{s}^{2} < 1/3$). The red solid  line represents radial velocity in the core and the envelope regions, respectively,. The solid blue line described tangential velocity in the envelope region, as well as grey upper band shows causality region.
  }
\end{figure}

In Fig. \ref{fig:vs_r} , \ref{fig:vs_rho}, the radial ($v^{2}_{sr}$) and tangential sound velocity ($v^{2}_{s \perp }$) as function of radii and density are shown.  From the computed results, we found that $v_{sr}$ as well as $v_{s \perp }$ is monotonically decreasing outward with the continuity at the interface for the $v_{sr}$ and discontinuity at interface for the $v_{s \perp }$. This discontinuity occurs at the core-envelope interface  actually comes from anisotropic pressure in the envelope region, and thus interferes with sound velocity. 

An important parameter which characterizes the stiffness of the EoS with respect to density perturbations is the \textit{adiabatic index} $\Gamma$.
The physical consistency of a relativistic anisotropic sphere is determined by  $\Gamma$, which is defined as the ratio of two specific heats \cite{bondi1964contraction}. The relativistic adiabatic indices  also are  important
parameters that affect the stability of any stellar
system. These are defined as

\begin{eqnarray*}
    \Gamma_{r}(r) = \frac{p_{r}(r)+\rho(r)}{p_{r}(r)} \frac{\partial p_{r}(r)}{\partial \rho(r)}
\end{eqnarray*}
\begin{eqnarray}
    \Gamma_{\bot}(r) = \frac{p_{\bot}(r)+\rho(r)}{p_{\bot}(r)} \frac{\partial p_{\bot}(r)}{\partial \rho(r)}
\end{eqnarray}

  

\begin{figure}[!tbp]
  \centering
  \subfloat[]{\includegraphics[width=0.9\textwidth]{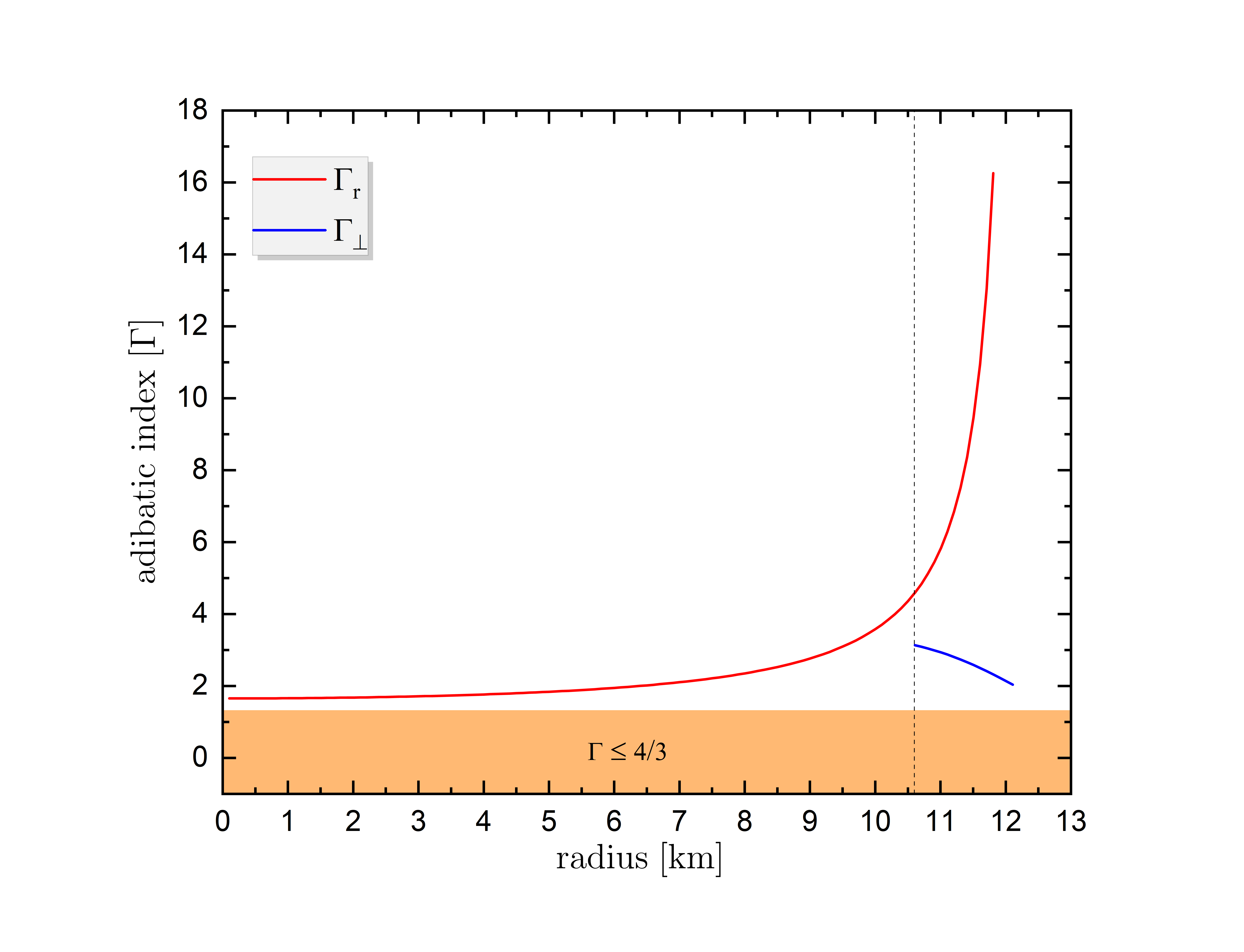}\label{fig:adia_r}}
  \hfill
  \subfloat[]{\includegraphics[width=0.9\textwidth]{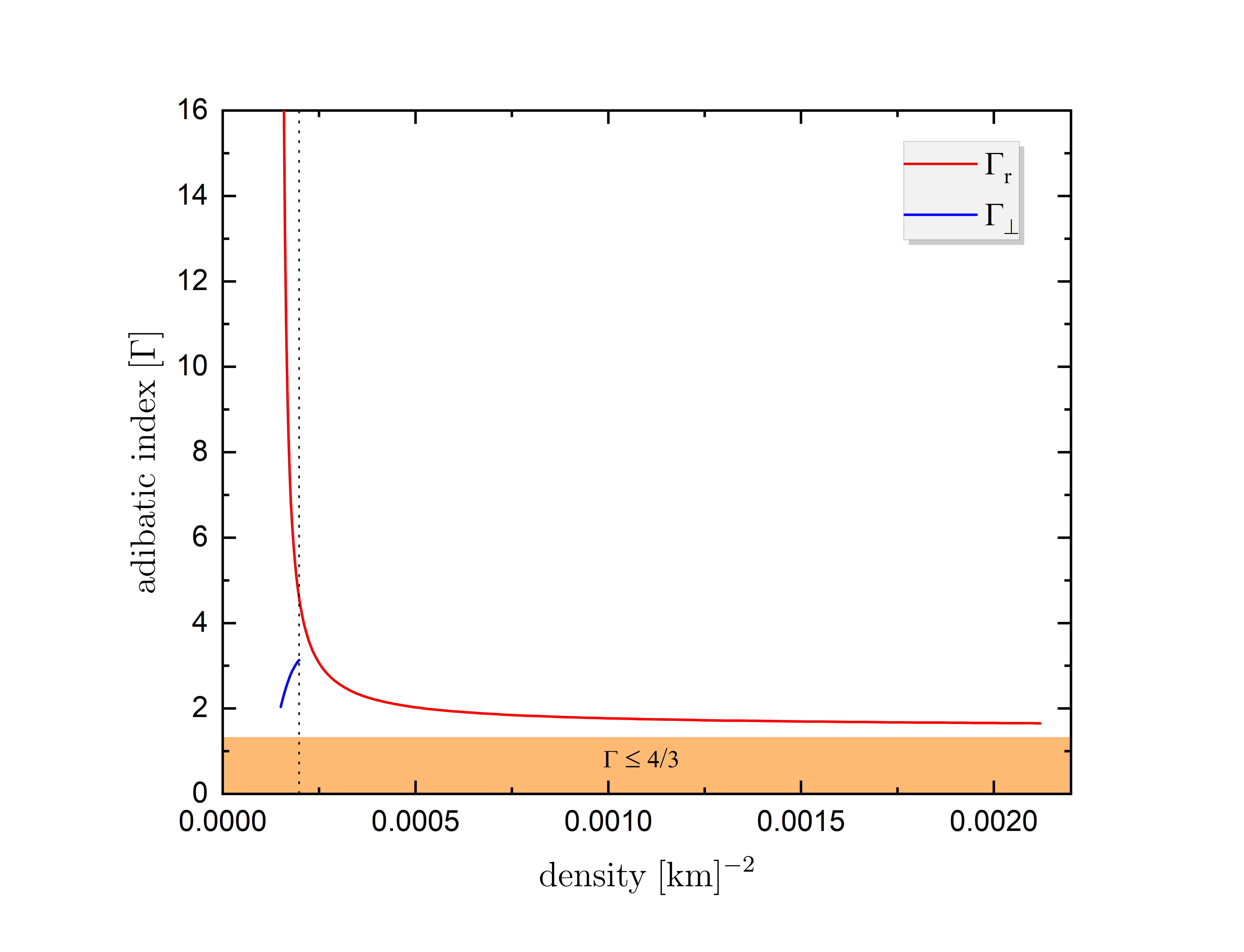}\label{fig:adia_rho}}
  \caption{ The adiabatic index $\Gamma$ versus radii and density(in geometrical unit) in a compact star core and envelope. The red-solid line is for radial $\Gamma$ in the core and evelope regions, respectively, the blue-solid line  corresponds to the case in which the tangential velocity the envelope region. The horizontal orange band is $\Gamma =4/3$ characteristic of a free ultra-relativistic Fermi gas.
  }
\end{figure}

The profile of $\Gamma$ of the core and envelope of both pressures are plotted in Fig.{\ref{fig:adia_r}}, {\ref{fig:adia_rho}}. Bondi \cite{bondi1964contraction} suggested that for a stable Newtonian sphere,  $\Gamma$ should be greater than $\frac{4}{3}$. The $\Gamma$ is a fundamental component of the instability criteria \cite{chandrasekhar_1964}.  In particular, the amount that contains all of the main attributes of the equation of state on the instability formulas as described in Ref. \cite{chandrasekhar_1964}, as a result, it serves as the link between the relativistic structure of a spherical static object and the equation of state of the internal fluid. Specifically, $\Gamma$ varies from 2 to 4 in most equations of state of neutron star matter. More specifically, in certain circumstances,  $\Gamma$ is a weak function of density, in other cases, however, the density dependency is more complicated \cite{haensel2007neutron}.

The results of  $v_{s}^2$  and $\Gamma$ for the compact star matter are presented in  Fig.{\ref{fig:vs_r},\;\ref{fig:vs_rho}}   and Fig.{\ref{fig:adia_r},\;{\ref{fig:adia_rho}}}, respectively. The results show that there are possible stable regions inside the anisotropic envelope of the star. Because $\delta \Delta / \delta \rho < 0$, the sound speed stability criterion, Eq.(\ref{condition:1}) suggests that in the TRV anisotropic model, no cracking will occur.
Fig.{\ref{fig:adia_r}} and {\ref{fig:adia_rho}} clearly indicate that the radial adiabatic indexes are continual  at the interface and satisfies the Bondi condition \cite{bondi1964contraction}, but in the tangential case it does not obey the Bondi condition. It will be the main reasons for the deformation and cracks for the Newtonian sphere.

\section{Estimates of stellar parameters : $\Delta R$, $ \Delta E_{g}$ and $ \Delta I $ }
\label{sec4_chap4}

What if the quark stars are solid? When strains reach a certain level, a star-quake occurs as a natural result. Inside solid stars, two kinds of stress forces might form: bulk-variable and bulk-invariable forces \cite{peng_2008}. Each of these factors would result in release of gravitational and elastic energy (to be in a same order), with an order of
 \begin{eqnarray}
     E \simeq \frac{G M^2}{R} \sim 10^{53} \frac{\Delta R}{R}\; \text{ergs}
 \end{eqnarray}

for $M \sim M_{\odot}$, where $\Delta R$ is the radius change caused by quakes. If the stellar radius of a solid quark star with a radius of $\sim$ 10 km changes dramatically by 10 m during a quake, an energy release of $\sim 10^{50}$ ergs is achievable \cite{xu_2009}.

\begin{figure}[H]
\centerline{ \includegraphics[scale=0.5]{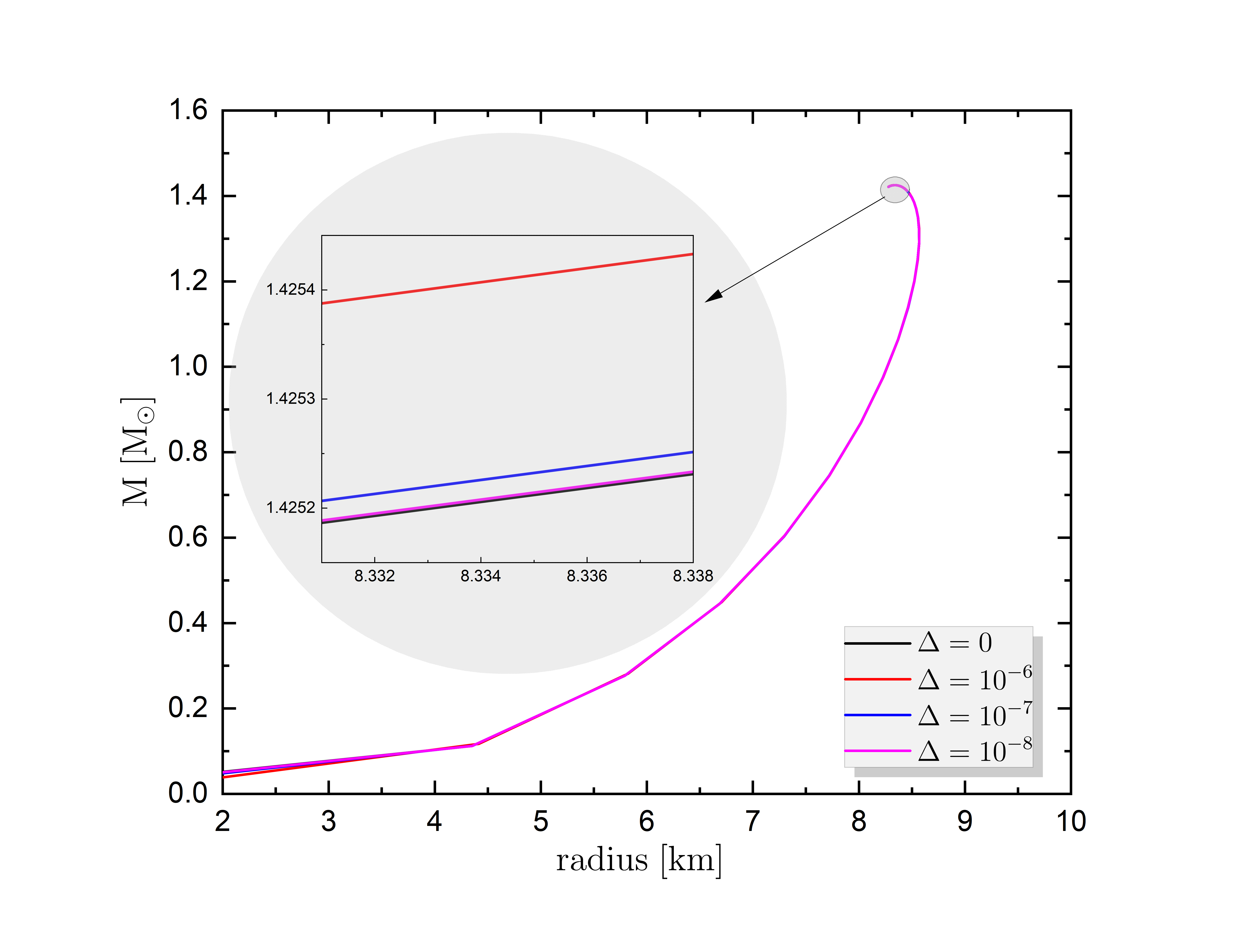}}
\caption{ The mass-radius relation with different choice of anisotropy. Where black-solid line for $\Delta =0$, which is for isotropy , red, blue and pink solid lines are for $\Delta =10^{-6}$, $\Delta =10^{-7}$ and   $\Delta =10^{-8}$, respectively.
\label{fig:mr_anis}}       
\end{figure}

This section will demonstrate how various anisotropy degrees affect the star structure. Fig. \ref{fig:mr_anis} shows how the M-R relation differs depending on the anisotropy value. In the M-R relation, we found that if we minor change the value of $\Delta$ within obtained limit, the mass and the radius also changed.  Here, we have demonstrated the difference in gravitational energy and moment of inertia caused by anisotropy based on the change in radius. Moreover, the mass difference, which is also seen as a result of the anisotropy variation, is between $10^{-5} M_{\odot}$ - $10^{-4} M_{\odot}$ . The mechanical energy that can be stored in this region , is about $\sim 2.5 \times 10^{49-50}$ erg
($10^{-5}-10^{-4} M_{\odot} c^2$).

To account for the small anisotropy parameter, $S$ of the order of $10^{-8}$ to $10^{-6}$, high numerical precision is required to arrive at the solution of the anisotropic TOV equation. Different anisotropic scales are presented here in the form of $\Delta$, and their values are chosen by the domain of anisotropy of Eq.(\ref{chap3_env_ani}) (for $\lambda=0.07$).  In the case of isotropic pressure, $S=0$, Eq.(\ref{hydro}) turn out to be the TOV equation. We have incorporated $\Delta$ with Eq.(\ref{hydro}) for solving the anisotropic TOV equations,  we employ the TRV equation of state deduced from our earlier study of a core-envelope model of superdense stars~\cite{thomas_2005,khunt2021distinct}. To begin with, the star's core is considered to be homogeneous, with the density, $\rho = \rho_{0}$. In the present study we have fixed the central density for the TRV EoS at $\rho _{0} = 1.34 \times 10^{15}$ g cm$^{-3}$, Eq.(\ref{mass}) and  (\ref{hydro}) are integrated numerically to determine the global structure of (e.g. mass and radius) of a compact star.

The difference in gravitational energy ($E_{g}(\Delta)-E_{g}(\Delta=0)$), moment of inertia ($I(\Delta)-I(\Delta=0)/I(\Delta=0)$), and radius ($R(\Delta)-R(\Delta=0)$) induced by anisotropy will be computed using the approximations below; we include the anisotropy and compare it to the value obtained without anisotropy, which are described below.

The total gravitational energy ($E_{g}$) as well as the stellar moment of inertia ($I$) are computed as follows :

(i) \textit{Gravitational Energy} \;($E_{g}(\Delta \neq 0$)): 

A refined formula to compute the $E_{g}$ containing the  compactness parameter $x_{GR}= r_{g}/R$, as proposed by \cite{lattimer2001neutron} is written as 

\begin{eqnarray}
 E_{g} (\Delta) \simeq 1.6 \times 10^{53} \Big[\frac{M(\Delta)}{M_{\odot
}}\frac{x_{GR}(\Delta)}{0.3}\Big] \frac{1}{1-0.25 ~ x_{GR}(\Delta)} \,\, \text{erg}.
\end{eqnarray}

Here, $x_{GR}  =2 G M/R c^{2}=r_{g}/R$, where $r_{g}$ is the Schwarzschild radius. It implicitly depends on the anisotropic parameters, $\Delta$ through $M$ and $R$.
The importance of relativistic effects for a neutron star mass $M$ and radius $R$ is characterized by the  compactness parameter $r_{g}/R$.

(ii) \textit{Moment of inertia versus $M$ and $R$} :

 To quantify moment of inertia, we have used approximate formula relating $I$ to stellar mass and radius \cite{ravenhall1994neutron}. They showed that the ratio $I/M R^{2}$ depends mostly on the compactness parameter ($x_{GR}$) and is expressed as \cite{thorne1977relativistic}
 \begin{alignat}{2}
 I(\Delta)\simeq 0.21 M(\Delta) R^{2}_{\infty} ; \;\;\;\;
 R_{\infty}= \frac{R}{\sqrt{1-x_{GR}(\Delta)}} .
\end{alignat}
 where $R_{\infty}$ is an apparent radius related to the circumferential radius 
 
\begin{figure}[h]
\centerline{\includegraphics[scale=0.5]{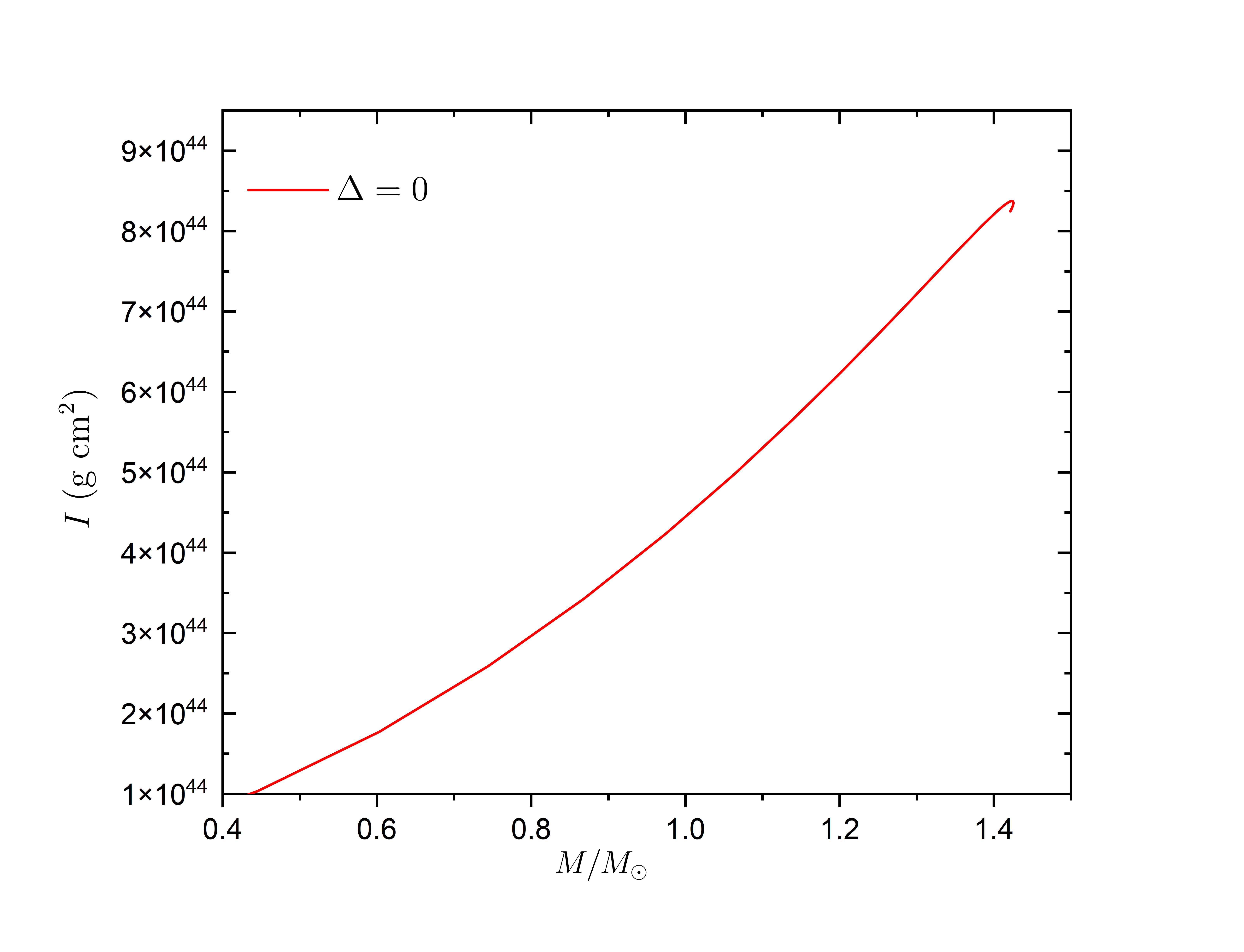}}
\caption{ Total moment of inertia versus total mass for $\lambda=0.07$ without anisotropy ($\Delta=0$).}
\label{fig:mi0}      
\end{figure}

In Fig.\ref{fig:mi0} we show the behaviour of the  moment of inertia with respect to the TOV mass for $\lambda=0.07$ with $\Delta=0$. The moment of inertia at maximum stable mass, $M_{max}$ is about $8.24 \times 10^{44}$ g cm$^2$.

In the computation, it is vital to enhance precision in order to accomplish the differences that we had in $ E_{g}(\Delta\neq 0)-E_{g}(\Delta=0)$ and $I(\Delta \neq 0)-I(\Delta=0)$. To optimize integration precision, we have divided the domain into smaller and smaller sections at the boundary, until the computed values are apparent.


\begin{figure}[H]
\centerline{ \includegraphics[scale=0.5]{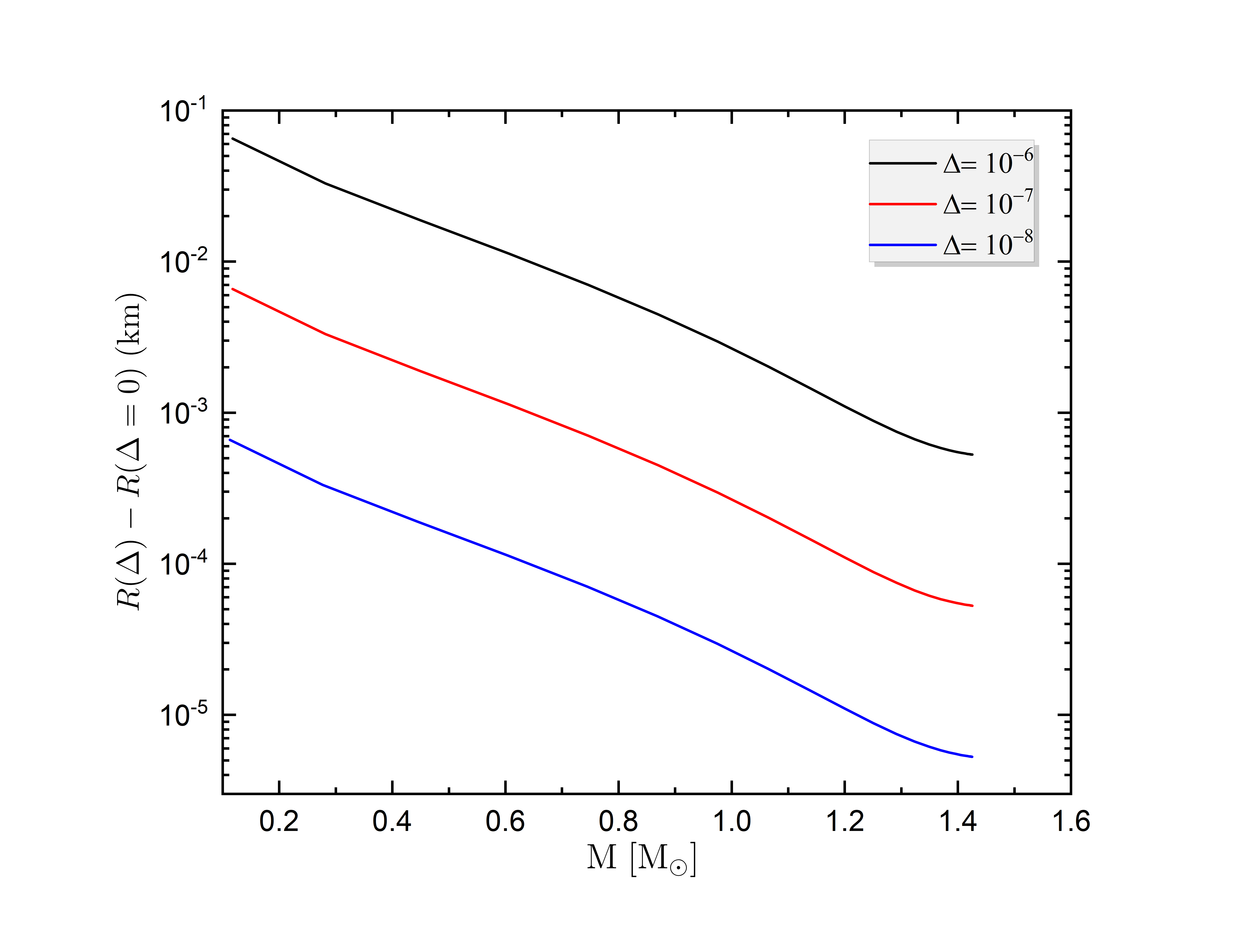}}
\caption{ The distinction $R (\Delta\neq 0)-R(\Delta = 0)$ in stellar radius as a function of stellar mass. Where black-dash line for $\Delta =10^{-6}$, red-dot line for $\Delta =10^{-7}$ and for $\Delta =10^{-8}$ is shown in blue dash-dot line.
\label{fig:mrdiff}}       
\end{figure}

\begin{figure}
\centerline{ \includegraphics[scale=0.5]{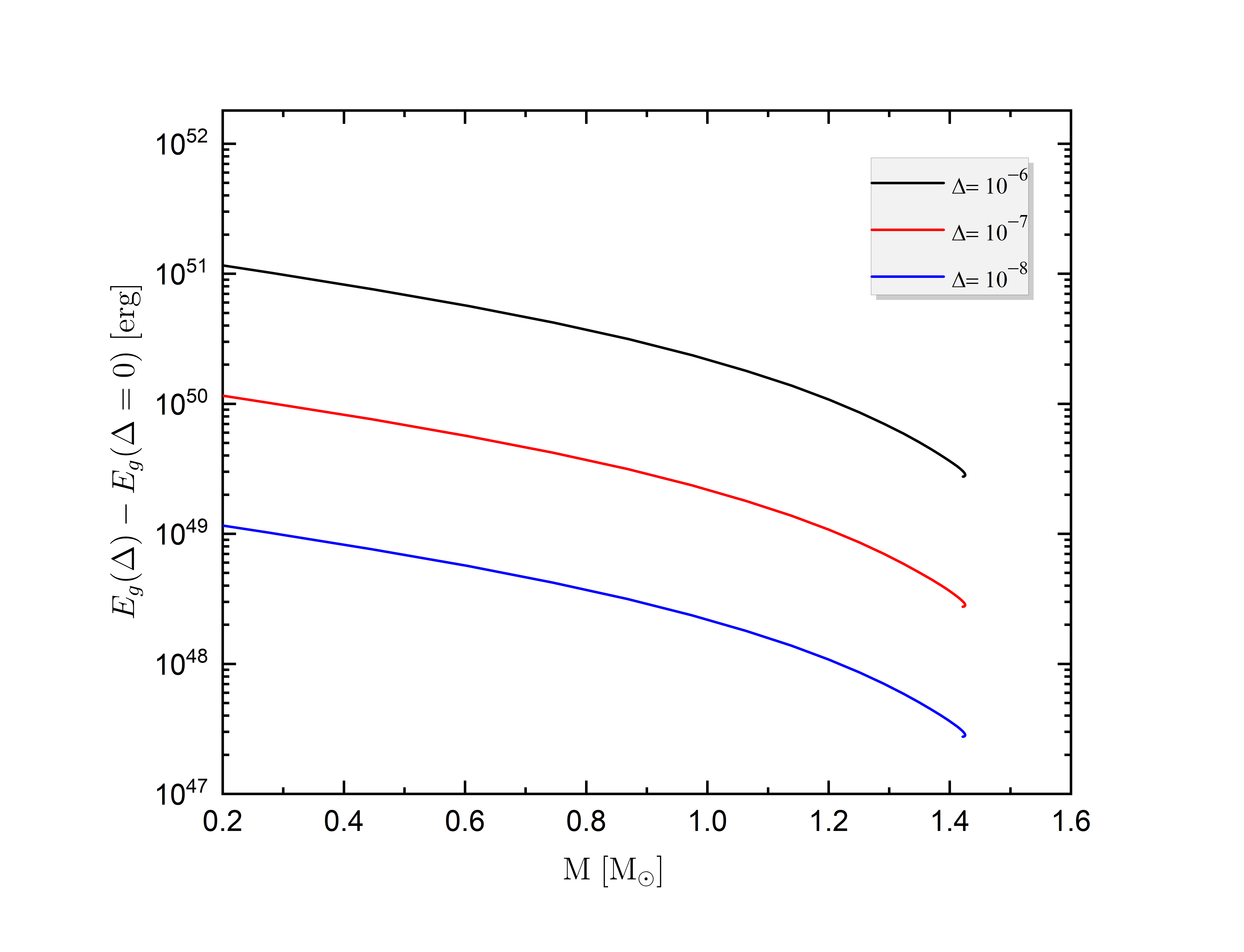}}
\caption{As in Fig. \ref{fig:mrdiff}, but for difference in gravitational energy.
\label{fig:engdiff}}      
\end{figure}

 \begin{figure}

\centerline{\includegraphics[scale=0.5]{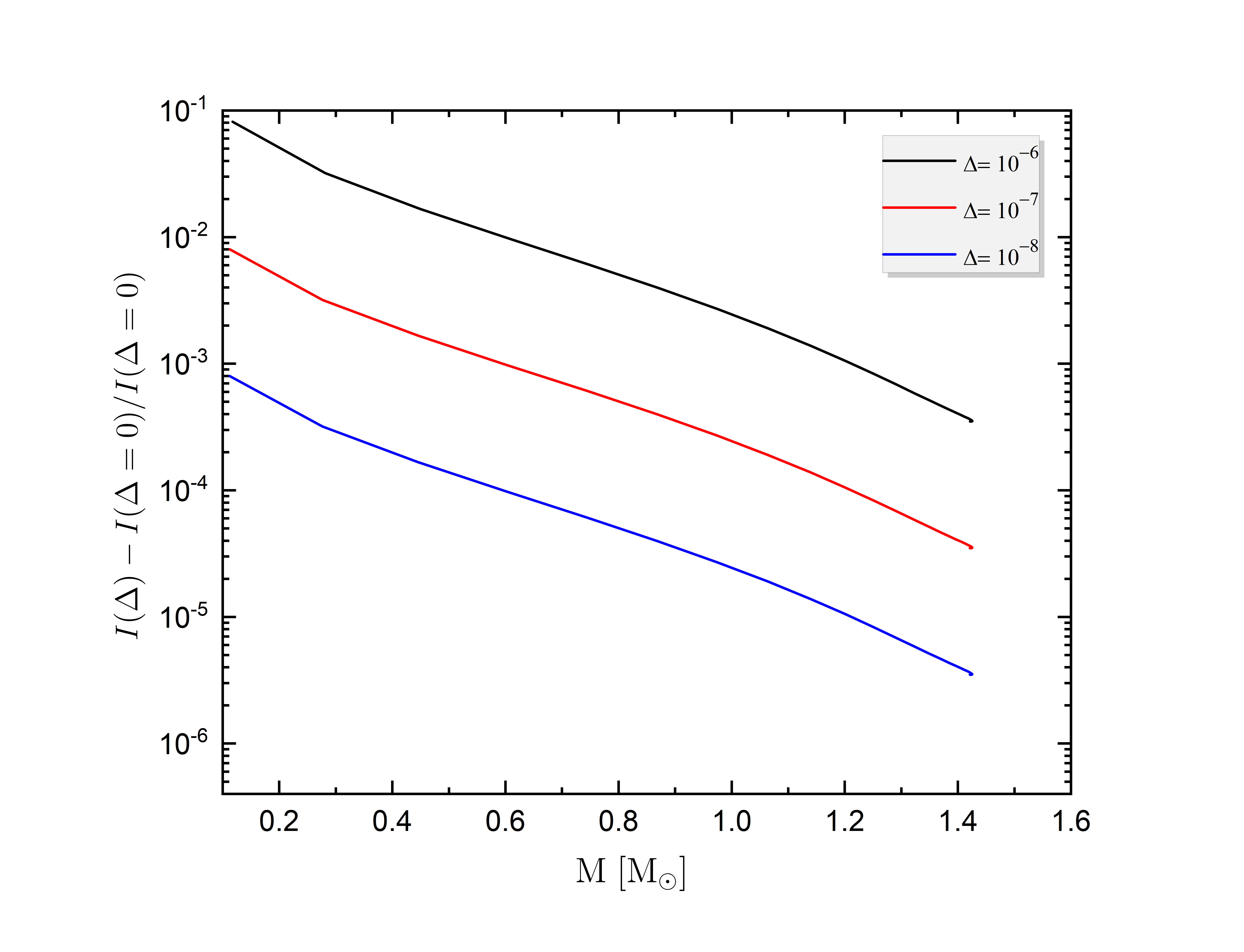}}
\caption{ As in Fig. \ref{fig:mrdiff}, but for the ratio difference in moment of inertia. 
\label{fig:midiff} }      
\end{figure}

The computed values of $R(M)$, $E_{g}(M)$ and $I(M)$ are shown in Fig. \ref{fig:mrdiff},~\ref{fig:engdiff} and \ref{fig:midiff}, respectively. In the mass-radius calculation, we considered $\Delta$ from $10^{-6}$ to $10^{-8}$. From Fig.\ref{fig:mrdiff},  it is seen that as $\Delta$  varies from $10^{-8}$ to $10^{-6}$, the radius of the star in this range varies by $10^{-5}$ to $10^{-1}$ km.

Starquakes can cause a drastic change in the evelope region, releasing both gravitational and tangential strain energy \cite{franco2000quaking,horvath2005energetics,xu2003solid}. In general, the changes in stellar radius, $E_{g}$, and $I$ have direct correlation to stellar mass and anisotropic parameter $\Delta$. This indicates that in a highly compact self-bound  star with a greater mass, a quake-like event should be more significant for  larger change in $\Delta$.  Moreover, the disruptive instability of the radius may lead to more chaotic events at the core-envelope interface and  can cause huge  energy release. Fig.\ref{fig:engdiff} shows the computed gravitational  energy difference corresponds to the anisotropy parameter $\Delta \sim 10^{-6} $ and $M_{max} \sim 1.43 M_{\odot}$. It is observed that the released energy is as high as $10^{50}$ erg. The energy released decreases with a decreasing magnitude of anisotropy (for $\Delta$ varies from $10^{-6}$ to $10^{-8}$). This scale of energy released  is comparable with the energy observed in  SGRs and GRBs (see in Table \ref{tabgrb:1}). From our calculation as shown in Fig.\ref{fig:engdiff}, we find that a giant starquake with $\Delta\leq 10^{-6}$ could produce such an energetic flare.  Based on different choices of the anisotropy parameters, for $\Delta=10^{-5}$, it is shown for quarks stars that the energy released becomes of  the order of $10^{49}$ erg \cite{xu2006superflares}. A  change in $\Delta$ might cause a change in its spin frequency, $\Delta\Omega /  \Omega$ $= - \Delta I / I$ leading to glitches. The giant energetic flare is considered to be associated with a high-amplitude glitch. We observed from Fig.\ref{fig:midiff} that such glitches with $\Delta \Omega \sim 10^{-5}$ to ${10^{-1}}$ could appear for mass variability of $M=0.2-1.43 M_{\odot}$ and $\Delta=10^{-8}$ to $10^{-6}$.

\section{ Discussion }
\label{sec:6_chap4}

To summarize, we have systematically investigated  the compact stars having a thin crust particularly changes in the gravitational energy, the moment of inertia as an implicit function of the anisotropy magnitude ($\Delta$) predicted by the TRV core-envelope model of compact star.
We modeled the compact star as a self-bound core of incompressible fluid with a fragile envelope.
The anisotropy variation as shown in  Fig.\ref{chap4_fig_1} indicates that the anisotropy starts at the core boundary  and increases up to a radial distance of 10 to 12 km . It  further decreases towards  the envelope boundary of the star. In Fig.\ref{fig:mrdiff}, we have shown the mass-radius relation for different values of the anisotropy parameter in the range $10^{-8} \leq \Delta \leq 10^{-6}$  and computed the difference in the radius of the star.  Further the gravitational energy difference and moment of inertia difference due to the  radial fluctuations causes by the anisotropy parameter  are also estimated (see Fig.\ref{fig:engdiff}). It is found that the magnitude of this energy is comparable to the energy produced by SGRs and GRBs.

Cracking is a concept that was first suggested for self-gravitating anisotropic matter configurations by L Herrera and collaborators \cite{herrera1992cracking,di1994tidal,di1997cracking}. This concept has been reintroduced in our analysis. In this study, we followed their approach and studied the cracking and stability of self-bound anisotropic stars. It has been revealed that, for certain dependent perturbation in particular, the ratio for fluctuations in anistropy to energy density, $\delta \Delta / \Delta \rho$, could be well understood in terms of the difference between the velocities of sound (i.e. $\delta \Delta / \delta \rho \equiv v_{s\bot}^{2}-v_{sr}^{2} $). Cracking points in a configuration that satisfies $\mid \delta \Delta/ \delta \rho \mid > 1$, would not lead to physically unstable models. From the present study, we are able to clearly determine, based on equation (\ref{condition:1}), the region within the matter distribution are more likely to be potentially stable/unstable. From the Fig.\ref{fig:pert}, we see that $ -1 \leq v_{s\bot}^{2}-v_{s r}^2 \leq 0 $
 condition is satisfied even in the anisotropic configuration. Our results indicate that anisotropic structure is potentially stable (as shown in Fig.\ref{fig:pert}). Moreover the work done by  \cite{abreu2007sound}  argued that when the system is entirely anisotropic  the system may be considered possibly stable and it can be essential for cracking like event. The faults generated due to the cracking events in the evelope region can lead to quakes. The superflares of $\gamma$-ray emission (e.g., GRBs, SGRs) could be caused by  quakes in highly compact self-bound stars.

Thus in conclusion, we are able to propose a viable model towards our understanding of the production of GRBs to the conventional framework. According to our theoretical calculation and numerical results : if the tangential pressure is slightly larger [only $(1+10^{-6})$ times] than the radial pressure in a self bound star with mass $1.43 M_{\odot}$, the released gravitational energy during a quake could be as high as $10^{50}$ erg, for spherically symmetric compact stars. We have examined the evolution of anisotropy in the envelope of highly compact star. 
 The envelope develops cracks under the pressure stress caused by  density fluctuations induced by the anisotropy. The amount of energy  emitted from this event is very remarkable.
It is possible for stresses to build in the envelope of a high compact star resulting  the crust being fractured (i.e., a starquake), which may have an impact on the star's spin development and generate  glitches.

A star-quake is a possible activation source for stellar instabilities, e.g., it might be linked to a pulsar glitch or a magnetar flare. The idea is that the star's evolution, such as magnetic field loss, produces strain in the elastic crust, and  the stored energy is abruptly released when the system reaches a critical threshold. The usual energy released in this process would consequently be somewhere of the order of the maximum mechanical energy that can be stored in the crust, which is estimated to be $10^{-9}-10^{-7} M_{\odot} c^{2}$ \cite{blaes1989neutron,mock1998limits}. This  is significant considering magnetars appear to be associated with rather regular flare outbursts \cite{watts2016colloquium}. Internal phase transitions are another possibility linked with the star's evolution.
From the point of view of the present study , it is important to note that the anisotropy largely at the envelope region that causes deformation at the envelope (star crust) leading to matter being thrown out and falling back. 
  
\chapter{Conclusions and Discussions}
\label{chap:5}

``\textit{A method is more important than a discovery, since the right method will lead to new and even more important discoveries.}''                                
\begin{flushright}
--Lev Landau
\end{flushright}

\bigskip

The mass and radius of a neutron star are very sensitive to the star's gravity theory and equation of state. Many attempts based on various equations of state through physical considerations have been employed to study neutron stars\cite{lattimer2001neutron,lattimer2012nuclear,potekhin_2010}. In this thesis, the correlation of matter and gravity theory has been employed to investigate the physical features of ultra-compact stars to some extent.

Some of the attempts  are based on hydrodynamics of NSs. In order to accumulate GTR that hydrodynamic equations are expressed in curved spacetime. The challenges involve rotation and/or NS seismology. However, using GR, we could forecast the physical state of a compact star without specifically requiring the EoS of nuclear matter. Various models of this kind have been constructed over the last several decades. Therefore, it is also crucial to understand how matter and energy effect gravity and vice versa.

In Chapter \ref{chap:2} we provided a concise introduction to theories of compact objects in the GR framework. The  relativistic correction in the TOV hydrostatic equation   elaborate with help of the SLy nuclear EoS. The total relativistic correction, the radial dependence of the relativistic corrections  $\mathcal{A} \equiv (1-2Gm/r c^2)^{-1}$, $\mathcal{B} \equiv 1+4\pi r^3 P/m c^2$ and $ \mathcal{C} \equiv 1+P/\rho c^2$. It is evident that no part of the star, not even its center, where $m\rightarrow 0$ occurs, possesses a precisely Newtonian gravity. In contrast to this, the total universal relativistic correction $\mathcal{A\;B\;C}$ continually moves closer to unity. This relativistic correction spell out that not only mass but all forms of energy act as a source of gravity in GR.

In the context of spacetime curvature, the curvature tensors $\mathcal{R}$, $\mathcal{J}$, $\mathcal{K}$, and $\mathcal{W}$ were primarily calculated as functions of radial distance in order to estimate the strength of gravity from the spacetime curvature both inside and outside of a neutron star. A typical neutron star has a radius of $10 - 11$ km and a mass of about $\sim 1.4 M_{\odot}$. A typical NS's inside and outside produces curvature  of about $\sim 1.18\times 10^{-12} M_{1.4} R^{-3}_{6} \;\text{cm}^{-2}$. The solar counterpart of this amount is 14 orders of magnitude smaller. These estimations on curvature indicate that relativistic gravity is crucial to characterize neutron stars due to their compactness. It is still questionable how the equation of state or gravity constrain measurements of a neutron star's mass and/or radius. The situation is challenged by the simple fact that the hydrostatic equilibrium equations of the gravity model and the equation characterizing the state existing at the core of neutron stars both have a role in determining the mass and radius of neutron stars. Neutron stars might therefore be used to test alternative or modified gravity theories or to search for general relativity-related deviations. After all, GR is more often regarded as the definitive theory of gravity, even at the deep gravitational well of neutron stars. In our study, we are also looking at the constraints enforced by strong gravity on the structural characteristics of compact stars.

In Chapter \ref{chap:3}, we focused on the computation of properties of compact objects, specifically neutron stars, based on an equation of state  derived from a core-envelope model of superdense stars. By solving Einstein's equations within a pseudo-spheroidal and spherically symmetric space-time geometry and spherical symmetry, we were able to analyze and compare the properties of these compact stars with those obtained using physical nuclear matter equations of state.

We were able to categorize compact stars into three distinct categories by analyzing the mass-radius relationship. The first category consisted of highly compact self-bound stars, characterized by exotic matter compositions, with radii measuring below 9 km. Such stars showcased extraordinary density and curvature of spacetime, indicating the presence of unique and exotic physical conditions.

The second category encompassed normal neutron stars with radii ranging between 9 to 12 km. These stars exhibited properties consistent with typical neutron stars found in astrophysical observations and theoretical models. They served as a reference point for comparison and validation of the computed results.

The third category comprised soft matter neutron stars, characterized by radii between 12 to 20 km. These stars exhibited relatively larger radii, indicating the presence of softer matter compositions compared to the highly compact self-bound stars. The distinction among these categories provides insights into the diversity of compact objects and the wide range of physical conditions that could exist within neutron stars.

Furthermore, we computed additional properties such as the Keplerian frequency, surface gravity, and surface gravitational redshift for all the three types of compact stars. These calculations contributed to a more comprehensive understanding of the physical characteristics and behavior of neutron stars with different matter compositions and sizes.

The outcomes of this study present significant relevance for the studies of exotically composed, highly compact neutron stars. The extensive analysis and classification offered here serve as an outline for deeper research in this field, providing for a more thorough examination of the unique properties and phenomena associated with such extreme dense objects.

We are expanding our understanding of the structure and composition of neutron stars by employing the core-envelope model and taking into account the equation of state obtained from superdense stars. This work elevates our understanding of compact objects while also progressing the study of strange forms of matter in extreme conditions and other aspects of relativistic astrophysics.

Overall, the study of compact objects, especially neutron stars presented in this this, based on the derived equation of state and core-envelope models, expands our knowledge of the rich diversity and physical properties of the fascinating compact bodies. The classification of compact stars into distinct categories and the computation of various properties enhanced our understanding of  compact stars, enabling us to examine deeper into the realms of extreme physics and exotic matter compositions. The findings presented here serve as a foundation for future research and investigations into highly compact neutron stars, pushing the boundaries of our knowledge and assisting new methods and discoveries in the field.

In Chapter \ref{chap:4}, we looked into the impact of local anisotropy and density perturbations on the stability of stellar matter structures within the framework of general relativity. By employing the concept of cracking, we have explored the properties and stability conditions of super-dense stars within the core-envelope model. Specifically, we have introduced anisotropic pressure to the envelope region and examined its impact on the stability of the structures.

Our findings suggest that self-bound compact stars with an anisotropic envelope could serve as potential progenitors for starquakes. By analyzing the difference in sound propagation between the radial and tangential directions, we have identified potentially stable regions within such configurations. It is observed that as the anisotropic parameter increases, strain energy accumulates in the envelope region, making it a candidate for the build-up of a quake-like situation.

We have quantified the stress-energy stored in the envelope region, which would be released during a starquake in a self-bound compact star, as a function of the magnitude of anisotropy at the core-envelope boundary. Our numerical studies on spherically asymmetric compact stars have revealed that the stress-energy can reach magnitudes as high as $10^{50}$ erg when the tangential pressure slightly exceeds the radial pressure. Remarkably, this energy level is comparable to the energy associated with giant $\gamma$-ray bursts.

The outcomes of our research have significant implications for the correlation studies between starquakes and GRBs. By understanding the mechanisms behind starquakes and the associated release of energy, we can establish valuable connections to the phenomena observed in GRBs. This study opens up avenues for future investigations and presents several exciting prospects for further research.

\section*{future scope :}
 Remarkable advancements are taking place in the field of compact star study.
There have been observations of massive neutron stars with masses of more than $2M_{\odot}$. The gravitational wave detectors LIGO and VIRGO have made the first direct observation of gravitational waves generated by the merging of two neutron stars. The ability of computational modelling programmes to predict the formation of neutron stars in core-collapse supernovas and the merging of neutron stars has advanced to a new level of maturity. The microphysical simulation of compact stars has made significant progress in recent years, revealing new information on the dynamics of matter under extreme conditions. In addition to new ground-based facilities like the X-ray satellite eROSITA, the James Webb Space Telescope (JWST), the Square Kilometre Array (SKA), The Neutron Star Interior Composition Explorer (NICER), and the Extremely Large Telescope (ELT), new space-based missions are being developed to explore compact stars in great detail.

\begin{itemize}

\item [\ding{182}] Further refinement of the core-envelope model: This study of core-envelope concept offers an invaluable foundation for studying the properties of superdense stars. Future studies could  focus on improving and expanding this model by incorporating new complexities. Consideration of more realistic compositions, such as hyperonic matter or quark matter, and examination of their effects on the characteristics of compact objects could be included .

\item [\ding{183}] Detailed stability analysis: A more thorough stability analysis can be pursued, even if the present work has provided light on the stability of compact stars based on the derived mass-radius relationship. Future studies could utilize the use of analytical methods or numerical simulations to examine the stability requirements and potential instabilities within various categories of compact stars. The mechanisms governing the stability of highly compact neutron-like stars might become better understood as a result.

\item [\ding{184}] Constraining the Equation of State: The comparison between the properties of compact stars obtained from the derived EoS and those based on nuclear matter equations of state opens route for refining our understanding of the equation of state at extreme densities. Future studies could aim to constrain the EoS further by combining observational data, such as mass and radius measurements from pulsars or gravitational wave events, with theoretical models. This would contribute to a more comprehensive understanding of the dense matter equation of state and its implications for compact stars.

\item[\ding{185}] Probing Observational Signatures: The estimated properties, such as the surface gravity, surface gravitational redshift, and Keplerian frequency, offer the potential to investigate the observable manifestations of various types of compact stars. Future studies can concentrate on determining how these characteristics correspond to observable quantities, such as gravitational wave signatures, X-ray emissions, or pulsar timing. We would be able to verify the accuracy of the predicted mass-radius connection and learn more about the physical composition of compact stars by comparing the predictions with observational data.

    \item[\ding{186}] Extension to other compact stellar models: Further research can explore the stability of different stellar models, such as hybrid stars or strange stars, under the influence of density perturbations and anisotropic pressures. Investigating a broader range of stellar configurations will provide a more comprehensive understanding of the phenomenon.
    \item [\ding{187}] Improved modeling techniques: We can forecast and study the behavior of self-bound compact stars and their envelope stability better if the cracking idea is improved and more advanced modeling methods are developed. A more accurate depiction of the astrophysical systems would be possible by incorporating more elements, like magnetic fields or rotational effects, in the analysis.

    \item [\ding{188}] Observational implications: It is crucial to look into how starquakes affect observations and how they relate to GRBs. Future research might concentrate on developing observational methods and approaches to spot starquake signs such as variations in electromagnetic radiation, GW emission, or other observable events. The nature of such events could be better understood by comparing theoretical predictions with future observations.

\end{itemize}

\cleardoublepage 



\clearpage
\appendix
\captionsetup{list=no}
\chapter{Appendix A :Geometrised System of Units}

\label{App:AppendixA}

The major reason for using a geometrised system of units in general relativity is that it turns out to be fairly natural to represent any quantity in terms of lengths by presenting a geometrical description of gravity. The way this is obtained in practice depends on the number
of quantities that one wants to ``geometrise''. In terms of time, length, and mass geometrisation, the most obvious choice is to convert the speed of light to pure numbers.


\begin{eqnarray}
 c=2.99792458 \times 10^{10} \;\;\text{cm s}^{-1}\;,
\end{eqnarray}
 and the gravitational constant
\begin{eqnarray}
G= 6.67384 \times 10^{-8}\;\; \text{cm$^{3}$ g$^{-1}$ s$^{-2}$}\;.
\end{eqnarray}

This implies that seconds and grams of the``cgs'' system can be written as
 \begin{eqnarray} \label{unit_sec}
     1\; \text{s} = 2.99792458\times 10^{10} \bigg(\frac{1}{c}\bigg) \;\;\text{cm}\; ,
\end{eqnarray}
 
\begin{eqnarray}\label{unit_G}
    1 \;\text{g} = 7.42565\times 10^{-29} \left({\frac{c^{2}}{G}}\right)\;\; \text{cm}\;.
\end{eqnarray}
 
 Within this general setup in which all quantities are converted to lengths, it may still be useful
to introduce a unit. In a gravitational context the most natural choice for such a unit is the
\textit{gravitational radius} $r_{g} \equiv G M /c^{2}$, with $M$ being the mass of the source of the gravitational
field. The mass itself can be expressed in units of the mass of the Sun, $M_{\odot}$, \textit{i.e.},
 
 \begin{eqnarray}
 M= n M_{\odot} \;,
  \end{eqnarray}
 with $n$ a real number. From the physical measurement of the solar mass, \textit{i.e.},
 \begin{eqnarray}
     M_{\odot}=1.9884\times 10^{33}\; \text{g}\;,
 \end{eqnarray}
 and from expressions (\ref{unit_sec})-(\ref{unit_G}), it is possible to find the relation between the cgs units and
the new unit of length $r_{g}$, \textit{i.e.},

\begin{eqnarray}
 1 \;\text{cm} = 6.77269\times 10^{-6} \left({\frac{M_{\odot}}{M}}\right)\;\; r_{g}\;, \\
 1 \;\text{s} = 2.03040\times 10^{5}\; \left( \frac{1}{c}\right) \left({\frac{M_{\odot}}{M}}\right)\;\; r_{g}\;, \\
  1 \;\text{g} = 5.02916\times 10^{-34}\; \left( \frac{c^{2}}{G}\right) \left({\frac{M_{\odot}}{M}}\right)\;\; r_{g}\;.
\end{eqnarray}

 It is also useful to write explicitly the conversion between the cgs and \textit{geometrised} systems
of units for a generic length $d$, a time $t$, a mass $m$, a frequency $f$, a velocity $v$, a rest-mass
density $\rho$, a pressures $p$, and a luminosity $L$, \textit{i.e.},

\begin{eqnarray}
d_{\text{cgs}}= 1.47651\times 10^{5} \left({\frac{M}{M_{\odot}}}\right)\; d_{\text{geo}}\;, \\
t_{\text{cgs}}= 4.92513\times 10^{-6} c  \left({\frac{M}{M_{\odot}}}\right)\; t_{\text{geo}}\;, \\
 m_{\text{cgs}}= 1.9884\times 10^{33}  \left(\frac{G}{c^{2}}\right) \left({\frac{M}{M_{\odot}}}\right)\; m_{\text{geo}}\;, \\
 f_{\text{cgs}}= 2.03040\times 10^{5}  \left(\frac{1}{c}\right) \left(\frac{M_{\odot}}{M}\right)\; f_{\text{geo}}\;, \\
v_{\text{cgs}}= 2.99792458\times 10^{10}  \left(\frac{1}{c}\right) \; v_{\text{geo}}\;, \\
   L_{\text{cgs}}= 3.62849\times 10^{59} \left(\frac{G}{c^{5}}\right) \; L_{\text{geo}}\;,
 \end{eqnarray}

 \begin{eqnarray}
   \rho_{\text{cgs}}= 6.17714\times 10^{17} \left(\frac{G}{c^{2}}\right) \left(\frac{M_{\odot}}{M}\right)^{2}\; \rho_{\text{geo}}\;, \\
    p_{\text{cgs}}= 5.55173\times 10^{38} \left(\frac{G}{c^{4}}\right) \left(\frac{M_{\odot}}{M}\right)^{2}\; p_{\text{geo}}\;.
   \end{eqnarray} 
 
where the indices ``cgs'' and ``geo'' refer to the pure numbers expressing the generic quantity
in the cgs system and in the geometrised system, respectively \par

It is customary to express all quantities in kilometres, $\text{km} = 10^{3} \text{m}$. The following
conversion factors are useful
\begin{eqnarray}
1\text{ s} = 2.99792458 \times 10^{5} \;  \text{km} \\
1\text{ kg} = 7.425648 \times 10^{-31} \;  \text{km} \\
1\text{ g cm}^{-3} = 7.425648 \times 10^{-19} \;  \text{km}^{-2} \\
1\text{ dyne cm}^{-2} = 8.262148\times 10^{-40} \;  \text{km}^{-2} 
\end{eqnarray}

As a final remark we note that, in the traditional geometrised system, $c$ and $G$ are set equal
to unity. However, this is just one of the possible choices and for specific physical applications,
\textit{i.e.}, where very low rest-mass densities are encountered, the corresponding value of $\rho_{\text{geo}}$ may
become excessively small to be used in numerical calculations. In these cases, it is more
convenient to assume a smaller value for the gravitational constant, \textit{i.e.}, $ G = 10^{-10}$.
\clearpage
\chapter{Appendix B :The curvature tensors  of the spherical symmetric spacetime}

\label{App:AppendixB}
The non-vanishing components of $\Gamma^{\rho}_{\mu\nu}$, $\mathcal{R}_{\mu \nu}$, $\mathcal{R}$ and $G_{\mu \nu}$ for the most general static spherical symmetric metric
\begin{equation} \label{appex_schw}
  \boxed{   \mathrm{d}s^{2}= -e^{\lambda(r)}\mathrm{d}r^{2}-r^{2}(\mathrm{d}\theta^{2}+\mathrm{sin}^{2}\theta \;\mathrm{d}\phi^{2})+e^{\nu(r)}\mathrm{d}t^{2}}
\end{equation}
where $\lambda$ and $\nu$ are functions of $r$ only.
 
Here
 \begin{align*}
x^{1}&=r,                            &  x^{2}&=\theta,              &  x^{3}&=\phi,            &x^{0}&=t \\
 g_{11}&=-e^{\lambda},              &g _{22}&=-r^{2} ,             &  g_{33}&=  -r^{2} \sin^{2}\theta,                     & g_{00}&=e^{\nu}\\
g^{11}&=-e^{-\lambda},          &  g^{22}&=- \frac{1}{r^2} ,                        & g^{33}&-\frac{1}{r^{2}\sin^{2}\theta},   & g^{00}&=e^{-\nu} \\
\text{det} g_{\mu \nu} &= -e^{(\lambda+\nu)} r^{4}\sin^{2} \theta
\end{align*}

The Christofell symbols $\Gamma^{\rho}_{\mu\nu}$ can be found from the metric $g_{\mu \nu }$ using
\begin{equation} \label{appendix_chris}
\boxed{\Gamma^{\rho}_{\mu\nu}=\frac{1}{2} g^{\rho \lambda} (\partial_{\mu}g_{\nu \lambda} +\partial_{\nu} g_{\mu \lambda}  - \partial_{\lambda} g_{\mu \nu}  )           }
\end{equation}
using  (\ref{appendix_chris}), all the non-vanishing components of Christoffel symbol of second kind are obtain as:
\begin{table}[htb]
\begin{center}
 \begin{tabular}{c c c c c c c c c c } 
 \hline
 $\Gamma^{t}_{tr}$  & $\Gamma^{r}_{tt}$  &$\Gamma^{r}_{rr}$& $\Gamma^{r}_{\theta \theta}$  & $\Gamma^{r}_{\phi \phi}$      & $\Gamma^{\theta}_{r \theta}$  & $\Gamma^{\theta}_{\phi \phi}$ & $\Gamma^{\phi}_{r\phi}$  & $\Gamma^{\phi}_{\theta \phi}$ \\ [1ex] 
 \hline
 $\frac{\nu'}{2}$      & $\frac{1}{2}e^{\nu-\lambda} \nu'$          & $\frac{1}{2}\lambda'$   &  $r e^{-\lambda} $  & $e^{-\lambda} r \sin ^2 \theta$  & $\frac{1}{r}$  & $-\cos \theta \sin \theta$  & $\frac{1}{r}$   & $\cot \theta $
  \\ [1.2ex] 
 \hline
  
\end{tabular}
\captionsetup{
	justification=raggedright,
	singlelinecheck=false
}
\caption{Christoffel symbols of the diagonal metric (\ref{appex_schw}).}
\label{table:appex_chris}
\end{center}
\end{table}

The Ricci tensor $\mathcal{R}_{\mu \nu}$ is obtained by contracting the Riemann curvature tensor $\mathcal{R}^{\sigma}_{\mu\lambda\nu}$ with an equal lower and upper index, such that
\begin{equation}
    \boxed{\mathcal{R}_{\mu \nu}= \mathcal{R}^{\sigma}_{\mu\lambda\nu} = ( \partial_{\sigma} \Gamma^{\sigma}_{\mu \nu} +  \Gamma^{\sigma}_{\kappa \sigma } \Gamma^{\kappa}_{\mu \nu}) -
    (\partial_{\nu} \Gamma^{\sigma}_{\mu \sigma} +  \Gamma^{\sigma}_{\kappa \nu } \Gamma^{\kappa}_{\mu \sigma}) }
\end{equation}

which leads to

\begin{align*}
    \mathcal{ R}_{rr}&= \frac{\partial \Gamma^{1}_{01}}{\partial x^{0}} +  \frac{\partial \Gamma^{2}_{02}}{\partial x^{0}} + \frac{\partial \Gamma^{3}_{03}}{\partial x^{0}} - \Gamma^{0}_{00}\Gamma^{1}_{01}
   - \Gamma^{0}_{00}\Gamma^{2}_{02} -\Gamma^{0}_{00}\Gamma^{3}_{03}
   - \Gamma^{1}_{01}\Gamma^{1}_{10} \\
   & + \Gamma^{2}_{02}\Gamma^{2}_{20}
   - \Gamma^{3}_{03}\Gamma^{3}_{30} \\
   & = \frac{\partial}{\partial r} \big(\frac{1}{r}\big)+\frac{\partial}{\partial r} \big(\frac{1}{r}\big) +\frac{\partial}{\partial r} \big(\frac{1}{2}\nu'\big) - \frac{1}{2}\lambda^{'}\frac{1}{r}- \frac{1}{2}\lambda^{'}\frac{1}{r} \\
   & - \frac{1}{2}\lambda^{'}\frac{1}{2} \nu'- \frac{1}{r^2}\frac{1}{r^2}+\frac{1}{4}\nu^{'2} \\
   & - \frac{1}{r^2} - \frac{1}{r^2} + \frac{\nu ''}{2} -\frac{\lambda^{'}}{2r}-\frac{\lambda^{'}}{2r}-\frac{1}{4}\lambda^{'} \nu{'}+ \frac{2}{r^{2}}+ \frac{1}{4}\nu^{'2} \\
   & = \frac{\nu^{''}}{2}- \frac{\lambda^{'}\nu^{'}}{4}+\frac{\nu^{'2}}{4}-\frac{\lambda^{'}}{r}
\end{align*}
In a similar way, we have
 \begin{align*}
    \mathcal{ R}_{\theta \theta}&= e^{-\lambda} \big[ 1-\frac{r}{2}(\lambda^{'}-\nu^{'})\big] -1 \\
     \mathcal{ R}_{\phi \phi}&= \big[e^{-\lambda} \big[ 1-\frac{r}{2}(\lambda^{'}-\nu^{'})\big] -1 \big]\sin^{2}\theta \\
     & =\mathcal{ R}_{\theta \theta}\sin^{2}\theta \\
     \mathcal{ R}_{tt}&=e^{\nu-\lambda} \left[ -\frac{\nu''}{2}+\frac{\nu'^{2}}{2} -\frac{\nu ' \lambda '}{2} - \frac{\nu'\lambda '}{4}-\frac{\nu'}{r}-\frac{\nu'^{2}}{4}
     \right]
 \end{align*}

The definition of Einstein tensor leads to

\begin{align*}
  G_{rr}&=\mathcal{R}_{rr}-\frac{1}{2}g_{rr}\mathcal{ R} \\
  & = \left( \frac{\nu''}{2} - \frac{\lambda' \nu '}{4} +  \frac{\nu '^{2}}{4}  -- \frac{\lambda'}{r}\right) \\
 &  -\frac{1}{2} \left[-e^{\lambda}  \left \{ -e^{\lambda }   \left (  \nu'' - \frac{\lambda' \nu '}{2}+\frac{\nu'^{2}}{2}-\frac{\lambda'}{r}+\frac{\nu'}{r} 
  \right)
 - \frac{2}{r^{2}} \left( e^{-\lambda} \left[1-\frac{r}{2} (\lambda'-\nu ')\right] -1 \right) \right\} \right]  \\
 & = \frac{\nu''}{2} -\frac{\lambda' \nu '}{4}+\frac{\nu'^{2}}{4}-\frac{\lambda'}{r}-\frac{1}{2} \left( \nu'' - \frac{\lambda' \nu '}{2} +\frac{\nu'^{2}}{2} -\frac{\lambda'}{r}-\frac{\nu '}{r}  \right) \\
 & - \frac{e^{\lambda}}{r^{2}} \left(  e^{-\lambda} \left[ 1-\frac{r}{2}(\lambda'-\nu') \right]  -1 \right) \\
  & = -\frac{\lambda'}{2r}+\frac{\nu'}{r}-\frac{1}{r^{2}} \left(  1-\frac{r}{2}(\lambda'-\nu') \right)+\frac{e^{\lambda}}{r^{2}}
 \end{align*}

\begin{align*}
    G_{\theta \theta} & = \left( e^{-\lambda} \left[ 1-\frac{r}{2}(\lambda'-\nu')        \right]-1   \right) \\
    & + \frac{1}{2}r^{2} \left[ -e^{-\lambda} \left(  \nu'' - \frac{\lambda' \nu '}{2} +\frac{\nu'^{2}}{2} -\frac{\lambda'}{r}+\frac{\nu '}{r}  \right) \frac{2}{r^{2}} \left(  e^{-\lambda} \left[  1-\frac{r}{2}(\lambda'-\nu')
    \right]
    -1\right)
    \right]
\end{align*}

\begin{align*}
    G_{\phi \phi} & = \left( e^{-\lambda} \left[ 1-\frac{r}{2}(\lambda'-\nu')        \right]-1   \right) \sin ^{2}\theta \\
    & + \frac{1}{2}r^{2}  \sin ^{2}\theta \left[ -e^{-\lambda} \left(  \nu'' - \frac{\lambda' \nu '}{2} +\frac{\nu'^{2}}{2} +\frac{\nu '}{r}  \right) \frac{2}{r^{2}} \left(  e^{-\lambda} \left[  1-\frac{r}{2}(\lambda'-\nu')
    \right]
    -1\right)
    \right]
\end{align*}

 \begin{align*}
     G_{tt} & = e^{\nu-\lambda} \left[- \frac{\nu'}{2} +\frac{\lambda' \nu'}{r} -\frac{\nu'^{2}}{4}-\frac{\nu'}{r} \right] \\
     & - \frac{1}{2}e^{\nu} \left[ -e^{-\lambda} \left(    \nu'' - \frac{\lambda' \nu '}{2} +\frac{\nu'^{2}}{2} -\frac{\lambda'}{r}+\frac{\nu '}{r} \right)\right] \\
     & + \frac{1}{2}e^{\nu} \left[ \frac{2}{r^{2}} \left(  e^{-\lambda} \left[  1-\frac{r}{2}(\lambda'-\nu')
    \right]
    -1\right)            \right]
 \end{align*}
\clearpage  
\chapter{Appendix C : Curvature scalars in the Compact star}

\label{App:AppendixC}
We derive the curvature scalars inside a spherically symmetric mass distribution in this appendix. With relation to pressure $P = P(r)$, density $\rho=\rho(r)$, and mass $m = m(r)$ within the radial coordinate $r$, we specifically calculate the Ricci scalar $\mathcal{R}$, the full contraction of the Ricci tensor $\mathcal{J}^2\equiv \mathcal{R}^{\mu \nu} \mathcal{R}_{\mu \nu}$, the full contraction of the Riemann tensor (specifically the Kretschmann scalar) $\mathcal{K}^{2}\equiv \mathcal{R}^{\mu \nu \alpha \beta}  \mathcal{R}_{\mu \nu \alpha \beta}$, and the full contraction of the Weyl tensor $\mathcal{W}^2\equiv \mathcal{C}^{\mu \nu \alpha \beta} \mathcal{C}_{\mu \nu \alpha \beta}$

We choose the coordinates as $t, r, \theta, \phi$, respectively. The most general spherically symmetric metric is given as in (\ref{schw_met_chap2})

Here 
\begin{eqnarray}\label{app3_2}
g_{11}= -e^{\lambda}= \left(1 -\frac{2 G m}{c^{2}r}  \right)^{-1}
\end{eqnarray}
but we refrain from using it until the end. 

The Einstein tensor with the mixed components. The mixed components of the Einstein tensor are obtained from the metric as 
\begin{eqnarray}
    G^{\mu}_{\nu}\equiv \mathcal{R}^{\mu}_{\nu} -\frac{1}{2}\delta^{\mu}_{\nu}\mathcal{R}
\end{eqnarray}

where $\mathcal{R}^{\mu}_{\nu}$ is Ricci tensor with mixed components. The mixed components of the Einstein tensor are obtained from the metric

\begin{eqnarray}\label{app3_G00}
    G^{0}_{0} = -\frac{e^{\lambda}-1+r \lambda'}{r^2 e^{\lambda}},
    \end{eqnarray}
   \begin{eqnarray} \label{app3_G11}
         G^{1}_{1} =  -\frac{e^{\lambda}-1-r \nu'}{r^2 e^{\nu}}, 
    \end{eqnarray}
 \begin{eqnarray}\label{app3_G22}
     G^{2}_{2} = -\frac{2(\lambda'-\nu')-(2\lambda''-\nu' \lambda'+(\lambda')^{2})r}
      {4 r e^{\lambda}},
 \end{eqnarray}
 \begin{eqnarray}
     G^{3}_{3}=G^{2}_{2}.
 \end{eqnarray}
From the metric given in (\ref{schw_met_chap2}) we derive the Ricci scalar as

\begin{eqnarray}\label{app3_ric_scal}
    \mathcal{R}= \frac{[ 4 r \lambda' - 2\nu'' + \lambda' \nu '- (\nu')^{2}] r^{2}+ 4(e^{\lambda}-1-r \nu') }
    {2r^2e^{\lambda}}.
\end{eqnarray}
the full contraction of the Ricci tensor, $\mathcal{J}^2\equiv \mathcal{R}^{\mu \nu} \mathcal{R}_{\mu \nu}$ as

\begin{align} \label{app3_J}
     \mathcal{J}^{2} &= \frac{1}{8 r^4 e^{2\lambda}} \biggl\{ 16 (1-e^{\lambda})^{2}+16r(\nu'-\lambda')(1-e^{\lambda})+ 4 r^2 [3(\nu')^2+3(\lambda')^2-2 \lambda' \nu'] \nonumber  \\
 &+4r^3\bigl[ 2\nu'' \nu ' +(\nu')^3 - 2\nu '' \lambda' +(\lambda')^2 \nu' -2(\nu')^2 \lambda' \bigr] \nonumber \\
 &+ 4r^4 \bigl[ 4\nu'' (\nu')^2 + (\nu')^2 (\lambda')^2 - 2\lambda'(\nu')^3+ 4(\nu'')^2+ (\nu')^4 -4 \nu'' \nu' \lambda'\bigr]\biggr\}
 \end{align}

the Kretschmann scalar,  $\mathcal{K}^{2}\equiv \mathcal{R}^{\mu \nu \alpha \beta}  \mathcal{R}_{\mu \nu \alpha \beta}$ (where $\mathcal{R}_{\mu \nu \alpha \beta}$ is Riemann curvature tensor), as

\begin{align} \label{app3_krets}
     \mathcal{K}^{2} &= \frac{1}{4 r^2 e^{2 \lambda}} \biggl \{ 16(1-e^{\lambda})^2 +8 r^2 [(\nu')^2+(\lambda')^2 ]
     \nonumber  \\
 & +r^4 \bigl[   4 (\nu'')^2 + 4 \nu'' (\nu')^2 - 4 \nu'' \nu' \lambda' +(\nu')^4- 2 \lambda' (\nu')^3 +(\nu')^2 (\lambda')^2
 \bigr]
 \biggr\}
 \end{align}

\cite{bronnikov2003possible}, and the full contraction of the Weyl tensor,  $\mathcal{W}^2\equiv \mathcal{C}^{\mu \nu \alpha \beta} \mathcal{C}_{\mu \nu \alpha \beta}$, as

\begin{eqnarray} \label{app3_weyl}
    \mathcal{W}^2= \frac{1}{3} \left(\frac{ r^2[-2 \nu'' -(\nu')^2 + \nu' \lambda'] +2r (\nu' \lambda')+4(e^\lambda-1)  }
    {2r^2 e^{\lambda}}
    \right)^2
\end{eqnarray}
we solve $\nu' , \nu'',$ and $\lambda'$ from Eqs. (\ref{app3_G00}), (\ref{app3_G11}) and (\ref{app3_G22}) to obtain
 
 \begin{align} 
      \nu' & = \frac{G^{1}_{1} r^2 e^{\lambda}+e^{\lambda}-1}{r}\label{app3_nu}, \\
     \lambda' & = \frac{G^{0}_{0} r^2 e^{\lambda}+e^{\lambda}-1}{r},\\
     \nu'' &= \frac{2(G^2_2  + G^3_3)r e^\lambda+(2+r \nu') (\lambda'-\nu') } 
     {2r}, \nonumber \\
     &=- \frac{r^2 e^\lambda (G^0_0 -G^1_1- 2G^2_2 -2 G^3_3)+ r^2 e^{\lambda} (3 G^1_1+G^0_0)+2 (e^{2\lambda}-1)+ r^4 e^{2\lambda}G^1_1 (G^0_0+G^1_1)   }
     {2r^2} \label{app3_lambda}.
\end{align}
 
where in the last step we used $\lambda'-\nu'$ obtained from the first two equations. Plugging these into Eq. (\ref{app3_ric_scal}) we obtain
the Ricci scalar in terms of the components of the Einstein tensor as

\begin{align}\label{app3_15}
    \mathcal{R} &=- G^0_0 -G^1_1-G^2_2-G^3_3 \nonumber \\
    &=-G^\mu_\mu.
\end{align}

Similarly, by plugging Eqs. (\ref{app3_nu})-(\ref{app3_lambda}) into Eq. (\ref{app3_J}) we obtain

\begin{align}\label{app3_16}
 \mathcal{R}_{\mu\nu} \mathcal{R}^{\mu\nu} = (G^0_0)^2 + (G^1_1)^2 +(G^2_2)^2+(G^3_3)^2.
  \end{align}

Again, by plugging Eqs. (\ref{app3_nu})-(\ref{app3_lambda}) into Eq. (\ref{app3_krets}) the Kretschmann scalar can be expressed as
\begin{align} \label{app3_17}
    \mathcal{K}^2&= 2 G^0_0G^1_1- 2G^0_0G^2_2-2G^0_0G^3_3 -2G^2_2G^1_1-2G^2_2G^3_3+3(G^0_0)^2+3(G^1_1)^2 \\
    & +2(G^3_3)^2 + 8\left(G^0_0 +G^1_1-\frac{1}{2}(G^2_2+G^3_3)\right)(1-e^{-\lambda})r^2+12(1-e^{\lambda})^2 e^{-2\lambda}
\end{align}

Finally, by plugging Eqs. (\ref{app3_nu})-(\ref{app3_lambda}) into Eq. (\ref{app3_weyl}) the full contraction of the Weyl tensor becomes

\begin{eqnarray}
    \mathcal{W}^2= \frac{4}{3}\left[G^0_0 +G^1_1 -\frac{1}{2} (G^2_2+G^3_3)+ \frac{3(e^{\lambda}-1)}{r^2 e^\lambda}       \right]^2.
\end{eqnarray}

The Einstein field equation with mixed components is 
\begin{eqnarray}\label{app3_19}
     G^\mu_\nu = \kappa T^\mu_\nu,
\end{eqnarray}
where $T^\mu_\nu$ are the mixed components of the energy-momentum tensor and $\kappa \equiv 8\pi G/c^4$. We assume that the energy-momentum
tensor is that of a perfect fluid,
\begin{eqnarray}
    T_{\mu\nu}=(\rho c^2+P)u_\mu u_\nu+Pg_{\mu\nu},
\end{eqnarray}
Contracting the $T_{\mu\nu}$ with $g^{\mu\nu}$, the mixed components
of the energy-momentum tensor can be written as
\begin{eqnarray}\label{app3_21}
    T^{\mu}_{\nu}=(\rho c^2+P)u^\mu u_\nu+ \delta^{\mu}_{\nu} P.
\end{eqnarray}

Thus $T^0_0=\rho c^2$ and $T^i_i=-P$. Using $u^\mu u_\mu=-1$ and $\delta^\mu _\mu=4$, the trace of the energy-momentum equation becomes

\begin{eqnarray}\label{app3_22}
    T^\mu _\mu =-(\rho c^2-3P).
\end{eqnarray}
Equation (\ref{app3_15}) and the Einstein equation (\ref{app3_19}) imply $\mathcal{R}=-\kappa T^\mu_\mu$,  and by referring further of Eq. (\ref{app3_22}) the Ricci scalar is obtained as 
\begin{eqnarray}
    \mathcal{R}=\kappa(\rho c^2-3P)
\end{eqnarray}
in terms of  the density $\rho$ and pressure $P$ within the star.  Similarly, the full contraction of the Ricci tensor in Eq. (\ref{app3_16}) becomes

\begin{eqnarray}
    \mathcal{R}_{\mu\nu} \mathcal{R}^{\mu\nu} = \kappa^2 [(\rho c^2)^2+3P^2].
\end{eqnarray}

We obtain the Kretschmann scalar provided in Eq. (\ref{app3_17}) by using the elements of the mixed energy-momentum tensor given in Eq. (\ref{app3_21}) and $g_{11}$ given in Eq. (\ref{app3_2}).

\begin{eqnarray}
    \mathcal{K}^2= \kappa^2 \left[  3(\rho c^2)^2 +3P^2+2P \rho c^2\right] - \kappa \frac{16 G m}{r^3 c^2} \rho c^2 + \frac{48G^2 m^2}{r^6c^4},
\end{eqnarray}

as well as the full contraction of the Weyl tensor provided in Eq. (\ref{app3_19})

\begin{eqnarray}
    \mathcal{W}^2= \frac{4}{3} \left( \frac{6 G m}{c^2 r^3} -\kappa \rho c^2\right)^2 .
\end{eqnarray}

Note that at the star's surface, $\rho = 0$, $P = 0$, and $m = M$, resulting in $\mathcal{K}^2=\mathcal{W}^2=48G^2 M^2/c^4r^6$ outside the star, a well-known Schwarzschild metric outcome. As a supplementary check, observe that they satisfy:

\begin{eqnarray}
    \mathcal{K}^2=\mathcal{W}^2+2 \mathcal{J}^2-\frac{1}{3}\mathcal{R}
\end{eqnarray}

\clearpage








\end{document}